\documentclass[iop]{emulateapj}
\usepackage{apjfonts}
\usepackage{natbib}
\usepackage{subfigure}
\usepackage{booktabs}

\newcommand{\xmm}{\hbox{\it XMM-Newton\/}}
\newcommand{\chandra}{{\it Chandra\/}}

\newcommand{\athena}{{\it Athena\/}}
\newcommand{\cdfs}{{\hbox{CDF-S}\/}}
\newcommand{\flux}{{erg~cm$^{-2}$~s$^{-1}$}}

\newcommand{\lum}{{erg~s$^{-1}$}}

\newcommand{\asca}{{\it ASCA\/}}

\newcommand{\xray}{{\hbox{X-ray}}}

\usepackage{amsmath}

\begin{document}
\title{The {{\em Chandra}} Deep Field-South Survey: 7 Ms Source Catalogs}
\author{
B.~Luo,\altaffilmark{1,2,3,4}
W.~N.~Brandt,\altaffilmark{4,5,6}
Y.~Q.~Xue,\altaffilmark{7}
B.~Lehmer,\altaffilmark{8}
D.~M.~Alexander,\altaffilmark{9}
F.~E.~Bauer,\altaffilmark{10,11,12}
F.~Vito,\altaffilmark{4,5}
G.~Yang,\altaffilmark{4,5}
A.~R.~Basu-Zych,\altaffilmark{13,14}
A.~Comastri,\altaffilmark{15}
R.~Gilli,\altaffilmark{15}
Q.-S.~Gu,\altaffilmark{1,2,3}
A.~E.~Hornschemeier,\altaffilmark{13}
A.~Koekemoer,\altaffilmark{16}
T.~Liu,\altaffilmark{7}
V.~Mainieri,\altaffilmark{17}
M.~Paolillo,\altaffilmark{18,19,20}
P.~Ranalli,\altaffilmark{21}
P.~Rosati,\altaffilmark{22}
D.~P.~Schneider,\altaffilmark{4,5}
O.~Shemmer,\altaffilmark{23}
I.~Smail,\altaffilmark{9}
M.~Sun,\altaffilmark{7}
P.~Tozzi,\altaffilmark{24}
C.~Vignali,\altaffilmark{25,15}
and J.-X.~Wang\altaffilmark{7}
}
\altaffiltext{1}{School of Astronomy and Space Science, Nanjing University, 
Nanjing, Jiangsu 210093, China}
\altaffiltext{2}{Key Laboratory of Modern Astronomy and Astrophysics 
(Nanjing University), Ministry of Education, Nanjing, Jiangsu 210093, China}
\altaffiltext{3}{Collaborative Innovation Center of Modern Astronomy and Space Exploration, 
Nanjing, Jiangsu 210093, China}
\altaffiltext{4}{Department of Astronomy \& Astrophysics, 525 Davey Lab,
The Pennsylvania State University, University Park, PA 16802, USA}
\altaffiltext{5}{Institute for Gravitation and the Cosmos,
The Pennsylvania State University, University Park, PA 16802, USA}
\altaffiltext{6}{Department of Physics, 104 Davey Lab,
The Pennsylvania State University, University Park, PA 16802, USA}
\altaffiltext{7}{CAS Key Laboratory for Research in
Galaxies and Cosmology, Department of Astronomy,
University of Science and Technology of China, Hefei, Anhui 230026, China}
\altaffiltext{8}{Department of Physics, University of Arkansas, 226 Physics
Building, 835 West Dickson Street, Fayetteville, AR 72701, USA}
\altaffiltext{9}{Centre for Extragalactic Astronomy, Department of Physics, Durham University, Durham DH1 3LE, UK}
\altaffiltext{10}{Instituto de Astrof\'{\i}sica, Facultad de F\'{i}sica, Pontificia Universidad Cat\'{o}lica de Chile, 306, Santiago 22, Chile}
\altaffiltext{11}{Millennium Institute of Astrophysics (MAS), Nuncio Monse{\~{n}}or S{\'{o}}tero Sanz 100, Providencia, Santiago, Chile}
\altaffiltext{12}{Space Science Institute, 4750 Walnut Street, Suite
205, Boulder, CO 80301, USA}
\altaffiltext{13}{NASA Goddard Space Flight Center, Code 662, Greenbelt, MD 20771, USA}
\altaffiltext{14}{Department of Physics, University of Maryland Baltimore County, Baltimore, MD 21250, USA}
\altaffiltext{15}{INAF -- Osservatorio Astronomico di Bologna, via Ranzani 1, 40127 Bologna, Italy}
\altaffiltext{16}{Space Telescope Science Institute, 3700 San Martin Drive,
Baltimore, MD 21218, USA}
\altaffiltext{17}{European Southern Observatory, Karl-Schwarzschild-Strasse 2, 85748, Garching bei M\"unchen, Germany}
\altaffiltext{18}{Dip.di Fisica Ettore Pancini, University of Naples
``Federico II'', C.U. Monte S. Angelo, Via Cintia, 80126, Naples, Italy}
\altaffiltext{19}{INFN - Sez.di Napoli, Via Cintia, 80126, Naples, Italy}
\altaffiltext{20}{Agenzia Spaziale Italiana - Science Data Center, Via del Politecnico 
snc, 00133 Roma, Italy}
\altaffiltext{21}{Lund Observatory, Department of Astronomy and Theoretical Physics, Lund University, Box 43, SE-22100 Lund, Sweden}
\altaffiltext{22}{Dipartimento di Fisica e Scienze della Terra, Universit\`a degli Studi di Ferrara, Via Saragat 1, I-44122 Ferrara, Italy}
\altaffiltext{23}{Department of Physics, University of North Texas, Denton, TX 76203, USA}
\altaffiltext{24}{INAF -- Osservatorio Astrofisico di Arcetri, Largo
E. Fermi 5, I-50125, Florence, Italy}
\altaffiltext{25}{Dipartimento di Fisica e Astronomia, Alma Mater Studiorum, Universit\`a degli Studi di Bologna, Viale Berti Pichat 6/2, 40127 Bologna, Italy}

\begin{abstract}
We present X-ray source catalogs for the $\approx7$~Ms exposure of
the \chandra\ Deep Field-South (\hbox{CDF-S}), which covers a total area of
484.2~arcmin$^2$. Utilizing {\sc wavdetect} for initial source detection and 
ACIS Extract for photometric extraction and significance assessment, we 
create a main source catalog containing 1008 sources that are detected in 
up to three X-ray bands: 0.5--7.0~keV, 0.5--2.0~keV, and 2--7~keV.
A supplementary source catalog is also provided including 47 lower-significance sources
that have bright ($K_s\le23$) near-infrared counterparts. 
We identify multiwavelength counterparts for 992 (98.4\%) of the main-catalog sources, and 
we collect redshifts for 986 of these sources, including 
653 spectroscopic redshifts and 333 photometric redshifts.
Based on the X-ray and multiwavelength properties, we identify 711 active galactic nuclei (AGNs) 
from the \hbox{main-catalog} 
sources. 
Compared to the previous $\approx4$~Ms \cdfs\ catalogs, 291 of the \hbox{main-catalog} sources are
new detections. 
We have achieved unprecedented X-ray sensitivity with
average flux limits over the central $\approx1$~arcmin$^2$ region
of $\approx1.9\times10^{-17}$, $6.4\times10^{-18}$, and $2.7\times10^{-17}$~\flux\
in the three X-ray bands, respectively.
We provide cumulative number-count measurements observing,
for the first time, that normal galaxies start to
dominate the X-ray source population
at the faintest 0.5--2.0~keV flux levels.
The highest X-ray source density reaches $\approx50\,500$~deg$^{-2}$, and 
$47\%\pm4\%$ of these sources are AGNs ($\approx23\,900$~deg$^{-2}$).
\end{abstract}

\section{INTRODUCTION}
Cosmic \xray\ surveys of the distant universe have made enormous advances over
the past two decades, since the launches of the
{\it Chandra \xray\ Observatory\/} (\chandra; e.g., \citealt{Weisskopf2000}) and
{\it \xray\ Multi-Mirror Mission\/} (\xmm; e.g., \citealt{Jansen2001}).
These surveys are a primary source of information about accreting supermassive
black holes (SMBHs), i.e., active galactic nuclei (AGNs), in the Universe,
providing insights about their demographics, physical properties, and
interactions with their environments (e.g., \citealt{Brandt2015} and
references therein). Furthermore, \xray\ surveys are an essential tool for
the study of clusters and groups (e.g., \citealt{Allen2011a} and references
therein) as well as \xray\ binary populations in starburst and normal galaxies
(e.g., \citealt{Mineo2014}; \citealt{Lehmer2016}; and references therein).
\xray\ surveys with a wide variety of sensitivities and solid angles are
required to gain a comprehensive understanding of \xray\ source populations
in the Universe (e.g., see Section 2.1 of \citealt{Brandt2015}). Such
surveys range from shallow, all-sky surveys, the lowest tier of
the \xray\ surveys ``wedding cake'', to the highest tier of ultradeep,
pencil-beam surveys. Ultradeep \xray\ surveys are particularly
important as cosmic ``time machines'', since fainter \xray\ sources of a
given type generally lie at higher redshifts and thus earlier cosmic epochs.
Furthermore, at a given redshift, such ultradeep surveys are capable of
probing objects with lower observable \xray\ luminosities that are generally
more typical members of source populations. Additionally, some intrinsically
luminous \xray\ sources may have low observable \xray\ luminosities owing
to strong obscuration (e.g., Compton-thick AGNs), and ultradeep \xray\
surveys are one of the key ways of identifying and characterizing such
important sources (e.g., see Section 3.3 of \citealt{Brandt2015}).

The deepest \xray\ surveys to date have been conducted in the \chandra\
Deep Field-South (\cdfs), which is arguably the most intensively
studied multiwavelength \hbox{deep-survey} region across the entire sky.
Currently, published \cdfs\ \xray\ catalogs exist for the 4~Ms \chandra\
exposure (covering 465~arcmin$^2$; e.g., \citealt{Xue2011}) and the
3~Ms \xmm\ exposure (covering 830~arcmin$^2$; e.g., \citealt{Ranalli2013}).
In 2013, we proposed to extend the 4~Ms \cdfs\ observations to a
total \chandra\ exposure of 7~Ms, and the new observations were conducted
during \hbox{2014--2016}. The very small \chandra\ point
spread function (PSF) and low background still allow for significant
gains in sensitivity near the field center, and thus the detection
of many new sources, even for such long exposures. Furthermore, all
previously detected \cdfs\ sources benefit greatly from the improved photon
statistics, which allow better measurements of \xray\ positions,
photometric properties, spectral properties, and variability; variability
studies also benefit from the significantly lengthened time baseline
of sensitive \cdfs\ \xray\ observations \citep[e.g.,][]{Yang2016}.
These better X-ray measurements advance significantly
the physical understanding of the sources producing
most of cosmic accretion power; e.g., a typical AGN in
the \cdfs\ will have $\approx 40$ times more counts than
the same AGN in the COSMOS Legacy Survey \citep[e.g.,][]{Civano2016}.

In this paper, we will present \cdfs\ point-source catalogs derived from the
7~Ms \chandra\ exposure for use by the community in advancing \xray\ deep-survey
science projects; previous \chandra\ Deep Fields catalogs of this type have
been widely utilized (e.g., \citealt{Alexander2003}; \citealt{Luo2008};
\citealt{Xue2011,Xue2016}). We will also present multiwavelength identifications,
basic multiwavelength photometry, and spectroscopic/photometric redshifts
for the detected \xray\ sources. \chandra\ source-cataloging methodology
has advanced greatly over the years since the \chandra\ launch, providing
substantially improved yields of demonstrably reliable sources
(e.g., \citealt{Xue2016}) and improved source characterization. Here we will
utilize ACIS Extract (AE; \citealt{Broos2010})\footnote{See
http://www.astro.psu.edu/xray/docs/TARA/ae\_users\_guide.html for details on ACIS Extract.} as a
central part of our point-source cataloging. AE is used as part of an
effective two-stage source-detection approach, and it allows for the optimal
combination of multiple observations with different aim points and roll
angles. 

Some key AGN science projects that should be advanced by the 7~Ms
\cdfs\ include investigations of (1) how SMBHs, including those in obscured systems,
grow and co-evolve with galaxies through the critical era at \hbox{$z\approx 1$--4}
when massive galaxies were largely assembling, and (2) how SMBHs grow within the
first galaxies at $z>4$. Starburst and normal galaxies are also detected
in abundance at the faintest \xray\ fluxes; their differential number counts
are comparable to those of AGNs at the faintest \hbox{0.5--2.0~keV} fluxes
reached by the 4~Ms \cdfs\ (e.g., \citealt{Lehmer2012}). The 7~Ms \cdfs\
will thus be a key resource for examining how the \xray\ binary populations
of starburst and normal galaxies evolve over most of cosmic time, both via
studies of the directly detected sources and via stacking analyses
\citep[e.g.,][]{Lehmer2016,Vito2016}. Owing to
its unique combination of great depth and high angular resolution, the 7~Ms
\cdfs\ should serve as a multi-decade \chandra\ legacy. For example, even
\athena, a \hbox{next-generation} \xray\ observatory aiming for launch in
$\approx 2028$ (e.g., \citealt{Barcons2015}), may not be able to reach the
flux levels probed in the central region of the 7~Ms \cdfs. \xray\ missions capable
of substantially surpassing the sensitivity of the 7~Ms \cdfs, such as the
{\it \xray\ Surveyor} (e.g., \citealt{Weisskopf2015}), are presently not funded
for construction.

The structure of this paper is the following. In Section~2 we present
the new \chandra\ observations and the reduction details for the full data
set. Section~3 describes the creation of observation images, exposure
maps, and the main and supplementary source catalogs. In Section~4, we 
present the main \chandra\ source catalog
in detail. Here we also present key aspects of the \xray\ source
characterization and the multiwavelength identifications. We briefly
compare the properties of the newly detected sources to those already
found in the 4~Ms \cdfs. Section~5 presents the supplementary catalog of
\xray\ sources that have lower detection significances but align spatially
with bright near-infrared (NIR) sources. In Section~6 we present an analysis of
source completeness and reliability, showing that we strike a reasonable
balance between these two criteria. Section~7 describes an analysis of
the background and sensitivity across the \cdfs. In Section~8 we present
cumulative number counts for the main \chandra\ source catalog, 
and in Section~9 we provide a
summary of the main results. The \chandra\ source catalogs and several
associated data products are being made publicly available along with this
paper.\footnote{http://www.astro.psu.edu/users/niel/cdfs/cdfs-chandra.html.}

We adopt a Galactic column density of
$N_{\rm H}=8.8\times 10^{19}$~cm$^{-2}$ along the line of sight to the
\cdfs\ \citep[e.g.,][]{Stark1992}. Coordinates are presented in the
J2000.0 system, and magnitudes are in the AB system \citep{Oke1983}.
We quote uncertainties at a 1$\sigma$ confidence level
and upper/lower limits at
a 90\% confidence level.
A cosmology with $H_0=67.8$~km~s$^{-1}$~Mpc$^{-1}$,
$\Omega_{\rm M}=0.308$, and
$\Omega_\Lambda=0.692$ (\citealt{Ade2016}) is used.

\section{OBSERVATIONS AND DATA REDUCTION}
\subsection{Observations of the 7 Ms CDF-S}
The 7 Ms \cdfs\ contains observations taken in four separate epochs of time.
The basic information on these observations, 102 in total,
is listed in Table~\ref{tbl-obs}.
There were 48 recent observations acquired between 2014 June 9 and 2016 March 24,
which constitute the last 3~Ms of exposure of the 7~Ms \cdfs. The first 1~Ms of
exposure consists of 11 observations taken between 1999 and
2000 \citep{Giacconi2002,Rosati2002,Alexander2003}, 
the next 1~Ms of exposure consists of 12 observations taken in 2007 \citep{Luo2008},
and another 2~Ms of exposure includes 31 observations in 2010 \citep{Xue2011}. 

All 102 \cdfs\ observations used the \chandra\ Advanced CCD Imaging 
Spectrometer imaging array (ACIS-I; \citealt{Garmire2003}), which
is optimized for \chandra\ surveys. \hbox{ACIS-I}
consists of four CCDs (I0-I3) with $1024\times1024$ pixels each;
the size of the CCD pixels is $0\farcs492\times0\farcs492$, 
and the ACIS-I array
size or the field of view of each observation
is $16\arcmin.9\times16\arcmin.9=285$~arcmin$^2$.
Of the 11 observations in the first 1~Ms of \cdfs\ exposure, ten (except observation 
1431-0) were taken in Faint mode, while the first observation (1431-0)
and all the observations in the later 6 Ms of \cdfs\ exposure were taken in Very Faint 
mode (see Table~\ref{tbl-obs}), which reduces \hbox{ACIS-I} particle background significantly and 
improves detection of
weak sources \citep{Vikhlinin2001}. During the first two observations (1431-0 and 
1431-1), the nominal focal-plane temperature was $-110\degr$C, while it was 
$-120\degr$C for the other observations.

The roll angles of the 102 observations (Table~\ref{tbl-obs}) were intentionally allowed to vary,
in order to obtain more uniform sensitivity across the field by averaging out some of the 
\hbox{CCD-gap} effects (e.g., see Figure~\ref{fig-rawimg} below) and to obtain
a larger areal coverage. The total area covered by the 7~Ms \cdfs\
is 484.2 arcmin${^2}$, substantially larger than the field of view 
of ACIS-I. The aim points of individual observations also differ slightly 
(Table~\ref{tbl-obs});
the average aim point for the merged observations,
weighted by the individual exposure times, is $\alpha_{\rm J2000.0}=
03^{\rm h}32^{\rm m}28\fs27$, $\delta_{\rm J2000.0}=-27\degr48\arcmin21\farcs8$.

\subsection{Data Reduction} \label{sec-datareduc}
The data for the 102 observations 
downloaded from the \chandra\ X-ray Center (CXC)
have gone through the CXC pipeline software 
for basic processing. The software versions 
are listed in Table~\ref{tbl-obs}, and 
the data for the previous 4~Ms \cdfs\ observations 
have been processed with newer versions of the software
compared to those presented in the previous catalog 
papers \citep{Luo2008,Xue2011}.
These data were then reduced and analyzed utilizing \chandra\ Interactive Analysis
of Observations (CIAO; v4.8)\footnote{See
http://cxc.harvard.edu/ciao/ for details on CIAO.} tools,
AE (version 2016may25), the MARX ray-tracing simulator (v5.3) that is used
in AE to model the \chandra\ ACIS-I PSF,\footnote{http://space.mit.edu/CXC/MARX/; 
this version of MARX 
fixed a PSF issue affecting the PSF simulations of off-axis sources
(see https://github.com/Chandra-MARX/marx/pull/21).}
and custom software. Most of the procedures are similar to those 
performed in \citet{Luo2008} and \citet{Xue2011,Xue2016}, and
the main steps are described below.

We adopted the CIAO tool {\sc acis\_process\_events} to reprocess level 1 event files,
applying Charge Transfer Inefficiency (CTI) corrections for observations
with nominal focal-plane temperatures of $-120\degr$C 
\citep{Townsley2000,Townsley2002}, flagging potential cosmic-ray background events
for Very Faint mode data ({\sc check\_vf\_pha=yes}),
and implementing a custom stripped-down bad-pixel file instead of the standard 
CXC one. A large fraction of the bad-pixel locations in the 
standard bad-pixel file appear to contain good $>0.7$~keV events that 
are appropriate for source searching and characterization; instead of
rejecting all events falling on these pixels, we 
chose to exclude manually those events below a
row-dependent energy of 0.5--0.7 keV that fall on ``hot'' soft
columns (see Section 2.2 of \citealt{Luo2008} for details). This approach allows us to recover
a significant number of additional good events ($\approx3.2\%$ of the total) 
compared to the standard 
level 2 data products from the CXC pipeline.

We employed the CIAO tool {\sc acis\_detect\_afterglow} to identify and flag cosmic-ray 
afterglow events. Following the procedure of \citet{Luo2008} and \citet{Xue2011},
we further employed custom software and removed 113 additional faint afterglow events
with three or more total counts falling on the same CCD 
pixel within 20~s (see Footnote 27 of \citealt{Xue2011}
for details about this choice).
We inspected the background light curve of each observation 
utilizing the {\sc event browser} tool in the Tools for ACIS Review \& Analysis
(TARA; \citealt{Broos2000}), and we removed background flares using
the CIAO tool {\sc deflare} with an iterative 3$\sigma$ clipping approach.
The 7~Ms \cdfs\ observations are not significantly affected by background flares.
Only four observations (1431-0, 16176, 16184, 17542) 
were affected by flares longer than 10\% of their durations (up to $\approx15\%$),
while the other observations have milder or no background flares. 
The cleaned exposure time for each observation is listed in Table~\ref{tbl-obs}.
In total, 1.2\% of the exposure was removed due to background flares;
the total cleaned exposure is 6.727~Ms. 

After generating a cleaned event file for each of the 102 observations, the next
steps were to register and align these observations to a common astrometric frame and 
merge them into a combined master event file.
The first action was fixing any astrometric offsets of individual observations.
For each observation, we created a 0.5--7.0 keV image using the standard \asca\
grade set (grades 0, 2, 3, 4, 5).
We then searched for X-ray sources in the image using {\sc wavdetect} \citep{Freeman2002}
with a ``$\sqrt{2}$~sequence'' of wavelet scales (i.e.,\ 1, 1.414, 2,
2.828, 4, 5.656, and 8 pixels)
and a false-positive probability threshold of 10$^{-6}$.
A PSF-size image was supplied to the {\sc wavdetect} run 
that was created with the CIAO tool {\sc mkpsfmap} assuming 
a power-law spectrum with a photon index of $\Gamma=1.4$ and 
an enclosed counts fraction (ECF) of 0.4.
Depending on the exposure times of the observations, \hbox{$\approx40$--170} X-ray sources 
are detected in the individual data sets. The initial \xray\ positions of these {\sc wavdetect}
sources were refined using AE, and the 
centroid positions determined by AE from its ``CHECK\_POSITIONS''
stage were adopted as the improved positions of these sources.
We registered the absolute astrometry of each observation to a common frame by matching 
the X-ray source positions to the NIR source positions in the 
Taiwan ECDFS Near-Infrared Survey (TENIS) $K_s$-band 
catalog \citep{Hsieh2012}, where 6651 bright ($K_s\le22$) 
TENIS sources within the field of view of 
the 7~Ms \cdfs\ were used. This NIR catalog was chosen as its astrometric frame
is consistent with those of other optical/NIR catalogs (e.g., see the list in Section~\ref{sec-ID} below)\footnote{The 
astrometry of the \citet{Xue2011} 4~Ms \cdfs\ observations was registered to the frame
of the Very Large Array catalog of \citet{Miller2013}, and there are $\approx0\farcs2$ offsets
in right ascension and declination between this frame and those of the optical/NIR catalogs,
${\rm median}({\rm RA_{TENIS}-RA_{VLA}})=-0\farcs193\pm0\farcs012$ and 
${\rm median}({\rm Dec_{TENIS}-Dec_{VLA}})=0\farcs268\pm0\farcs014$.
Therefore, the X-ray source positions in the current 7~Ms \cdfs\ catalogs have the same systematic
offsets from the 4~Ms source positions inherited from the different choices of the astrometric
systems. We caution that the TENIS astrometric frame is off from that of
the first Gaia data
release \citep{Gaia2016} by ${\rm median}({\rm RA_{TENIS}-RA_{Gaia}})=-0\farcs154\pm0\farcs004$ and
${\rm median}({\rm Dec_{TENIS}-Dec_{Gaia}})=0\farcs290\pm0\farcs005$. The
\citet{Miller2013} Very Large Array astrometric frame thus agrees better with the Gaia frame, although there
are not a sufficient number of sources in common for a direct comparison.
\label{foot-astrometry}}
and the fractions of X-ray sources with a bright NIR counterpart are high (ranging 
from $\approx50$--80\%) in the \cdfs\ observations.

The matching of the X-ray and TENIS sources as well as the 
World Coordinate System (WCS) update were performed
using the CIAO tool {\sc reproject\_aspect} with a 3$\arcsec$ matching radius and a
0\farcs6 residual limit;\footnote{This is a parameter used in {\sc reproject\_aspect}
to remove source pairs with residual positional offsets larger than the given limit.} 
\hbox{$\approx30$--110} matches are found in the individual observations,
and the positional offsets were used to obtain the WCS transformation for each observation.
The WCS transformation matrices range from  0\farcs007 to 1\farcs011 in linear translation,
$-0\fdg051$ to 0\fdg014 in rotation, and 0.9996 to 1.0012 in scaling; the resulting
registrations are accurate to $\approx$0\farcs3. Although the astrometric offsets are small,
registering the observations and correcting for the offsets is a necessary step 
for detecting very faint
X-ray sources and obtaining 
the best-possible \xray\ source positions as well as reliable photometric properties. 
The astrometry registered event files were produced with the 
CIAO tool {\sc reproject\_events}. We then projected the event files to the astrometric 
frame of observation 2406\footnote{The choice of this astrometric frame is to be consistent 
with our previous \cdfs\ analyses \citep{Luo2008,Xue2011};
choosing the frame of another observation 
does not affect any of the results.} using {\sc reproject\_aspect} and {\sc reproject\_events}.
Finally, we merged these 102 event files into a master event file using the CIAO tool {\sc dmmerge}.
We note that X-ray source detection was performed in the merged observation while source
characterization was carried out mainly using the individual astrometry registered observations
(e.g., the AE photometric extraction).

\section{Images, Exposure Maps, and X-ray Source Detection}
\subsection{Image and Exposure Map Creation} \label{sec-imagesandemaps}
\begin{figure}
\centerline{
\includegraphics[scale=0.48]{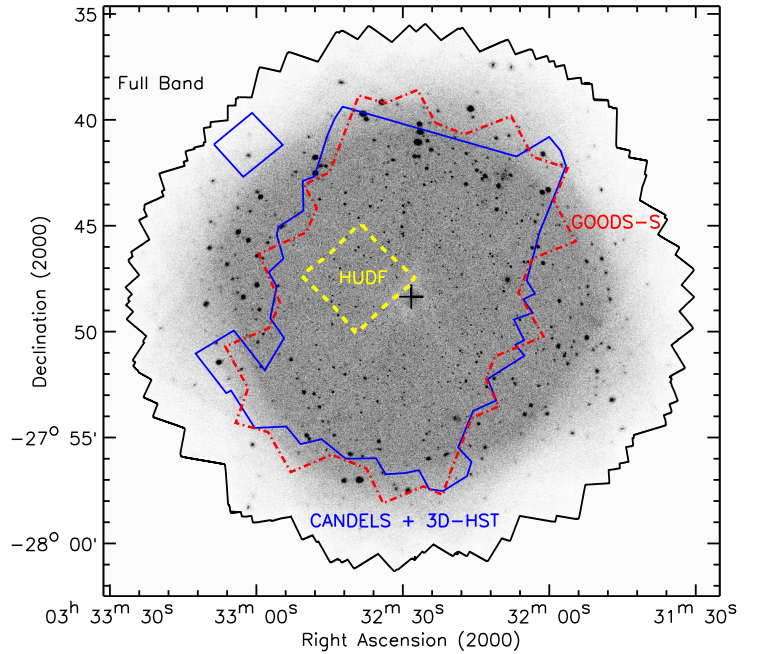}
}
\caption{
Full-band raw image of the 7 Ms CDF-S in 
linear gray scale. The black outline surrounding the image indicates the
coverage of the entire \cdfs. The blue solid, red dash-dotted, and
yellow dashed regions show the coverage of the
CANDELS $+$ 3D-HST fields,
GOODS-S survey, and HUDF, respectively.
The central black plus sign marks the average aim point.
The apparent lightening of the area surrounding the black plus sign
is caused by the relatively low effective exposure in this region due
to the ACIS-I CCD gaps (see Figure~\ref{fig-expimg}).
The apparent scarcity of sources near the field center is largely
due to the small PSF size at that location, 
which makes sources difficult to  
identify visually in this figure
(see Figures~\ref{fig-clrimg} and \ref{fig-srcdist} for clarification).
}
\label{fig-rawimg}
\end{figure}

We created \xray\ images from the merged event file using the standard \asca\
grade set in three bands: \hbox{0.5--7.0~keV} (full band; FB), \hbox{0.5--2.0~keV}
(soft band; SB), and \hbox{2--7~keV} (hard band; HB).\footnote{The upper energy bound of the
bands has been
changed from 8~keV in our previous \cdfs\ analyses \citep{Alexander2003,Luo2008,Xue2011} 
to 7~keV; see Footnote 16 of \citet{Xue2016} for detailed reasoning.}
The full-band raw image is shown in Figure~\ref{fig-rawimg};
also illustrated are the fields of view of some of the deepest optical--NIR surveys
from the {\it Hubble Space Telescope} ({\it HST}) within the \cdfs, including the
Hubble Ultra Deep Field \citep[HUDF;][]{Bechwith2006}, 
the Great Observatories Origins Survey Southern field
\citep[GOODS-S;][]{Giavalisco2004}, and the combined field of 
the Cosmic Assembly \hbox{Near-infrared} Deep Extragalactic Legacy
Survey (CANDELS; \citealt{Grogin2011,Koekemoer2011}) and the 3D-HST survey \citep{Skelton2014}.

\begin{figure}
\centerline{
\includegraphics[scale=0.46]{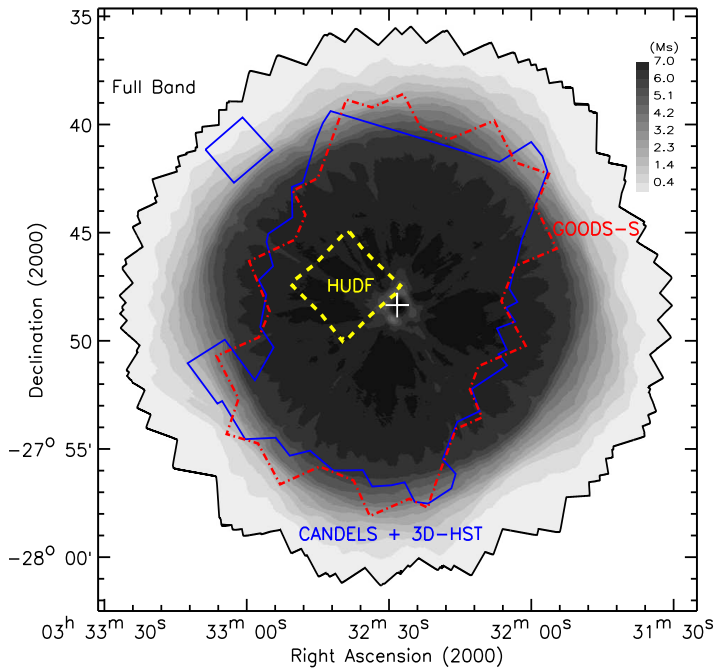}
}
\caption{
Full-band effective-exposure map of the 7 Ms \hbox{CDF-S}.
The linear gray scale bar is shown in the upper right; the displayed
effective exposure times are in units of Ms.
The maximum effective exposure time is 6.651~Ms, which is smaller than
the total cleaned exposure of 6.727~Ms as the aim points of 
individual observations differ.
The radial trails and central ring-like structure with
relatively low effective exposures are caused by the
ACIS-I CCD gaps.
The regions and the plus sign are the same as those in
Figure~\ref{fig-rawimg}.
}
\label{fig-expimg}
\end{figure}

Due to the effects of vignetting, gaps between the CCDs,
bad-pixel filtering, bad-column filtering, spatially
dependent degradation in quantum efficiency
due to contamination on the ACIS optical-blocking filters, and slightly different aim points 
between observations, 
the effective exposure time of the combined observation cannot reach the 
total cleaned exposure of 6.727~Ms, and it varies across the \cdfs\ field. 
Therefore, we constructed
effective-exposure maps in the three bands following the basic approach described in
Section 3.2 of \citet{Hornschemeier2001}, taking into account the above effects.
A power-law spectrum with a photon index of $\Gamma=1.4$ was assumed when creating the exposure maps,
which is approximately the slope of the cosmic \xray\ background spectrum in the full
band \citep[e.g.,][]{Marshall1980,Gendreau1995,Hasinger1998}.
The full-band \hbox{effective-exposure} map is shown in Figure~\ref{fig-expimg}, with a
maximum effective exposure time of 6.651~Ms.

\begin{figure}
\centerline{
\includegraphics[scale=0.48]{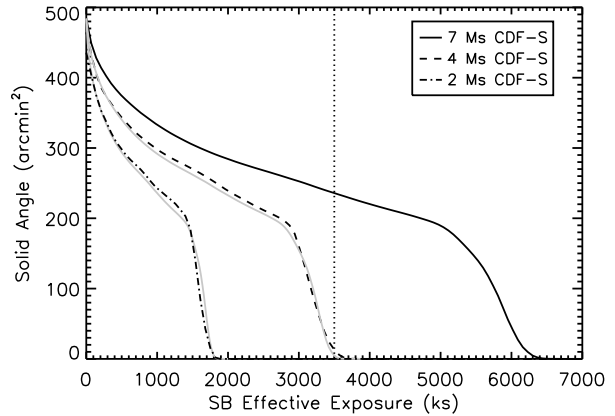}
}
\caption{
Survey solid angle as a function of minimum soft-band
effective exposure for the 7~Ms CDF-S (black solid curve). The vertical dotted
line indicates an effective exposure of 3.5 Ms, and 235.9 arcmin$^2$ (49\%)
of the 7~Ms CDF-S and 13.0~arcmin$^2$ (3\%) of the 4~Ms \cdfs\ 
have $>3.5$~Ms effective exposure.
For comparison, the dashed and dash-dotted curves display the
4~Ms CDF-S \citep{Xue2011} and 2~Ms CDF-S \citep{Luo2008}
solid angles, which were derived following
the same procedures as in this paper. The 7~Ms curve can 
be roughly rescaled to 4~Ms and 2~Ms curves with scaling factors of 1.8 and 3.5 in effective exposure
(gray solid curves),
respectively.
}
\label{fig-soildangle}
\end{figure}

From the effective-exposure maps, we can derive the
survey solid angle as a
function of the minimum effective exposure. For the purpose of comparing to the previous 2~Ms
and 4~Ms \cdfs\ results (the definition of the full band and hard band have changed), we show in 
Figure~\ref{fig-soildangle} such a relation in the soft band.
Approximately 49\% of the 7 Ms \cdfs\ field has \hbox{$>3.5$~Ms} effective exposure,
while $3.5$~Ms is
close to the deepest effective exposure achieved in the 4 Ms
\cdfs\ with only 3\% of the field having longer exposure times.
In the 7~Ms \cdfs, $\approx45\%$, 39\%, and 9.4\% of the field have $>4$~Ms, $>5$~Ms,
and $>6$~Ms effective exposure, respectively.
The survey solid-angle curves for the 4 Ms CDF-S \citep{Xue2011} 
and 2~Ms CDF-S \citep{Luo2008} are similar to the 7~Ms curve in the sense that 
they are approximately the scaled-down versions of the 7~Ms curve
with scaling factors of 1.8 and 3.5 in effective exposure, respectively. The overall larger solid angles 
of the 7~Ms \cdfs\ compared to the 4~Ms \cdfs\ suggest that we are not only able to detect
new sources below the 4~Ms \cdfs\ sensitivity limit, but also detect new sources
above that limit owing to the significantly 
increased solid angle coverage at any given exposure time.

\begin{figure*}
\centerline{
\includegraphics[scale=0.48]{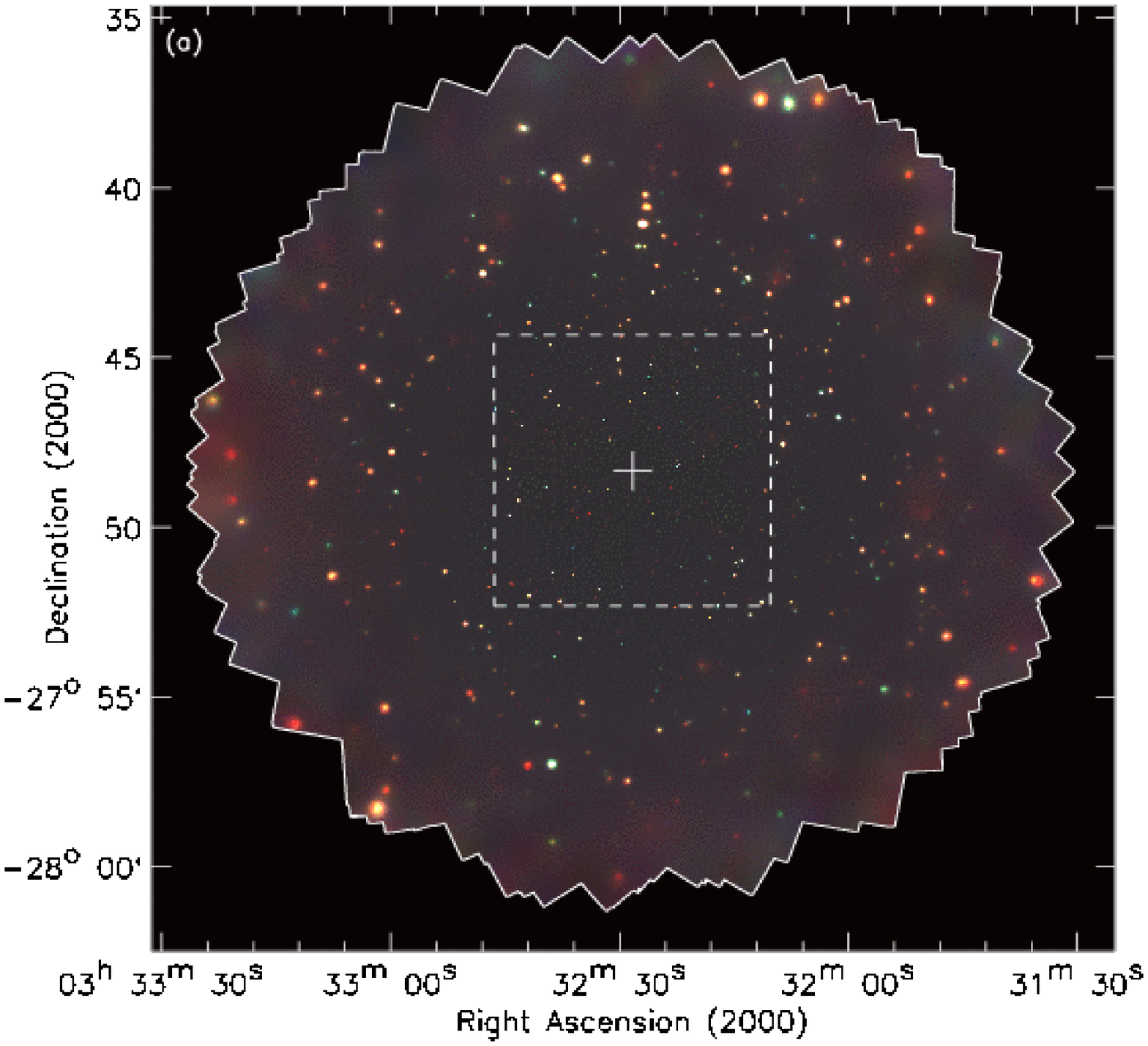}
\includegraphics[scale=0.48]{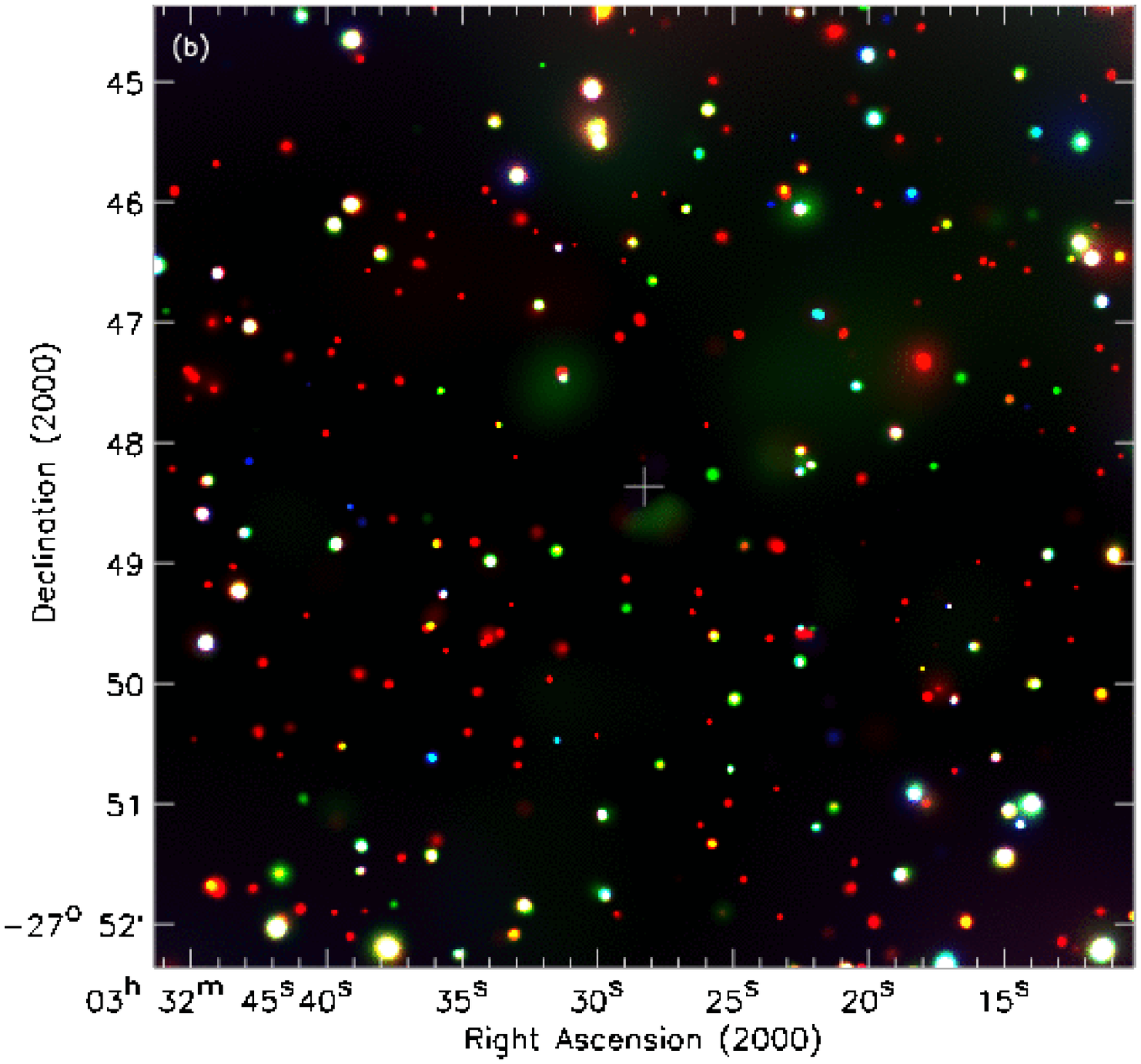}
}
\caption{
(a) "False-color" image of the 7 Ms CDF-S.
The image is a color composite of the exposure-corrected and adaptively
smoothed images in the \hbox{0.5--2.0~keV} (red), 2--4 keV (green),
and 4--7 keV (blue) bands. The smoothed images have uneven weights in the composite 
for the purpose of 
enhanced display, and thus the source color in the image 
does not reflect accurately the X-ray color of the source.
The apparent smaller size and lower brightness
of the sources near the field center are due to the smaller PSF size at
that location. The CDF-S boundary and the average aim point are shown,
as was done in Figure~\ref{fig-rawimg}. An expanded view of the 
central $8\arcmin\times8\arcmin$ region (dashed square region) is displayed in 
panel (b). Extended faint color halos in panel (b) are usually
artifacts instead of real extended sources.
}
\label{fig-clrimg}
\end{figure*}

We created exposure-corrected smoothed images in the 0.5--2.0~keV, 
2--4~keV, and 4--7~keV bands following
Section 3.3 of \citet{Baganoff2003} using the CIAO tool {\sc csmooth}.
The images and effective-exposure maps were adaptively smoothed with the same
scale map in each band, and the smoothed images were divided by the 
corresponding smoothed exposure maps. These exposure-corrected smoothed
images were combined 
to produce a color composite, as shown in Figure~\ref{fig-clrimg}; an expanded view of the
central $8\arcmin\times8\arcmin$ region is also shown to illustrate the large abundance of 
sources near the field center. Although many of the X-ray sources are clearly visible
in the adaptively smoothed images, our source searching was performed on the raw images
(e.g., Figure~\ref{fig-rawimg}), as detailed in Section~\ref{sec-detection} below.
Our {\sc csmooth} processes were not optimized to enhance the visibility of 
extended sources (which requires external background
files for proper computation of the source significance), and extended faint color halos 
which appear in Figure~\ref{fig-clrimg}b are usually
artifacts
rather than real extended sources \citep[e.g.,][]{Finoguenov2015}.

\subsection{X-ray Source Detection} \label{sec-detection}
X-ray source detection for the 7~Ms \cdfs\ follows the same two-stage approach as was employed
in the \citet{Xue2011} 4~Ms \cdfs\ catalog, which maximizes the number of reliable sources
detected \citep[e.g.,][]{Lehmer2009,Xue2011}. A candidate source list 
was initially generated by {\sc wavdetect} source detection, and it was 
then filtered by AE to produce our main source catalog, which
includes significant \xray\ sources that are unlikely to be
false detections caused by background fluctuations. 
A supplementary source
catalog was also produced that contains \hbox{lower-significance} \xray\ sources with
bright ($K_s\le23$) NIR counterparts.

To generate the candidate source list, we ran
{\sc wavdetect} on the combined raw images in the full band, soft band, and hard band,
using a
``$\sqrt{2}$~sequence'' of wavelet scales (i.e.,\ 1, 1.414, 2,
2.828, 4, 5.656, 8, 11.314, and 16 pixels)
and a \hbox{false-positive} probability threshold of 10$^{-5}$.
The {\sc wavdetect} process made use of a merged PSF map, 
created 
by choosing the minimum PSF size at each pixel location among all the PSF maps of
individual observations (e.g., see Section 2.2.2 of \citealt{Xue2016}); such a process
is optimized for \hbox{point-source} detection.\footnote{See
http://cxc.harvard.edu/ciao/threads/wavdetect\_merged/index.html\#min.}
We merged the three source lists for the three bands into the candidate source list
by cross matching them with a matching radius of $2\farcs5$ for sources
within $6\arcmin$ of the average aim point and $4\farcs0$ for sources at
larger \hbox{off-axis} angles (the distance
between the source position and the average aim point); we also visually inspected all the sources beyond 
$8\arcmin$ of the average aim point and removed nine candidate sources which are likely the same
as their companion detections \hbox{$\approx4$--7\arcsec} away. 
The X-ray source positions in the merged source list were adopted from, in
order of priority, the \hbox{full-band}, \hbox{soft-band}, and hard-band positions.
The resulting candidate source list includes 1121 sources. 

The relatively loose {\sc wavdetect} source-detection threshold of 10$^{-5}$
introduces a non-negligible number of spurious detections. 
We also performed {\sc wavdetect} source searching with the more stringent 
false-positive probability thresholds
of 10$^{-6}$, 10$^{-7}$, and 10$^{-8}$. 
We then
assigned a minimum {\sc wavdetect} false-positive probability 
to each of the 1121 candidate sources according to the 
minimum {\sc wavdetect} threshold value at which the source was detected. Of the 1121 sources,
644, 58, 102, and 317 have minimum {\sc wavdetect} \hbox{false-positive} probabilities of 10$^{-8}$, 
10$^{-7}$, 10$^{-6}$, and 10$^{-5}$, respectively.
Candidate sources
with smaller minimum {\sc wavdetect} false-positive probabilities
are more likely 
real detections and most of the spurious detections will have minimum false-positive probabilities
of 10$^{-5}$
(e.g., see Figure~\ref{fig-pbvssig} below).

Before filtering the candidate source list with AE, we 
improved the source positions through the AE ``CHECK\_POSITIONS'' procedure. 
As was done in our previous \cdfs, Extended \chandra\ Deep Field-South (\hbox{E-CDF-S}),
and \chandra\ Deep Field-North (\hbox{CDF-N}) catalogs 
\citep{Luo2008,Xue2011,Xue2016}, we adopted AE centroid positions for sources 
within $8\arcmin$ of the average aim point and matched-filter positions for sources located
at larger off-axis angles.\footnote{The matched-filter position
is the position found by correlating the merged image in the vicinity of a source 
with the combined source PSF (see Section~5.3 of the AE User's Guide).} 
We further visually inspected the raw and adaptively smoothed
images for each source and manually chose centroid or matched-filter positions for 
$\approx60$ sources which
align better with the apparent source centers 
(mostly sources located within 6--$8\arcmin$ of the average aim point where the matched-filter
positions are preferred). 

We then utilized AE to extract photometric properties of the 
candidate sources. The details of the AE photometric extraction
are described in the AE User's Guide; 
a short summary is presented in Section 3.2 of \citet{Xue2011}.
Briefly, AE performed source and background extractions for each source in 
each observation and then merged the results.
A polygonal extraction region that 
approximates the $\approx90\%$ ECF contour of the local PSF was utilized to extract 
source counts; smaller extraction regions ($\approx40\%$--75\% ECFs) were used 
in crowded areas where sources have overlapping $\approx90\%$ ECF apertures.
The AE ``BETTER\_BACKGROUNDS'' algorithm was adopted for background extraction
(Section 7.6.1 of the AE User's Guide),
which seeks to obtain a single background region plus a background scaling that 
simultaneously models all background components, including the background
that arises from the PSF wings of neighboring sources.
A minimum number of 100 counts in the merged background spectrum is required
to ensure photometric accuracy, which was achieved through the AE
``ADJUST\_BACKSCAL'' stage. The extraction results from individual observations 
were then merged to produce photometry for each source through the AE
``MERGE\_OBSERVATIONS'' procedure.

One important output parameter from AE is the binomial no-source probability, $P_{\rm B}$,
which is the probability of still observing the same number of source counts or more
under the assumption that there is no real source at the relevant location and the observed 
excess number of counts over background is purely due to background fluctuations.
The formula to obtain $P_{\rm B}$ is given by:
\begin{equation}
P_{\rm B}(X\ge S)=\sum_{X=S}^{N}\frac{N!}{X!(N-X)!}p^X(1-p)^{N-X}~. \label{eq-pb}
\end{equation}
In this equation, $S$ is the total number of counts in the
\hbox{source-extraction} region and $B$
is the total number of counts in the background-extraction
region; $N$ is the sum of $S$ and $B$; $p=1/(1+BACKSCAL)$,
with $BACKSCAL$ being the area ratio of the background and source-extraction regions.
A smaller $P_{\rm B}$ value indicates that a source has a larger probability of being real.
For each source, AE computed a $P_{\rm B}$ value in each of the three bands, and
we adopted the minimum of the three as the final $P_{\rm B}$ value for the source.

Although $P_{\rm B}$ is a classic confidence level, it is usually not a good indicator of  
the fraction of spurious sources (e.g., a cut at $P_{\rm B}=0.01$
does not correspond to a 1\% spurious rate), mainly because 
the extractions were performed on a biased sample of candidate sources that already survived a
filtering process by {\sc wavdetect}. 
Furthermore, given its definition, the value of $P_{\rm B}$ also depends on the 
choices of source and background extraction regions.
Therefore, we cannot reject spurious sources simply based on
the absolute value of $P_{\rm B}$ itself. Fortunately, from past experience \citep[e.g.,][]{Luo2010,Xue2011},
we found that the superb multiwavelength coverage
in the \cdfs\ allows us to identify counterparts for the majority ($\approx96\%$) of the 
X-ray sources, and the accurate X-ray and optical/NIR/radio source positions also ensure high-confidence
associations (false match rate $\approx2\%$). These combined factors indicate that X-ray sources
having a multiwavelength counterpart (down to the magnitude limits of the multiwavelength
catalogs) are extremely likely to be 
real detections. Thus we proceeded to choose a 
$P_{\rm B}$ threshold that retains a large number of sources with multiwavelength counterparts
while removing most of the sources without counterparts. The multiwavelength catalogs used for 
identification and the identification procedure are the same as those described in Section \ref{sec-ID}
below. After evaluation of the matching results at several possible 
$P_{\rm B}$ threshold values (0.001--0.01), 
we adopted $P_{\rm B}<0.007$ as the criterion to prune the candidate source list and generate a main source 
catalog, which includes 1008 sources with a $\approx97\%$ multiwavelength-identification rate.
The detailed properties of the main-catalog sources are presented in Section 4 below.

The choice of the $P_{\rm B}$ threshold is an empirical decision, optimized to balance the 
needs of recovering a large number
of real sources (high completeness) 
while keeping the fraction of potential spurious sources small (high reliability). A slightly 
different choice of the threshold value will affect the final source catalog as well
as the catalog completeness and reliability slightly; in fact, most of the main-catalog sources, 922 out
of the 1008 (91.5\%), have $P_{\rm B}<0.001$ and are highly reliable ($>98\%$ identification rate).
The $P_{\rm B}$ threshold value was 0.004
for the 2~Ms CDF-N \citep{Xue2016} and 4~Ms \cdfs\ \citep{Xue2011}, and it was
0.002 for the 250~ks \hbox{E-CDF-S} \citep{Xue2016}. As reasoned above, the absolute $P_{\rm B}$ values 
are not directly comparable, but these choices were also made based on the multiwavelength 
identification results to optimize the balance between completeness and reliability, 
consistent with our current approach.

Our adopted $P_{\rm B}$ threshold will have inevitably rejected real X-ray sources. To recover
some of these real sources, we created a supplementary source
catalog that contains \hbox{lower-significance} X-ray sources which have bright optical/NIR counterparts, as
has also been done for the 4~Ms \cdfs\ \citep{Xue2011}; the chance of a bright optical/NIR source being associated with
a spurious X-ray detection is quite small. A total of 47 candidate \cdfs\ sources having 
$0.007\le P_{\rm B}<0.1$ are associated with bright, $K_s\le23$, TENIS sources, where the false
match rate is only 1.7\%, and these 47 sources constitute the supplementary catalog.
The basic X-ray and multiwavelength properties of the supplementary catalog sources are 
presented in Section 5 below.

The distributions of $P_{\rm B}$ for sources in the candidate-list catalog with different
minimum {\sc wavdetect} false-positive probabilities 
are displayed in Figure~\ref{fig-pbvssig}. Sources that are 
detected by {\sc wavdetect} at smaller false-positive probability thresholds are also considered more 
significant by AE with smaller $P_{\rm B}$ values in general. Most (99.7\%) of the candidate sources
detected with minimum {\sc wavdetect} false-positive 
probabilities of $10^{-8}$ remain in the main catalog, while
a substantial fraction ($\approx30\%$) of the candidate sources with 
minimum {\sc wavdetect} false-positive probabilities of $10^{-5}$
were rejected by the AE filtering, likely being spurious detections.
Of the 47 supplementary
catalog sources, 1, 9, and 37 have minimum {\sc wavdetect} false-positive
probabilities of $10^{-8}$, $10^{-6}$, and $10^{-5}$, respectively.

\begin{figure}
\centerline{
\includegraphics[scale=0.5]{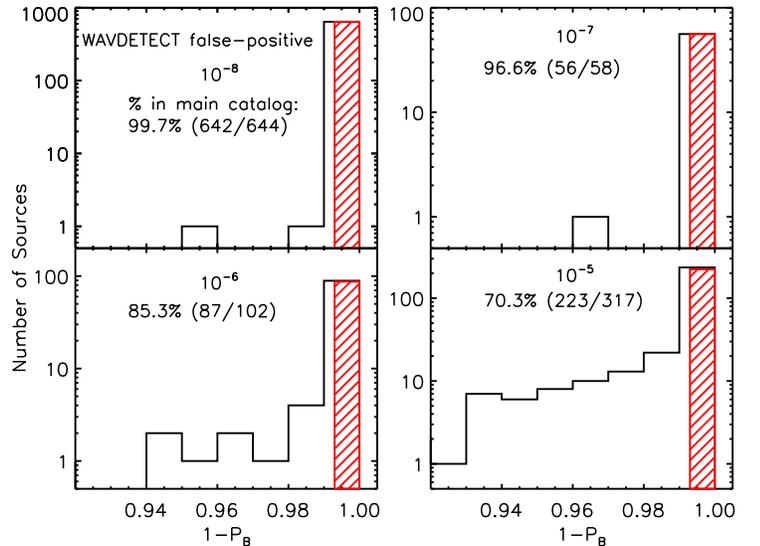}
}
\caption{
Distribution of 1 minus the AE binomial no-source probability ($P_{\rm B}$) for
sources in the candidate-list catalog with different 
minimum {\sc wavdetect} false-positive probabilities.
Sources having $P_{\rm B}<0.007$ were included in the
main source catalog, and they are indicated by the red shaded bars,
which have a slightly smaller width than the rightmost $P_{\rm B}$ bin
(0.007 vs.\ 0.01).
The fraction of main-catalog sources among each
minimum {\sc wavdetect} false-positive bin is annotated, with the numbers of sources
shown in the parentheses. The fraction drops from
99.7\% at a minimum {\sc wavdetect} false-positive probability of $10^{-8}$ to
70.3\% at a minimum {\sc wavdetect} false-positive probability of $10^{-5}$.
}
\label{fig-pbvssig}
\end{figure}

\section{Main {\emph{Chandra}} Source Catalog}

After determining the main and supplementary catalog sources, we performed another AE 
photometric-extraction procedure on these catalog sources instead of the candidate sources;
the exclusion of those rejected candidate sources affected the photometry of several sources slightly,
due to changes in the extraction regions and/or background levels.
The characterization of the X-ray and multiwavelength source properties, including X-ray positional uncertainties,
multiwavelength counterparts, redshifts, X-ray photometric and basic spectroscopic
properties, and AGN classification, follows similar approaches as were used
in the \citet{Xue2011} 4~Ms \cdfs\ catalog, and these 
are described in detail in the following subsections.
A summary of the main catalog data columns can be found in Section~4.8 below.

\subsection{X-ray Source Positional Uncertainty} \label{sec-dpos}
We investigated the accuracy of the X-ray source positions by comparing them to the 
positions of the 6651 bright ($K_s\le22$)
TENIS sources that were used in Section \ref{sec-datareduc} to register the astrometry of the 
\cdfs\ observations. We matched the \xray\ sources to the $K_s$-band sources using a 
matching radius of 1\farcs5; we removed manually two matches where the TENIS
positions are significantly affected by source blending.
There are 662 matches found, with a median positional offset of 0\farcs30. The expected number of false matches
is small, $\approx27$ ($\approx4.1\%$) estimated by shifting the \xray\ source positions manually
and recorrelating them to the TENIS \hbox{$K_s$-band} sources (e.g., see Section~3.3.1 of \citealt{Luo2008}); 
such a small false-match rate 
does not affect our
analysis of the \xray\ source positional uncertainties below. The 1\farcs5 matching radius
was used here only to obtain \xray\ source positional uncertainties; later 
we adopt a more sophisticated 
\hbox{likelihood-ratio} matching technique, which takes the derived \xray\ source positional uncertainties
as input parameters,
to identify multiwavelength counterparts for the X-ray
sources (Section~\ref{sec-ID}).

\begin{figure*}
\centerline{
\includegraphics[scale=0.45]{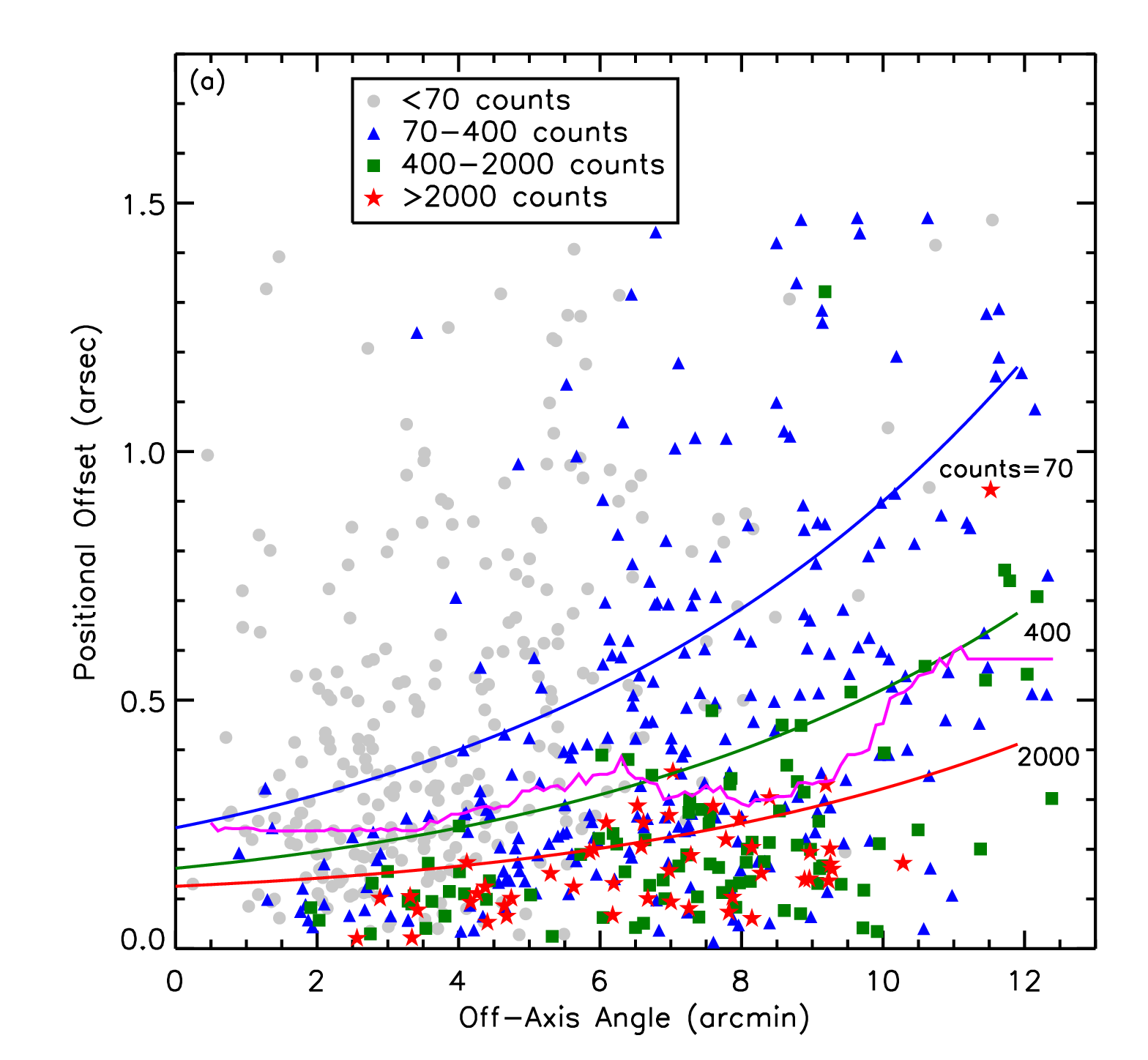}
\includegraphics[scale=0.45]{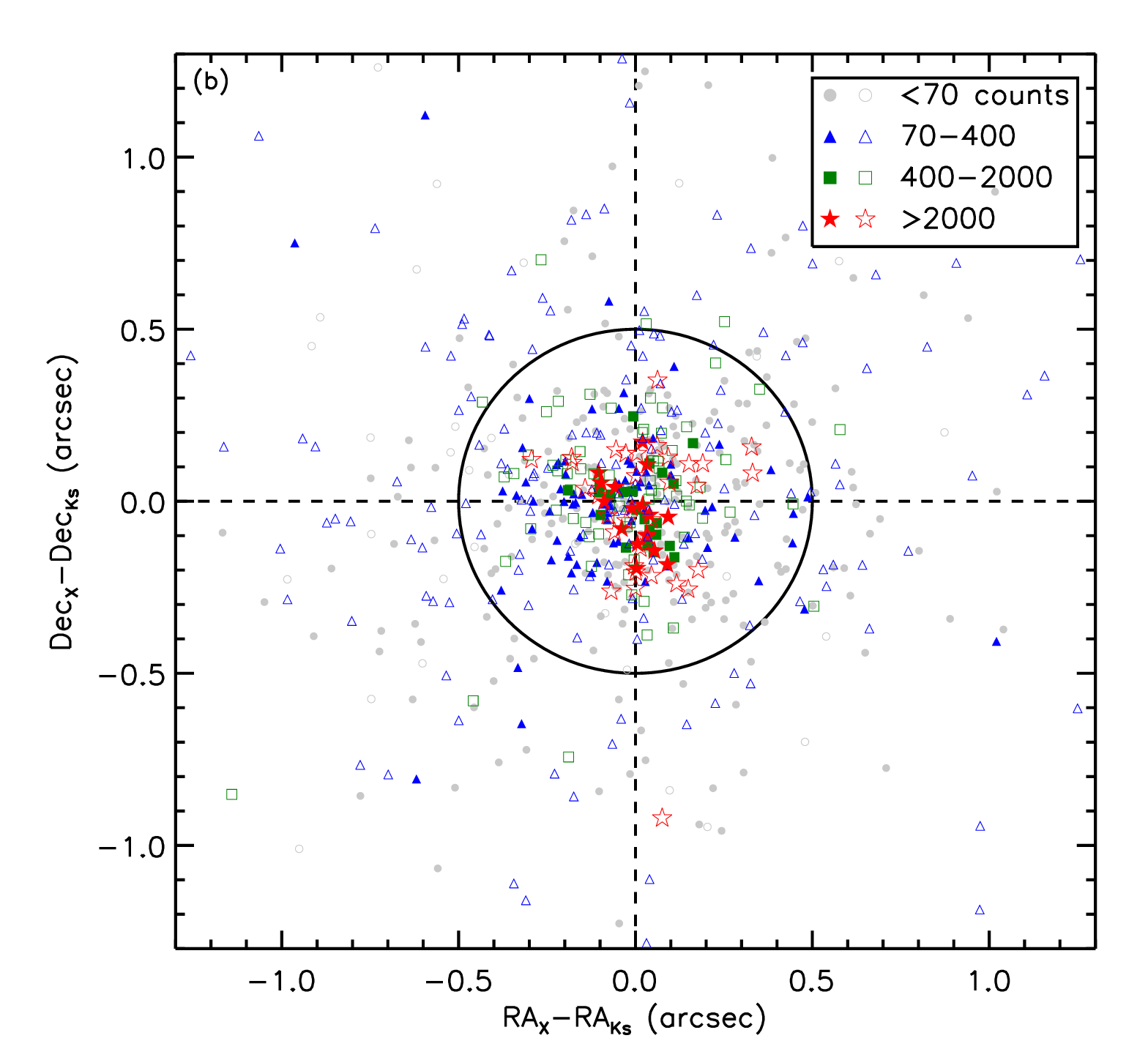}
}
\caption{
(a) Positional offset vs.\ off-axis angle for the 662 main-catalog sources
that have a bright ($K_s\le22$) TENIS
counterpart with a matching radius of 1\farcs5.
The gray circles, blue triangles, green squares, and red stars
represent X-ray sources with $<70$, 70--400, 400--2000, and $>2000$
counts,
respectively. The magenta solid curve displays the running median
of the positional offsets for all these sources
in off-axis angle bins of 3\arcmin.
These data were used to derive the 1$\sigma$
positional uncertainties of the X-ray sources (see Equation~\ref{eq-dpos}).
The blue, green, and red solid curves represent the
quadratic sum of the 1$\sigma$ positional errors
($\sqrt{\sigma_{\rm X}^2+\sigma_{Ks}^2}$, where $\sigma_{Ks}=0\farcs1$)
for sources with 70, 400, and 2000 counts, respectively;
there are still broad ranges of counts for sources marked as blue triangles (70--400) and green squares (400-2000),
and thus the blue and green curves are lower limits on the expected 1$\sigma$
positional offsets.
Approximately, $68\%$ (1$\sigma$) of the blue triangles, green squares,
and red stars lie below their corresponding solid curves, respectively.
(b) Distribution of the positional offsets in the right ascension (RA) vs.\ declination (Dec) 
plane for
the 662 main-catalog sources having a bright TENIS counterpart.
The symbols have the same meaning as in panel (a); in addition,
filled and open symbols represent sources having an off-axis angle
of $\le6\arcmin$ and $>6\arcmin$, respectively.
The majority of the sources, especially those that are on axis and
have a large number of counts, lie within the black circle, which
has a radius of 0\farcs5. For each of the four groups of sources in different
count bins, the mean positional offsets in right ascension and declination are
consistent with zero within the uncertainties.
}
\label{fig-posoff}
\end{figure*}

\begin{figure}
\centerline{
\includegraphics[scale=0.5]{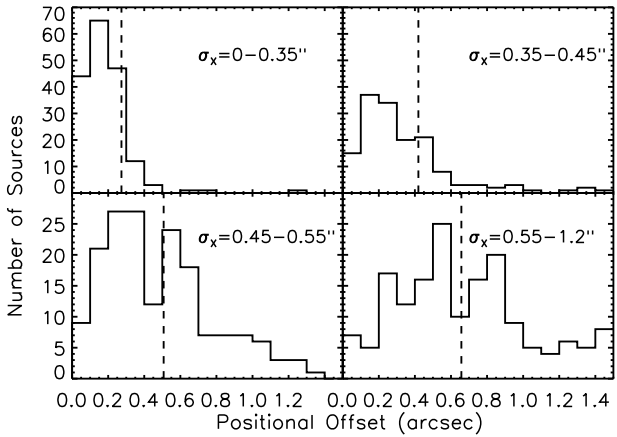}
}
\caption{
Histograms of the distributions of positional offsets for the 662
main-catalog sources
that have a bright TENIS
counterpart. These sources were divided into four bins according to their positional uncertainties,
and each bin contains approximately the same number of sources.
The vertical dashed line in each panel indicates the expected
positional offset for each group of sources, which is the quadratic sum of the
median X-ray positional uncertainty and the TENIS source positional
uncertainty ($\sqrt{\sigma_{\rm X,median}^2+\sigma_{Ks}^2}$,
where $\sigma_{Ks}=0\farcs1$), and $\approx68\%$ (1$\sigma$) of the
sources
have a positional offset smaller than this value.
}
\label{fig-poshist}
\end{figure}

The positional offsets between the X-ray and $K_s$-band sources have clear off-axis angle 
and source-count dependencies, as illustrated in
Figure~\ref{fig-posoff}a. The former is caused by the broader \chandra\ PSF sizes at larger off-axis angles,
and the latter is due to the fact that locating the centroid of a faint \hbox{X-ray} source is difficult.  
As was done in our previous \cdfs\ and E-CDF-S catalogs
\citep{Luo2008,Xue2011,Xue2016}, we derived an empirical relation for the X-ray positional uncertainty
adopting the basic functional form proposed by \citet{Kim2007}:
\begin{equation}
\log \sigma_{\rm X}=0.0606 \theta-0.320\log C-0.064~.  \label{eq-dpos}
\end{equation}
In the above equation, $\sigma_{\rm X}$ is the 1$\sigma$ positional uncertainty in units of arcseconds,
$\theta$ is the off-axis angle in arcminutes, and $C$ is number of source counts in the 
energy band where the source position was determined (Section~\ref{sec-detection}); an upper limit of 2000 was
set on $C$ as the positional accuracy does not improve significantly with larger numbers of counts.
The coefficients of Equation~\ref{eq-dpos} were determined so that for a given sample of X-ray--$K_s$
matches, the fraction of sources having positional offsets smaller than expectations 
($\sqrt{\sigma_{\rm X}^2+\sigma_{Ks}^2}$, where $\sigma_{K_s}=0\farcs1$ is the adopted TENIS source positional
uncertainty) is $\approx68\%$;\footnote{Based on similar practices, 
the 90\% and 95\% confidence-level 
positional uncertainties are approximately 1.6 and 2.0 times the 1$\sigma$ positional uncertainties.} a few such examples are displayed in Figure~\ref{fig-posoff}a.
In Figure~\ref{fig-posoff}b, we display the positional offsets in right ascension and declination.
Most of the X-ray sources with large numbers of counts (e.g., $>400$) or small 
\hbox{off-axis} angles
(e.g., $<6\arcmin$) have their positions determined reliably to within
$\approx0\farcs5$. Considering the entire X-ray source sample, there is no systematic offset in right ascension or 
declination when compared to the $K_s$-band sources, 
as required by our astrometric registration (Section~\ref{sec-datareduc}).
For the \hbox{main-catalog} 
sources, the positional uncertainties range from $0\farcs11$ to $1\farcs28$,
with a median value of $0\farcs47$.
Figure~\ref{fig-poshist} presents the distributions of X-ray--$K_s$ positional offsets
in four bins of X-ray positional uncertainties; the offsets are consistent with 
expectations from the positional uncertainties.

\subsection{Multiwavelength Source Identifications} \label{sec-ID}
We searched for optical, NIR, infrared (IR), and radio counterparts for the X-ray sources,
following the 
\hbox{likelihood-ratio} matching technique described in \citet{Luo2010}. 
This technique computes a likelihood ratio for each potential counterpart, 
taking into account the positional uncertainties of the \xray\ and optical/NIR/IR/radio sources
as well as the expected magnitude distribution of counterparts. A threshold value for
the likelihood ratio that maximizes the matching completeness and reliability 
was chosen to determine the final matches (see Section~2.2 of \citealt{Luo2010} for
details); in cases where multiple counterpart candidates satisfy the 
threshold cut for a single X-ray source ($\approx3\%$ of the total main-catalog sources on average),
we selected the candidate with the highest likelihood ratio.
The false-match probability is estimated based on the \citet{Broos2011}
shift-and-recorrelate Monte Carlo method, and our approach is described   
in detail in Section 4.3 of \citet{Xue2011}. This approach does not account for any potential
false matches introduced when only the highest likelihood-ratio counterpart was selected 
in the cases of multiple candidates. For example, if the observed X-ray emission comes from
a high-redshift obscured AGN, and it is gravitationally
lensed by a foreground AGN/galaxy, the foreground object could potentially have a higher matching 
likelihood ratio
and would be falsely selected as the counterpart. Lensing has affected the identification 
of submillimeter galaxies (SMGs), and $\approx5\%$ of bright SMGs are lensed
\citep[e.g.,][]{Blain1996,Chapman2002,Simpson2015,Danielson2017}. Our \xray\ sources are located 
at a smaller median redshift ($\approx1.6$ for AGNs) than typical SMGs ($\approx2.5$), and the lensed 
fraction is likely much lower.  
We estimate that such potential
false matches have a negligible contribution to the overall false-match rate derived from
the \hbox{likelihood-ratio} matching approach.

Multiwavelength identifications were performed with the following seven optical--radio 
catalogs (in order of increasing wavelength),\footnote{We also
examined the GOODS-S MUlticolour Southern Infrared Catalog (MUSIC) v2
$K$-band catalog \citep{Grazian2006} and the Multiwavelength Survey by Yale-Chile (MUSYC)
$K$-band catalog \citep{Taylor2009}. These two catalogs 
do not provide any additional useful counterpart information,
and thus we do not list them here.} and a primary counterpart was chosen from one of these catalogs when available.
\begin{enumerate}

\item
The Wide Field Imager (WFI) $R$-band catalog, with a 5$\sigma$ limiting magnitude
of 27.3 \citep{Giacconi2002,Giavalisco2004}.

\item
The GOODS-S {\it HST} version r2.0z $z$-band catalog, with a 5$\sigma$ limiting magnitude
of 28.2 \citep{Giavalisco2004}.\footnote{See
http://archive.stsci.edu/pub/hlsp/goods/catalog\_r2/.} This catalog covers a solid angle
of \hbox{$\approx160$~arcmin$^2$} in the center of the \cdfs\ (Figure~1).

\item
The Galaxy Evolution from Morphologies and SEDs (GEMS)
{\it HST} $z$-band catalog, with a 5$\sigma$ limiting magnitude of 
27.3 \citep{Caldwell2008}. The GEMS survey complements the GOODS-S survey and 
covers the entire remaining 
area of the \cdfs\ that is not covered by GOODS-S.

\item
The CANDELS $+$ 3D-HST {\it HST} WFC3 F125W$+$F140W$+$F160W combined
catalog (hereafter the CANDELS catalog; \citealt{Skelton2014}).
The magnitudes in the F125W band were used, which has
a 5$\sigma$ limiting magnitude of 28.3.

\item
The TENIS $K_s$-band catalog, with a 5$\sigma$ limiting magnitude of
25.0 in the inner 400 arcmin$^2$ region \citep{Hsieh2012}.

\item
The Spitzer Extended Deep Survey (SEDS) Infrared Array Camera (IRAC)
3.6~$\mu$m catalog, with a 3$\sigma$ limiting magnitude of $\approx26$
\citep{Ashby2013}.

\item
The Very Large Array (VLA) 1.4 GHz catalog from \citet{Miller2013}, with
a 5$\sigma$ limiting flux density of $\approx40$~$\mu$Jy.

\end{enumerate}

The absolute astrometry for each of the above catalogs was registered to the TENIS astrometric
frame before the matching; the systematic offsets were negligible except
for the VLA catalog, where $\approx0\farcs2$ shifts
in right ascension and declination were required (see Footnote~\ref{foot-astrometry}). 
Compared to the multiwavelength catalogs used for source identification
in the 4~Ms \cdfs\ catalog \citep{Xue2011}, additional deep NIR and 
IR survey catalogs have become available, including the TENIS, CANDELS,
and SEDS catalogs, which aided greatly with the 7~Ms source identification.
For each X-ray source that has at least one match from these catalogs, we chose a primary counterpart
from, in order of priority, the CANDELS, GOODS-S, GEMS,
TENIS, VLA, WFI, and SEDS catalogs. This order was empirically determined based on the combined
factors of positional accuracy, sensitivity, and potential source-blending problems (e.g., in the SEDS IRAC
catalog).
We identified primary counterparts for
982 (97.4\%) of the 1008 main-catalog sources, and 701, 26, 186, 49, 4, and 16 of them are from the
CANDELS, GOODS-S, GEMS, TENIS, WFI, and SEDS catalogs, respectively. There were no primary counterparts selected from 
the VLA catalog.
The false-match probabilities
for the matches found in the seven catalogs range from 0.14\% (VLA) to 4.0\% (WFI).
For each of 982 matched \xray\ sources, we consider its false-match probability to be the minimum
one among the false-match probabilities of the optical through radio catalogs where the X-ray source has a match.
For example, if the counterpart of an \hbox{X-ray} source is in the WFI catalog only, its false-match probability
is 4.0\%; if the counterpart is in both the WFI and VLA catalogs, 
its \hbox{false-match} probability is 0.14\%; if the counterpart is in all seven catalogs, 
its false-match probability is 0.14\%.
The
mean false-match rate for the entire sample, derived by averaging the false-match 
probabilities of 
individual sources, is 1.6\%.  

\begin{figure}
\centerline{
\includegraphics[scale=0.5]{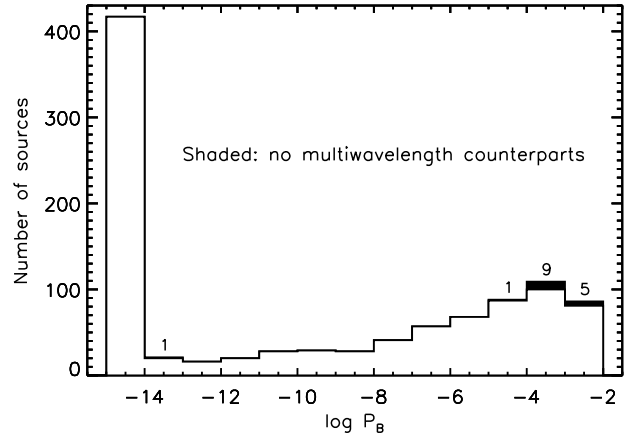}
}
\caption{
Distribution of the AE binomial no-source probability ($P_{\rm B}$)
for the main-catalog sources;
sources with $\log P_{\rm B}<-15$
are plotted in the $\log P_{\rm B}=-15$ to $-14$ bin.
The shaded regions indicate sources that have no multiwavelength
counterparts, with the numbers of such sources shown on the top of
each bin. A significant fraction of the 16 sources which lack counterparts
are likely spurious detections.
}
\label{fig-pbhist}
\end{figure}

We
examined the 26 X-ray sources which lack counterparts, and manually assigned multiwavelength matches to 
10 of them. Six of these X-ray sources have a CANDELS companion $\approx1\farcs0$--$1\farcs6$ away, and their
likelihood ratios fell slightly below our threshold value for matches. The other four \xray\ sources are within the
extent ($2\farcs1$--$4\farcs7$) of \hbox{low-redshift} galaxies ($z=0.038$--0.215) upon visual inspection, and they
are probably off-nuclear sources
associated with the galaxies (e.g., ultraluminous X-ray sources; ULXs) that
are \hbox{$\approx4$--12~kpc} away from the nuclei \citep[e.g.,][]{Lehmer2006}. 
These 10 manually matched sources are noted in
Column~20 of the main-catalog table.
In total, multiwavelength counterparts were identified 
for 992 (98.4\%) of the main-catalog sources. The false-match 
rate is around 1.6\% or slightly higher, considering any possible additional false matches in the 10 manually matched
cases, and it could be
as large as $\approx2.5\%$ in the extremely unlikely case that all the 10 manual matches are incorrect. 
For the 16 X-ray sources without counterparts, we expect that a significant fraction or even most of them
are spurious detections, as discussed in our AE source-filtering stage (Section~\ref{sec-detection}). 
The locations of these 16 sources in the $P_{\rm B}$ distribution for the main-catalog
sources are shown in Figure~\ref{fig-pbhist}, and they indeed have large $P_{\rm B}$ values in general, indicating
less-significant detections. We discuss the one source that is detected
significantly ($\log P_{\rm B}\approx-13$) but has no counterpart in Section~\ref{sec-notes-blank} below.

There are three pairs of X-ray sources (XIDs 431, 432, 556, 558, 649, 653) 
that were matched to the same counterparts, which are galaxies at redshifts of 0.038, 0.075, and 0.579,
respectively. The pair sources further away from the galactic centers 
could be off-nuclear sources associated with the galaxies \citep[e.g.,][]{Lehmer2006}.
For one pair of sources (XIDs 556 and 558), a strong radio counterpart is observed 
(1.4 GHz flux density 452~$\mu$Jy), and the two X-ray sources could be X-ray emission associated with 
extended radio jets/lobes.
These pair sources are noted in Column~20 of the main-catalog table,
and we further noted another possible off-nuclear source (XID 761) in this column.

For the 992 main-catalog sources with primary counterparts, we further searched for their 
multiwavelength photometric properties by matching the primary counterparts to the other optical--radio catalogs
above
with a matching radius of $0\farcs75$ or $1\farcs0$ (for the VLA
catalog only).
The multiwavelength information for the 992 sources in the
WFI, \hbox{GOODS-S}, GEMS, CANDELS,
TENIS, SEDS, and VLA catalogs are presented in \hbox{Columns~21--41} of the main-catalog table.

\subsection{Spectroscopic and Photometric Redshifts} \label{sec-redshift}
The superior multiwavelength data in the \cdfs\ allow collection of
spectroscopic and photometric redshifts for the majority of the identified X-ray sources.
We matched the primary counterparts to spectroscopic and photometric
catalogs using a matching radius of $0\farcs5$ or $0\farcs75$ (if the primary counterpart
is from the SEDS IRAC catalog).
Spectroscopic redshifts were searched for in $\approx30$ public catalogs including a few
X-ray spectroscopic redshift results, and redshifts for 
665 of our 
main-catalog sources were collected from 26 of these catalogs, listed below: 
(1) \citet{Colless2003}, (2) \citet{Szokoly2004}, (3) \citet{Zheng2004}, 
(4) \citet{Doherty2005}, (5) \citet{Mignoli2005}, (6) \citet{Ravikumar2007},
(7) \citet{Kriek2008}, (8) \citet{Vanzella2008}, (9) \citet{Treister2009},
(10) \citet{Wuyts2009}, (11) \citet{Balestra2010}, (12) \citet{Silverman2010},
(13) \citet{Casey2011}, (14) \citet{Feruglio2011}, (15) \citet{Xia2011}, 
(16) \citet{Cooper2012}, (17) \citet{Iwasawa2012}, (18) \citet{Mao2012},
(19) \citet{Kurk2013}, (20) \citet{LeFevre2013}, (21) \citet{Rauch2013}, 
(22) \citet{Delmoro2014},
(23) \citet{Hsu2014}, (24) \citet{Morris2015}, (25) \citet{Santini2015}, and 
(26) \citet{Tasca2016}. The spectroscopic redshifts were flagged as ``Secure'' or ``Insecure''
depending on whether they were obtained from several reliable spectral features 
with $\gtrsim95\%$ confidence level.
The spectroscopic redshifts for Galactic stars were set to zero;
there are 12 stars in our catalog. 
The spectroscopic redshifts, quality flags, and the catalogs that the redshifts were collected from
(numbered 1--26 as cited above) are presented in Columns~42--44 of the main-catalog table. 
We collected photometric redshifts from the following public catalogs: (1) \citet{Luo2010},
(2) \citet{Rafferty2011}, (3) \citet{Hsu2014}, (4) \citet{Skelton2014}, (5) \citet{Santini2015},
and (6) \citet{Straatman2016}.
Unlike spectroscopic redshifts, photometric redshifts from different catalogs 
sometimes do not agree with each
other, and thus we present all the available photometric redshifts in Columns~45--50 of the main-catalog table.
In total, 985 of our main-catalog sources have at least one photometric redshift.\footnote{The 
counterpart of XID 679
is blended with a brighter optical/NIR source $\approx1\farcs6$ away; the two sources are only
resolved in the CANDELS catalogs among the optical through radio catalogs we examined.
The spectroscopic and photometric redshifts collected for XID 679 are likely based on the 
spectroscopic and photometric properties of the companion object.}

For each source, we adopted a preferred redshift from, in order of priority, (1) a secure spectroscopic redshift,
(2) an insecure spectroscopic redshift that agrees with at least one of its photometric redshifts to within
15\% ($|z_{\rm phot}-z_{\rm spec}|/(1+z_{\rm spec})\le0.15$; an empirical choice driven by
experience), (3) a \citet{Hsu2014} photometric redshift,
(5) a \citet{Luo2010} photometric redshift, (5) a \citet{Straatman2016} photometric redshift,
(6) a \citet{Skelton2014} photometric redshift,
(7) a \citet{Santini2015} photometric redshift, (8) a \citet{Rafferty2011} photometric redshift, and (9) an
insecure spectroscopic redshift (when it is the only redshift available).
The \citet{Hsu2014} photometric redshifts were preferred in general among all the 
photometric redshifts because of the combined factors that 
(1) \citet{Hsu2014} is a dedicated study of the 
\cdfs\ photometric redshifts, (2) it utilized intermediate-band photometric data, 
(3) the resulting photometric redshifts have good accuracy overall, 
(4) the details of the SED fitting of individual sources are publicly available, and (5) 
the highest fraction (94\%) of our main-catalog sources have matches in this catalog.
In addition, for 16 sources, we adopted the \citet{Luo2010} photometric redshifts instead of the
\citet{Hsu2014} photometric redshifts after reviewing the SED fitting plots.

\begin{figure*}
\centerline{
\includegraphics[scale=0.5]{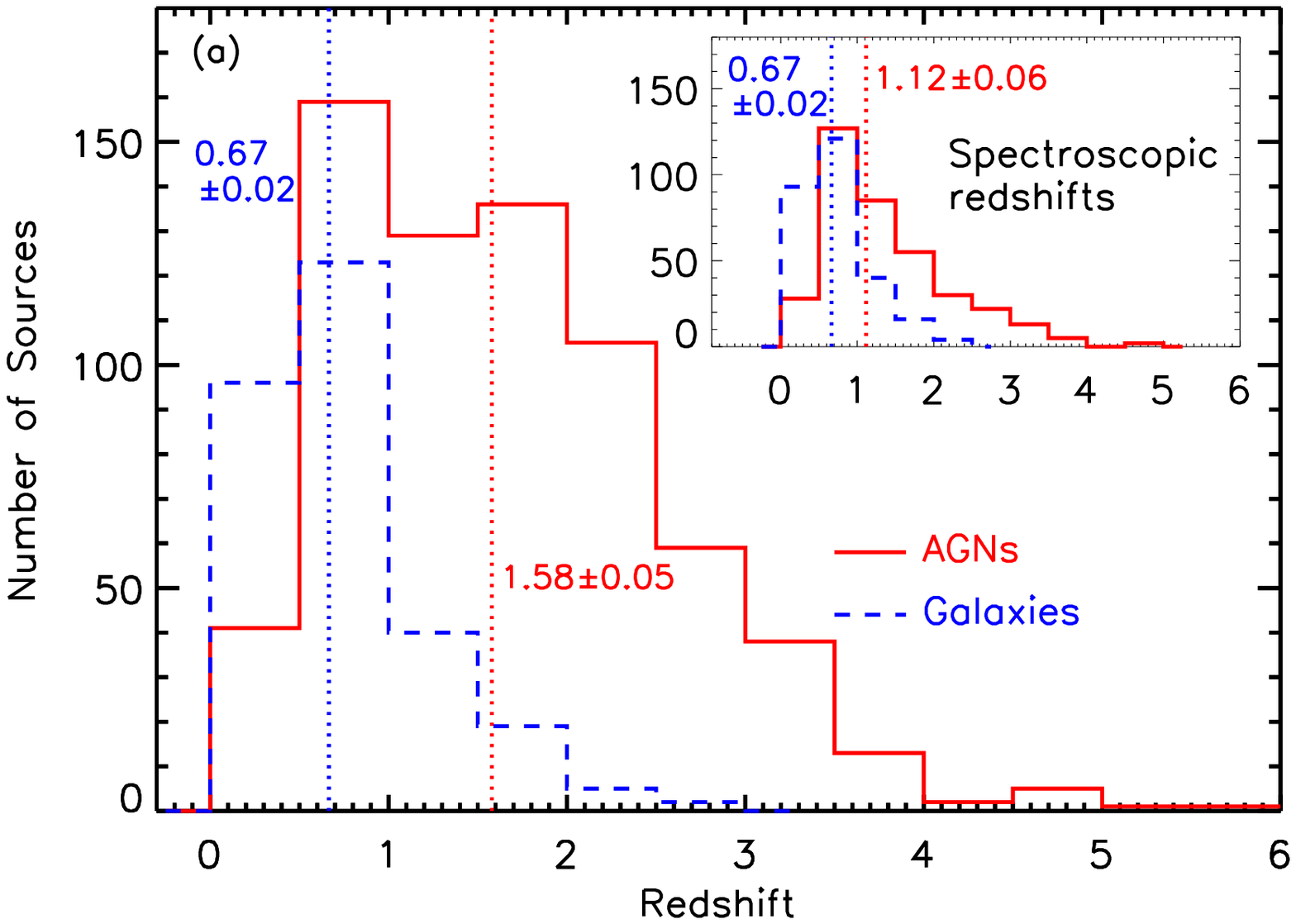}
\includegraphics[scale=0.5]{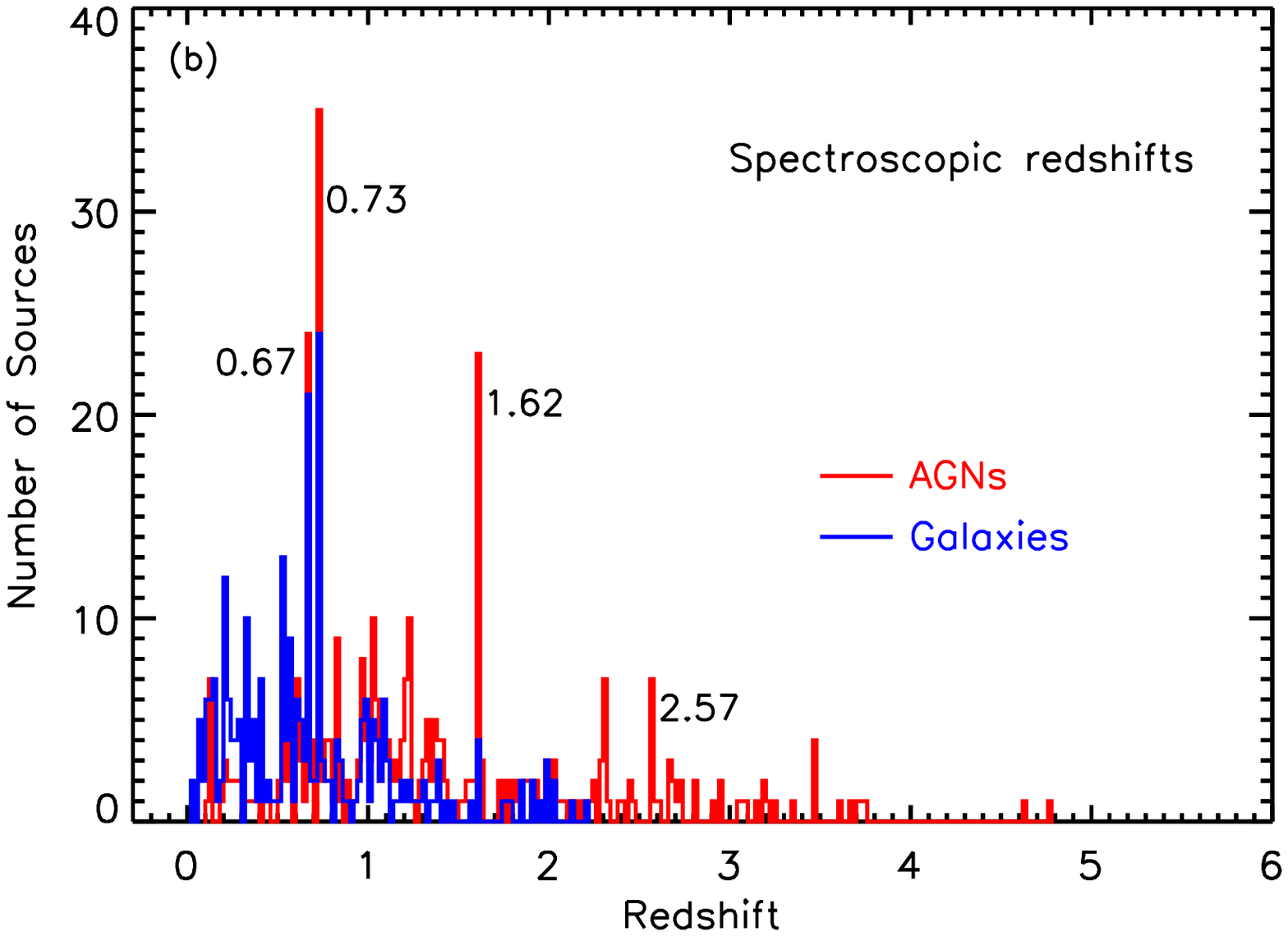}
}
\caption{
(a) Redshift distributions
for the 689 AGNs (red histogram) and 285 galaxies
(blue histogram) in bins of $\Delta z=0.5$.
The inset displays the redshift distribution for the 367 AGNs 
and 274 galaxies with final adopted spectroscopic redshifts.
The vertical dotted lines indicate the
median redshift for every distribution, and the corresponding
median values and their
1$\sigma$ uncertainties (derived via bootstrapping) are listed.
(b) Redshift distributions for the 367 AGNs
and 274 galaxies with final adopted spectroscopic redshifts
in bins of $\Delta z=0.02$.
Some of the prominent redshift spikes are noted.
}
\label{fig-zhist}
\end{figure*}

Out of the 992 main-catalog sources with a primary counterpart,
986 (99.4\%) have final adopted redshifts, and we present these preferred redshifts and their origins
in Columns~51--52 of the main-catalog table. For those adopted photometric redshifts,
we also quote their $1\sigma$ uncertainties in Columns~53--54 of the table, although
we caution that these uncertainties often underestimate the real errors 
(e.g., see Section~3.4 of \citealt{Luo2010}). The redshift distributions for the main-catalog sources
are shown in Figure~\ref{fig-zhist}a. 
The median
redshift for all the X-ray sources is $1.12\pm0.05$
with an interquartile range of 0.67--1.95,
where
the 1$\sigma$ uncertainty on the median value was derived via
bootstrapping.\footnote{See http://www.stat.wisc.edu/\textasciitilde larget/math496/bootstrap.html.}
In Figure~\ref{fig-zhist}b we display the distributions of spectroscopic redshifts 
in fine redshift bins 
($\Delta z=0.02$); there are some prominent redshift spikes indicative of X-ray
large-scale structures (e.g., $z=0.67$, 0.74, 1.62, and 2.57; e.g., \citealt{Gilli2003,Silverman2010,Dehghan2014,Finoguenov2015}).
In the main catalog, there are two sources having $z>5$, and both are photometric redshifts.
XID 172 is at $z\approx5.2$ from \citet{Hsu2014}, and it has $z\approx5.7$ from \citet{Luo2010}
and $z\approx7.7$ from \citet{Straatman2016}; it is outside the CANDELS region and has 
TENIS, SEDS, and VLA counterparts. XID 238 is at $z\approx5.8$ from \citet{Skelton2014};
it only has CANDELS and VLA counterparts.\footnote{The VLA counterpart is 1\farcs3 away from
the CANDELS counterpart, farther than our adopted 1\farcs0 matching radius. We manually assigned the match.} 
XID 172 is likely X-ray absorbed ($\Gamma_{\rm eff}=1.1$), and XID 238 appears to be a soft \xray\ source
($\Gamma_{\rm eff}=2.3$).
It is probable that both sources are high-redshift AGNs.

\begin{figure*}
\centerline{
\includegraphics[scale=0.5]{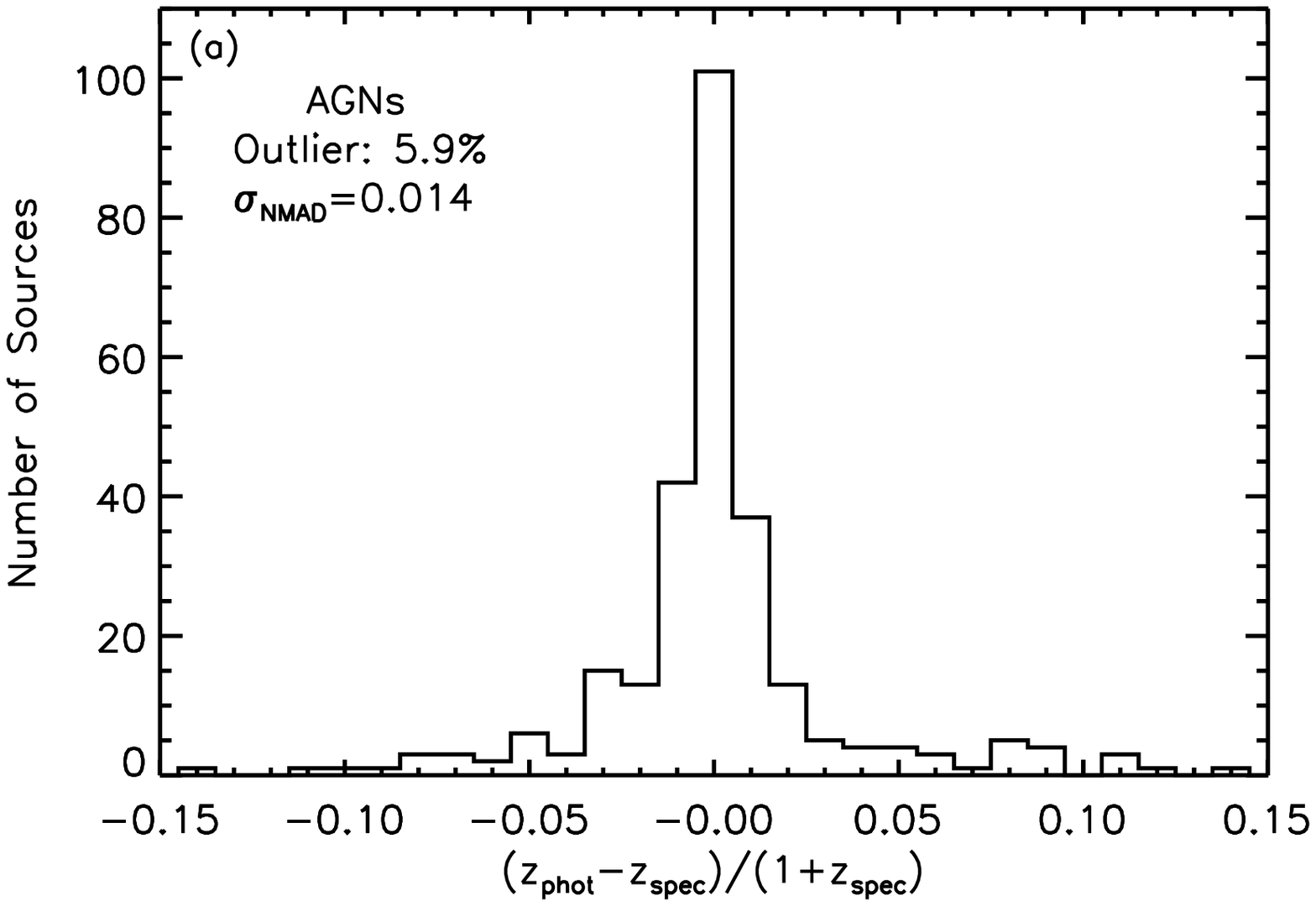}
\includegraphics[scale=0.5]{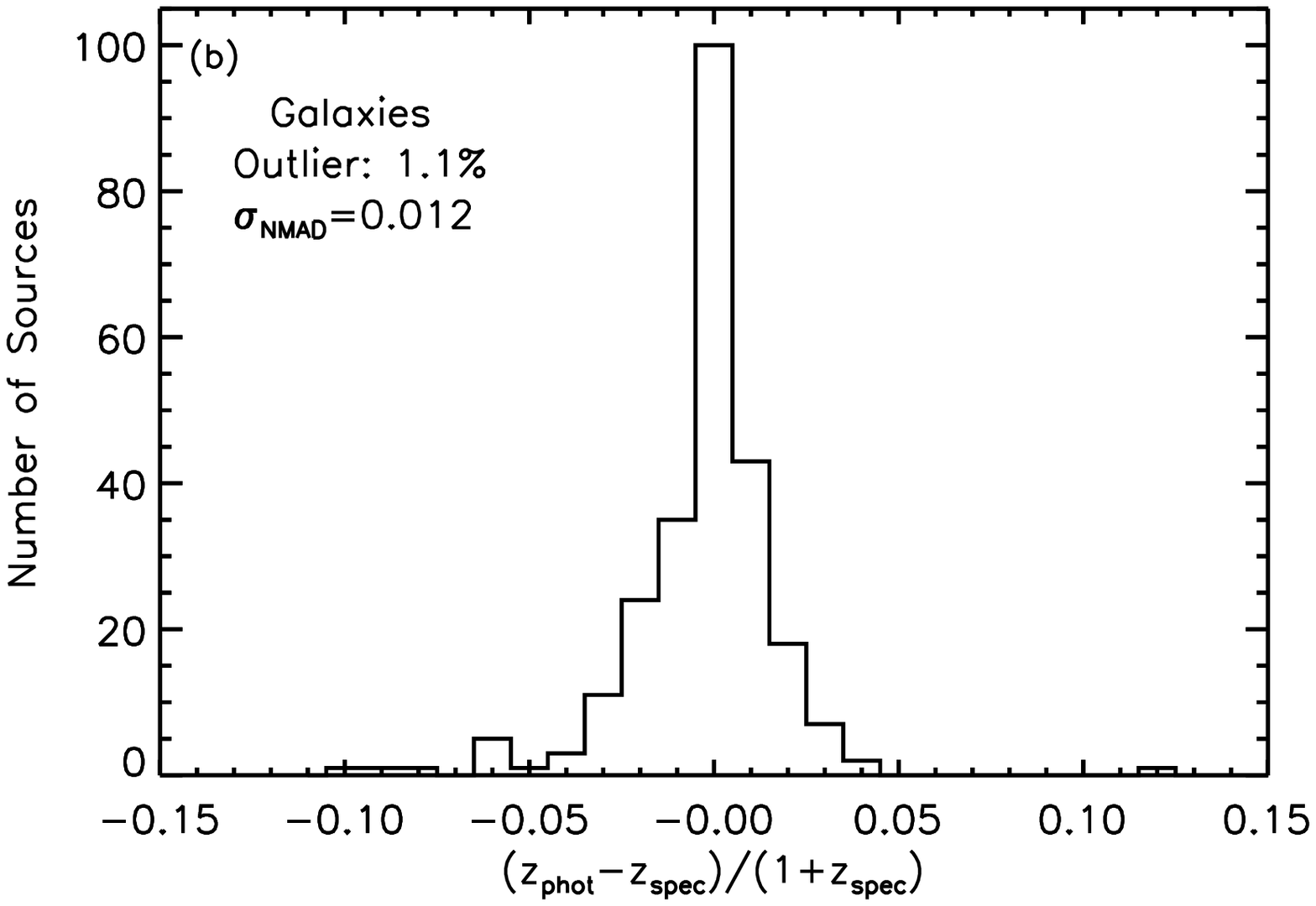}
}
\caption{
Distribution of the accuracy of the photometric redshifts
for the (a) 290 AGNs and (b) 256 galaxies in the main catalog
that have photometric redshifts from \citet{Hsu2014} and also 
secure spectroscopic redshifts.
The outlier fractions and the redshift accuracy indicators ($\sigma_{\rm NMAD}$) 
are displayed 
in each panel.
}
\label{fig-zphot}
\end{figure*}

\begin{figure}
\centerline{
\includegraphics[scale=0.5]{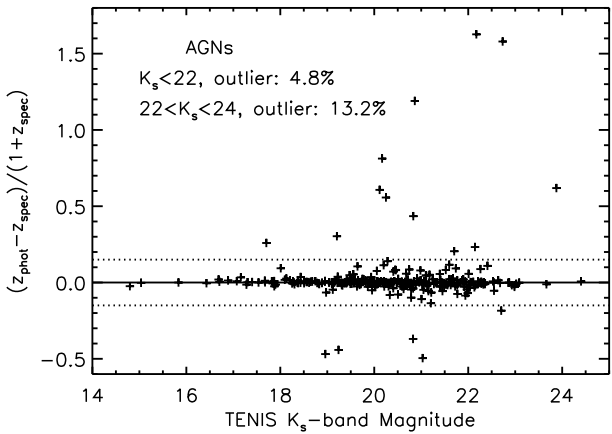}
}
\caption{
Photometric redshift accuracy vs.\ TENIS $K_s$-band magnitude for the 
AGNs in the main catalog
that have photometric redshifts from \citet{Hsu2014} and also
secure spectroscopic redshifts.
The solid and dashed lines indicate $(z_{\rm phot}-z_{\rm spec})/(1 + z_{\rm spec})$
values of 0, 0.15, and $-0.15$, respectively.
The outlier fraction of the photometric redshifts increases 
toward larger \hbox{$K_s$-band} magnitudes. 
}
\label{fig-zphotks}
\end{figure}

Of the 986 final adopted redshifts, 653 are spectroscopic redshifts (including the 12 Galactic stars) 
and 
333 are photometric redshifts.  
Most (284/333) of the photometric redshifts are from \citet{Hsu2014}.
We assess the quality of the \citet{Hsu2014} photometric redshifts
by comparing them to the available secure spectroscopic redshifts for our sources.
The comparison was performed for the 290 AGNs and 256 galaxies (see Section~\ref{sec-classify} below
for AGN classification) in the main catalog
which have both photometric redshifts from \citet{Hsu2014} and
secure spectroscopic redshifts.
We calculated the fraction of outliers defined
as having $|z_{\rm phot}-z_{\rm spec}|/(1 + z_{\rm spec})>0.15$, and we estimated the 
accuracy of the photometric redshifts by computing   
the normalized median absolute deviation of the redshift differences,
defined as
$\sigma_{\rm NMAD}=1.48\times {\rm median}\left({|z_{\rm phot}-z_{\rm spec}-
{\rm median}(z_{\rm phot}-z_{\rm spec})|}/(1+z_{\rm spec}
)\right)$ \citep[e.g.,][]{Luo2010}. 
The results are presented in Figure~\ref{fig-zphot}. The photometric redshifts
are of high quality in general, especially for the galaxies. 
Note that some of the spectroscopic redshifts were used to
train the SED templates in
\citet{Hsu2014}, which may bias the results toward better accuracy.
The quality of the photometric redshifts also appears to depend on 
source brightness; Figure~\ref{fig-zphotks} shows that 
the outlier fraction for the AGN photometric redshifts
increases toward larger $K_s$-band magnitudes (see also, e.g., 
Section~3.4 of \citealt{Luo2010}).
Among the 333 sources with adopted photometric redshifts (322 AGNs),
93 have $K_s\le22$ ($\approx5\%$ outlier fraction according to Figure~\ref{fig-zphotks}),
146 have $22<K_s\le24$ ($\approx13\%$ outlier fraction), and 94 have $K_s>24$.
There is not a sufficient number of spectroscopic redshifts at $K_s>24$
for assessing the quality of these photometric redshifts.
Assuming arbitrarily a 15\% (30\%) outlier fraction for sources with $K_s>24$,
the average outlier fraction for the 333 sources with adopted photometric redshifts
is $\approx11\%$ (16\%).

\subsection{X-ray Photometric and Basic Spectroscopic Properties} \label{sec-xphotmetric}
The aperture-corrected net source counts were derived from the AE extraction results.
For each source in each of the three X-ray bands, if the $P_{\rm B}$ value is less than 
our adopted threshold (0.007), we consider it as being detected and present in the main catalog the
number of source counts along with the associated $1\sigma$ statistical uncertainties 
computed by AE following \citet{Gehrels1986}, otherwise it is considered undetected and we 
present the 90\% confidence-level upper limit on the source counts following the \citet{Kraft1991}
Bayesian method.
The aperture-correction factors were derived from the energy-dependent correction factors estimated by
AE for individual observations
(see Section 3.2 of \citealt{Xue2011} for
details), and the median correction factors for the full, soft, and hard bands are 0.885, 0.907, 
and 0.843, respectively.
The net source counts and their uncertainties are presented in Columns~8--16 of 
the main-catalog table.
Sources near the edge of the \cdfs\ field have relatively low effective exposure times, 
large PSF sizes, non-uniform local background,
and sometimes substantially varying (up to a factor of a few) 
effective exposure times within the extraction apertures. 
Additional photometric uncertainties for these sources
might arise besides the statistical uncertainties presented in the current catalog, especially
for faint sources. We noted 45 such sources in Column~17 of the main-catalog table (marked with
``E'') that are covered by less than 20 of the 102 \cdfs\ observations (4--19 observations); these
sources all have large off-axis angles (9\farcm9--12\farcm4) and relatively low effective exposure times
(37--847~ks).
We also noted in Column~17 another 34 sources that are in crowded regions 
(marked with ``C'') and were extracted using $\approx40\%$--74\% ECF apertures instead of the
standard $\approx90\%$ ECF apertures; the photometry of these sources might still have some mild
contamination from the companion sources.

\begin{figure}
\centerline{
\includegraphics[scale=0.5]{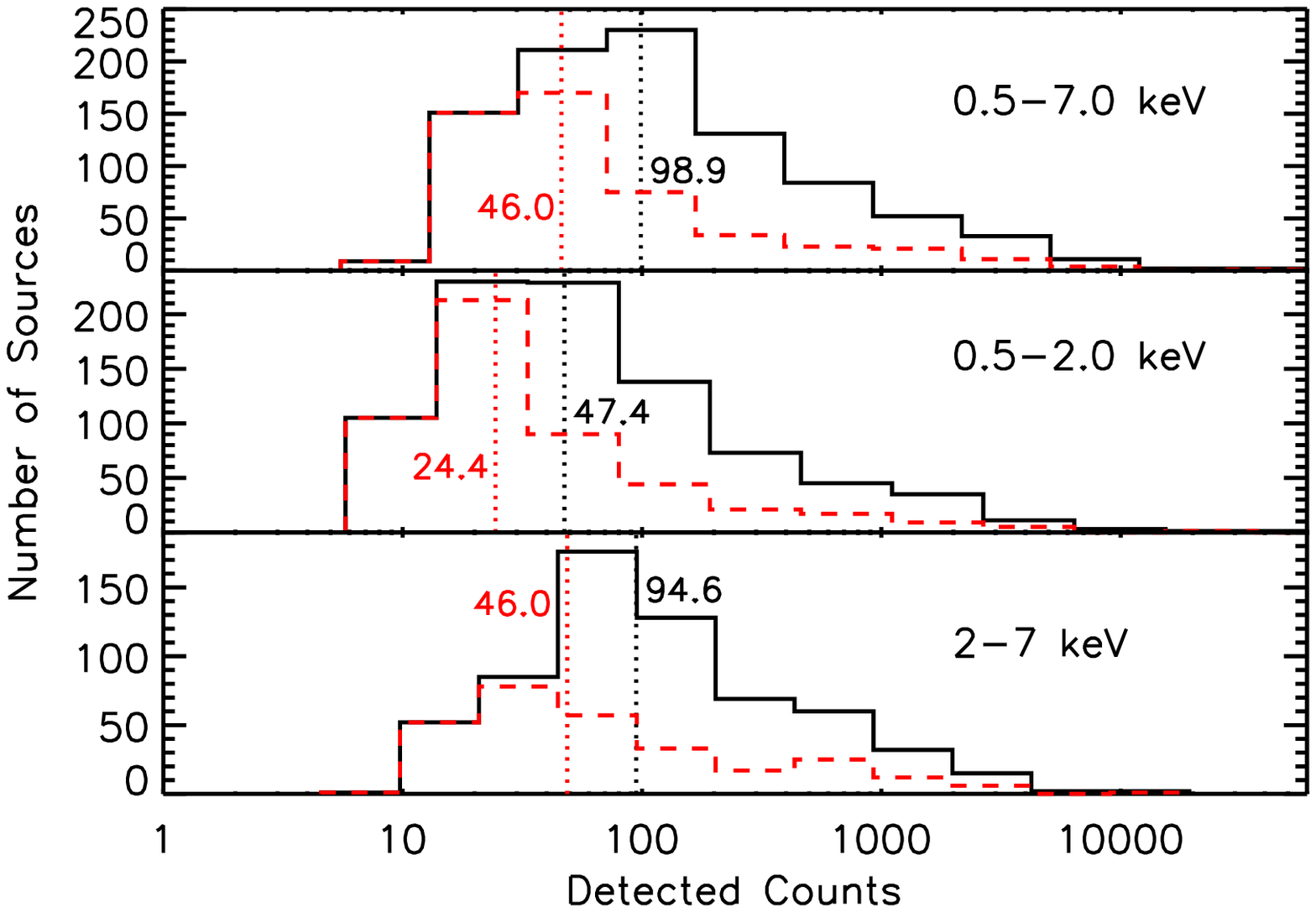}
}
\caption{
Distributions of source counts for the main-catalog
sources in the full (top), soft (middle), and hard (bottom) bands.
The black-solid histograms show the count
distributions for all the sources in the \cdfs\ field, while
the red-dashed histograms show the
distributions for the subgroups of
sources within $6\arcmin$ of the average aim point.
Sources with upper limits on the counts are not included in the plots.
The vertical dotted lines indicate the
median numbers of counts for every distribution, and the corresponding
median values are listed.
}
\label{fig-cnthist}
\end{figure}

The distributions of the source counts in the three bands are displayed in
Figure~\ref{fig-cnthist}. 
Table~\ref{tbl-cnt} summarizes the basic statistics of the source counts in the three bands;
the median numbers of detected counts in the full, soft, and hard bands are 
$98.9\pm6.1$, $47.7\pm2.0$, and $94.6\pm6.0$, respectively,
where
the 1$\sigma$ uncertainties on the median values were derived via
bootstrapping.
In Table~\ref{tbl-det}, we provide the numbers of sources 
detected in one band but not another; there are 22 sources detected only in the full band,
84 only in the soft band, and 8 only in the hard band.
There are 456 sources with $>100$ full-band counts, allowing basic spectral analyses, and there
are 90 sources with $>1000$ full-band counts.

While detailed spectral analyses of the X-ray sources are beyond the scope of the current
study and will be presented in additional papers (e.g., \citealt{Yang2016}; T.~Liu et al., in preparation),
we still derived basic spectroscopic properties for the catalog sources.
Assuming the \hbox{0.5--7.0~keV} spectra of the X-ray sources are power laws modified by only 
Galactic absorption, we derived the effective power-law photon indices 
($\Gamma_{\rm eff}$) from the hard-to-soft band ratios.
The band ratio is defined as the ratio between the hard-band and soft-band count rates,
and the count rate was computed by dividing the aperture-corrected net counts by the effective exposure time.  
For the 502 sources detected in both the soft and hard bands, we computed the $1\sigma$ 
uncertainties of the band ratios using the Bayesian code {\sc behr} \citep{Park2006}.
For the 479 sources detected in either the soft or hard band, but not both, we adopted
the mode values of the \hbox{band-ratio} probability density distributions 
calculated using {\sc behr} as the band ratios; these are not upper or lower limits but 
best-guess estimates.\footnote{Although considerably uncertain, 
these values appear more appropriate for computing the source fluxes than 
assuming simply a uniform spectral shape (e.g., $\Gamma_{\rm eff}=1.4$) for
all such sources.} These band ratios 
were only used for estimating $\Gamma_{\rm eff}$, source fluxes,
and intrinsic absorption column densities, and
their uncertainties were not calculated.
For the 22 sources detected only in the full band, the band ratios cannot be constrained, and 
$\Gamma_{\rm eff}=1.4$ was adopted for them.
For each source, we calibrated the relation between the effective photon index and band ratio
using simulated spectra produced by the {\sc fakeit}
command in XSPEC (version 12.9.0; \citealt{Arnaud1996}) with the AE-generated 
merged spectral response files for this source. 
The \hbox{band-ratio-to-$\Gamma_{\rm eff}$} conversion factors differ slightly for different sources.
The uncertainties on $\Gamma_{\rm eff}$ were calculated following the error propagation 
method in Section 1.7.3 of \citet{Lyons1991}.
The band ratios and effective photon indices are presented in Columns~58--63 of the main-catalog table.

\begin{figure}
\centerline{
\includegraphics[scale=0.5]{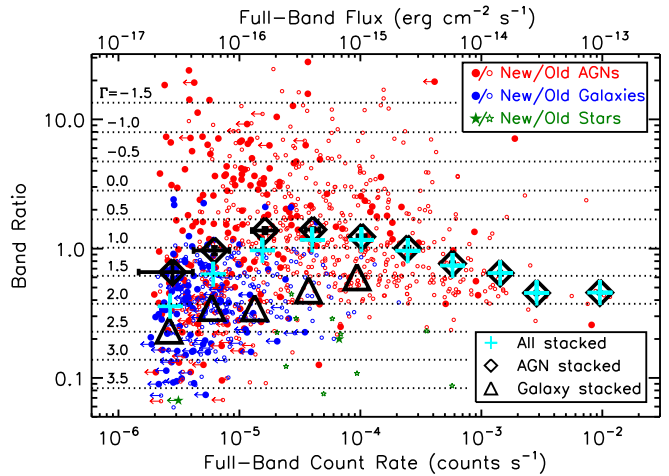}
}
\caption{
Band ratio vs.\ full-band count rate for the main-catalog sources.
Sources having upper limits on the count rates are indicated
by the arrows. Sources detected only in the full band (22 sources)
are not included
in this plot.
Red, blue, and green symbols represent AGNs, galaxies, and stars,
respectively. Filled symbols indicate newly detected sources, while
open symbols are sources that were present in the \citet{Xue2011} 4~Ms \cdfs\ catalogs.
The horizontal dotted lines show the band ratios corresponding to
given effective photon indices; these were computed using the
average of the XSPEC-derived $\Gamma_{\rm eff}$-to-band-ratio conversions.
The top $x$-axis displays the full-band fluxes at the corresponding
count rates, derived assuming $\Gamma_{\rm eff}=1.4$.
The cyan plus signs represent the stacked count rates and band ratios
for all the sources within logarithmic count-rate bins of 0.4, while
the black diamonds and triangles are the stacked values for the
AGNs and galaxies, respectively. The error bars for the stacked AGN data points
are shown to illustrate the typical uncertainties of these stacked values;
they become smaller than the symbol size in high count-rate bins.
}
\label{fig-bratio}
\end{figure}

Figure~\ref{fig-bratio} illustrates the band ratio as a function of \hbox{full-band} count rate
for the main-catalog sources; the corresponding average $\Gamma_{\rm eff}$ and full-band flux values are 
also shown. We split the sources into several count-rate bins and present their stacked count rates and 
band ratios, and we investigated these average values for
the AGN and galaxy populations respectively (see Section~\ref{sec-classify} below for source 
classification).
For the AGNs, the average band ratio rises when the count rate declines from $\approx10^{-2}$
to $\approx10^{-4}$~counts~s$^{-1}$, and it drops as the count rate decreases further below
$\approx10^{-5}$~counts~s$^{-1}$. A similar trend is present if we compute the average band ratios
using the median values instead of stacking. The rise in band ratio
(decrease in $\Gamma_{\rm eff}$) toward lower fluxes (harder when fainter)
at high count rates
has been observed in 
previous deep surveys \citep[e.g.,][]{Alexander2003,Lehmer2005,Luo2008,Xue2011}, and it is
caused by an increase 
in the fraction of absorbed AGNs detected at lower fluxes. The decline 
at the lowest count rates (softer when
fainter)
was also weakly present in the 4~Ms \cdfs\ (Figure 18 of \citealt{Xue2011}), and it is likely due to
the bias of preferentially detecting soft-band sources at the lowest flux levels. 
At high count rates (high fluxes), the numbers of \hbox{soft-band} and \hbox{hard-band} detected sources are comparable,
while at a count rate of $10^{-5}$~counts~s$^{-1}$, the \cdfs\ area that is sensitive for detecting such 
a soft-band
source is $\approx1.5$ times the area that is sensitive for detecting a hard-band source
(see Section~7 below for the sensitivity analysis), and a similar 
ratio is observed between the detected numbers of soft-band and \hbox{hard-band} sources. The sensitivity difference
is more pronounced at even lower count rates, and thus it could cause the apparent softer-when-fainter trend.
If we stacked only sources that are detected in both the soft and hard bands, 
or if we stacked only sources
within the innermost 3\arcmin-radius region where the \hbox{soft-band} 
and hard-band sensitivity difference is small,
the average band ratio does not appear to drop below the count rate of $10^{-5}$~counts~s$^{-1}$.
This bias should also be responsible,
at least partially, for a similar trend observed for galaxies in the low count-rate regime
in Figure~\ref{fig-bratio}.

\begin{figure}
\centerline{
\includegraphics[scale=0.5]{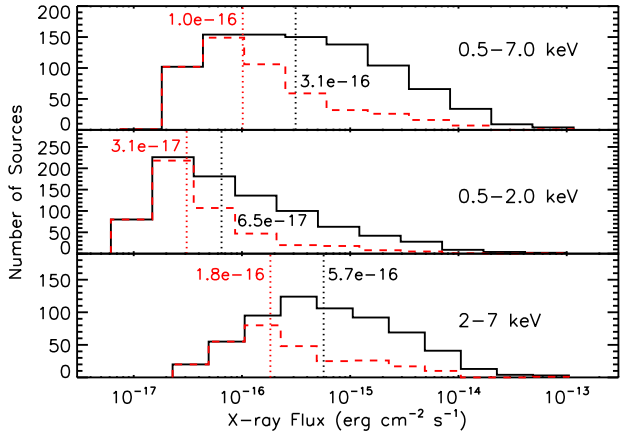}
}
\caption{
Distributions of X-ray fluxes for the main-catalog sources
in the
full (top), soft (middle), and hard (bottom) bands.
The black-solid histograms show the flux
distributions for all the sources in the \cdfs\ field, while
the red-dashed histograms show the
distributions for the subgroups of
sources within $6\arcmin$ of the average aim point.
Sources with upper limits on the fluxes are not included in the plots.
The vertical dotted lines indicate the median fluxes
for every distribution, and the corresponding
median values are listed.
}
\label{fig-fluxhist}
\end{figure}

\begin{figure}
\centerline{
\includegraphics[scale=0.5]{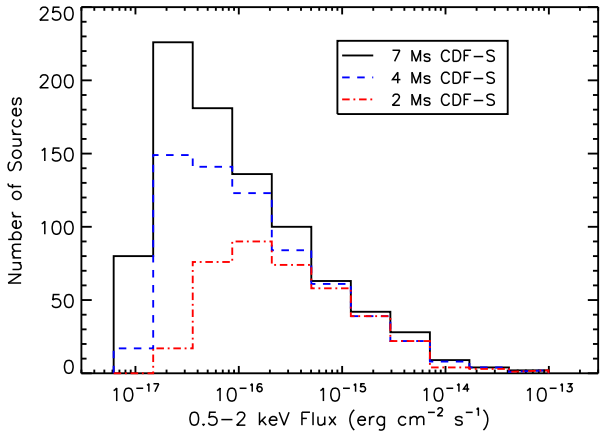}
}
\caption{
Soft-band flux distribution for the 7~Ms \cdfs\ compared to those
for the 4~Ms \cdfs\ \citep{Xue2011} and 2~Ms \cdfs\ \citep{Luo2008},
showing the improvement of source detection from deeper exposures.
The same flux binning was used for all three histograms. 
}
\label{fig-fluxhistcom}
\end{figure}

Using XSPEC and the AE-generated
merged spectral response files for each source,
we converted the count rate or upper
limit on the count rate to the corresponding flux or flux upper limit, assuming
that the spectrum is a power law having a photon index of $\Gamma_{\rm eff}$
modified by Galactic absorption.
The distributions of the source fluxes in the three bands are displayed in
Figure~\ref{fig-fluxhist}; the median fluxes in the full, soft, and hard bands are
$3.1\times10^{-16}$, $6.5\times10^{-17}$, and $5.7\times10^{-16}$~\flux, respectively.
In Figure~\ref{fig-fluxhistcom}, we present the soft-band flux distributions for the 
7~Ms \cdfs, 4~Ms \cdfs\ \citep{Xue2011}, and 2~Ms \cdfs\ \citep{Luo2008}, respectively.
It is clearly visible that significant numbers of new and fainter sources are detected
with increased exposure times.

\begin{figure}
\centerline{
\includegraphics[scale=0.5]{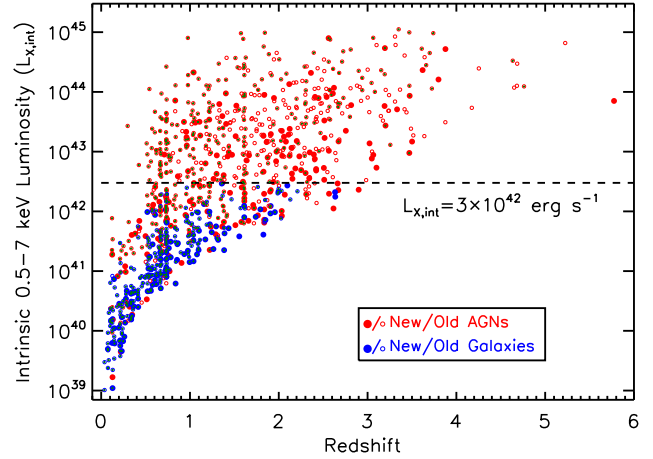}
}
\caption{
Intrinsic rest-frame 0.5--7.0 keV luminosity (in \lum)
vs.\ redshift for the
main-catalog sources.
Red and blue symbols represent AGNs and galaxies,
respectively; filled symbols indicate newly detected sources, while
open symbols are sources that were present in the \citet{Xue2011}
4~Ms CDF-S catalogs.
Tiny green dots mark sources 
with spectroscopic redshifts.
The horizontal dashed line indicates $L_{\rm X,int}=3\times10^{42}$~\lum,
which is one of the criteria utilized to classify AGNs.
The 22 sources which lack redshifts and
the 12 stars are
not included in this plot.
}
\label{fig-lz}
\end{figure}

Applying the $K$ correction assuming a power-law spectrum
and also correcting for Galactic absorption, we computed the  
apparent rest-frame \hbox{0.5--7.0~keV} luminosity ($L_{\rm X}$) from, in order of priority, the observed
full-band, soft-band, or \hbox{hard-band} 
flux; the luminosities derived from the fluxes in different energy bands
are actually consistent with each other as the same spectral shapes were adopted throughout all these 
calculations.
These \xray\ luminosities have not been corrected for any intrinsic
absorption and hence are referred to as ``apparent''. We further used 
the Portable, Interactive, Multi-Mission Simulator
(PIMMS)\footnote{\url{http://cxc.harvard.edu/toolkit/pimms.jsp}.} to estimate 
intrinsic absorption by
assuming that the intrinsic power-law spectrum has a fixed photon index of 1.8 and any 
effective photon index smaller than this value is caused by intrinsic absorption. In this manner, 
we estimated
intrinsic absorption column densities ($N_{\rm H,int}$) for 701 sources, which range from 
$2.3\times 10^{19}$~cm$^{-2}$ to $1.9\times 10^{24}$~cm$^{-2}$
with a median value of $4.9\times 10^{22}$~cm$^{-2}$. 
For sources with effective photon indices greater than 1.8, the intrinsic absorption
column densities were set to zero.
With the estimated intrinsic column densities, the absorption-corrected intrinsic 
\hbox{0.5--7.0~keV} luminosities ($L_{\rm X,int}$) were 
computed, and the correction factors range from 1--240 with a median value of 2.8 for the 
701 sources having $N_{\rm H,int}>0$.
The distribution of $L_{\rm X,int}$
as a function of the 
source redshift is displayed in Figure~\ref{fig-lz}. 
There are 613 sources having $L_{\rm X,int}>10^{42}$~\lum\ and 108 sources having
$L_{\rm X,int}>10^{44}$~\lum.
The fluxes and luminosities are presented in Columns~64--69 of the main-catalog table;
we did not compute X-ray luminosities for the 12 stars or the 22 sources which lack redshifts.

\subsection{AGN Classification, Source Spatial Distribution, and Postage-Stamp Images} \label{sec-classify}

We classified AGNs from the detected X-ray sources by selecting sources having X-ray
and/or multiwavelength properties significantly different from those of typical normal galaxies.
Besides AGNs and Galactic stars, the other X-ray sources are considered to be normal galaxies,
although it is
possible that some of these galaxies host low-luminosity and/or heavily obscured AGNs
where the AGN signatures are not evident based on our selection criteria; some of these
missed AGNs could 
be identified by other means such as \hbox{X-ray} variability (e.g., \citealt{Young2012}; 
P.~Shao et al.,
in preparation).

\begin{figure*}
\centerline{
\includegraphics[scale=0.5]{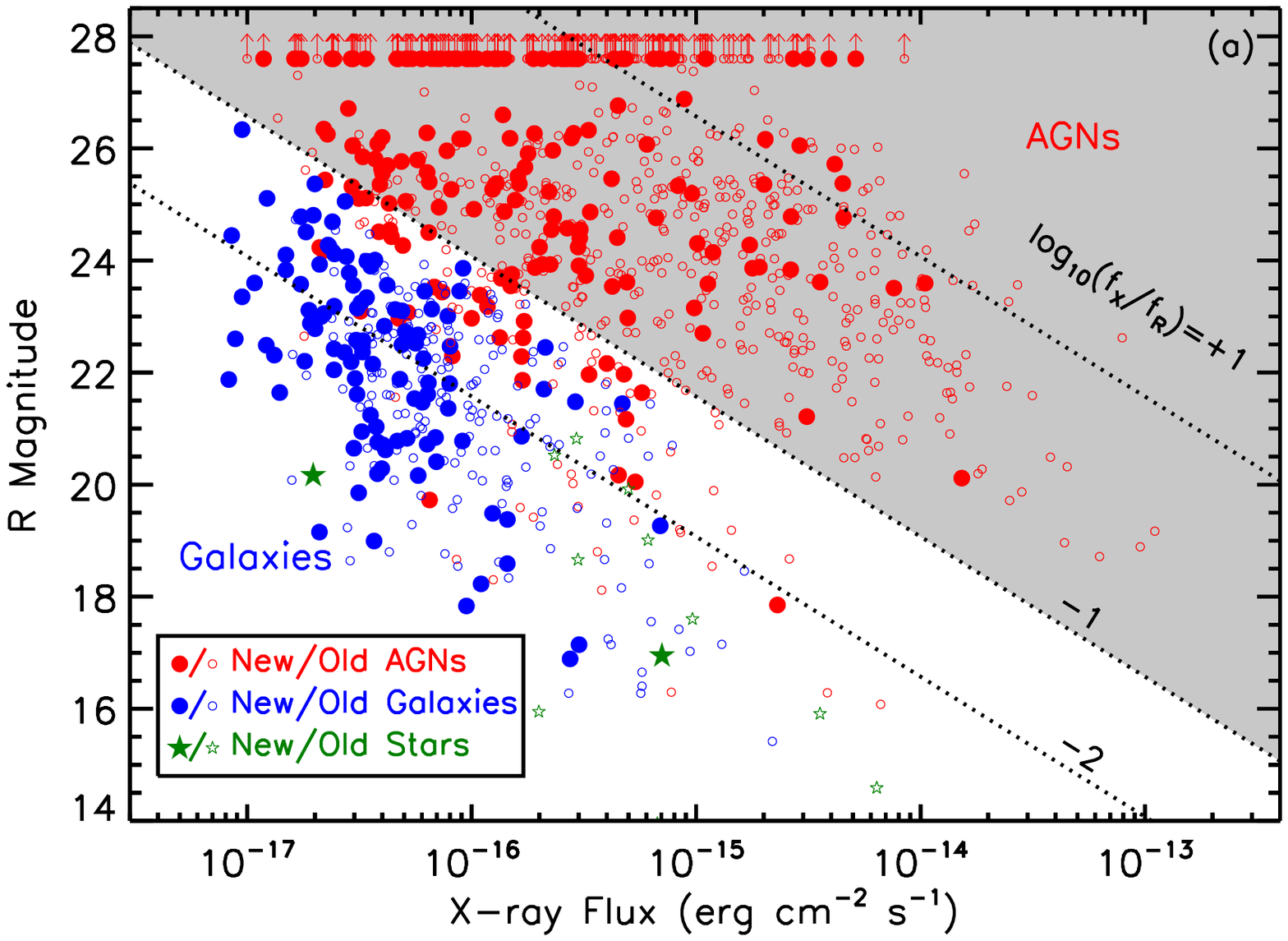}
\includegraphics[scale=0.5]{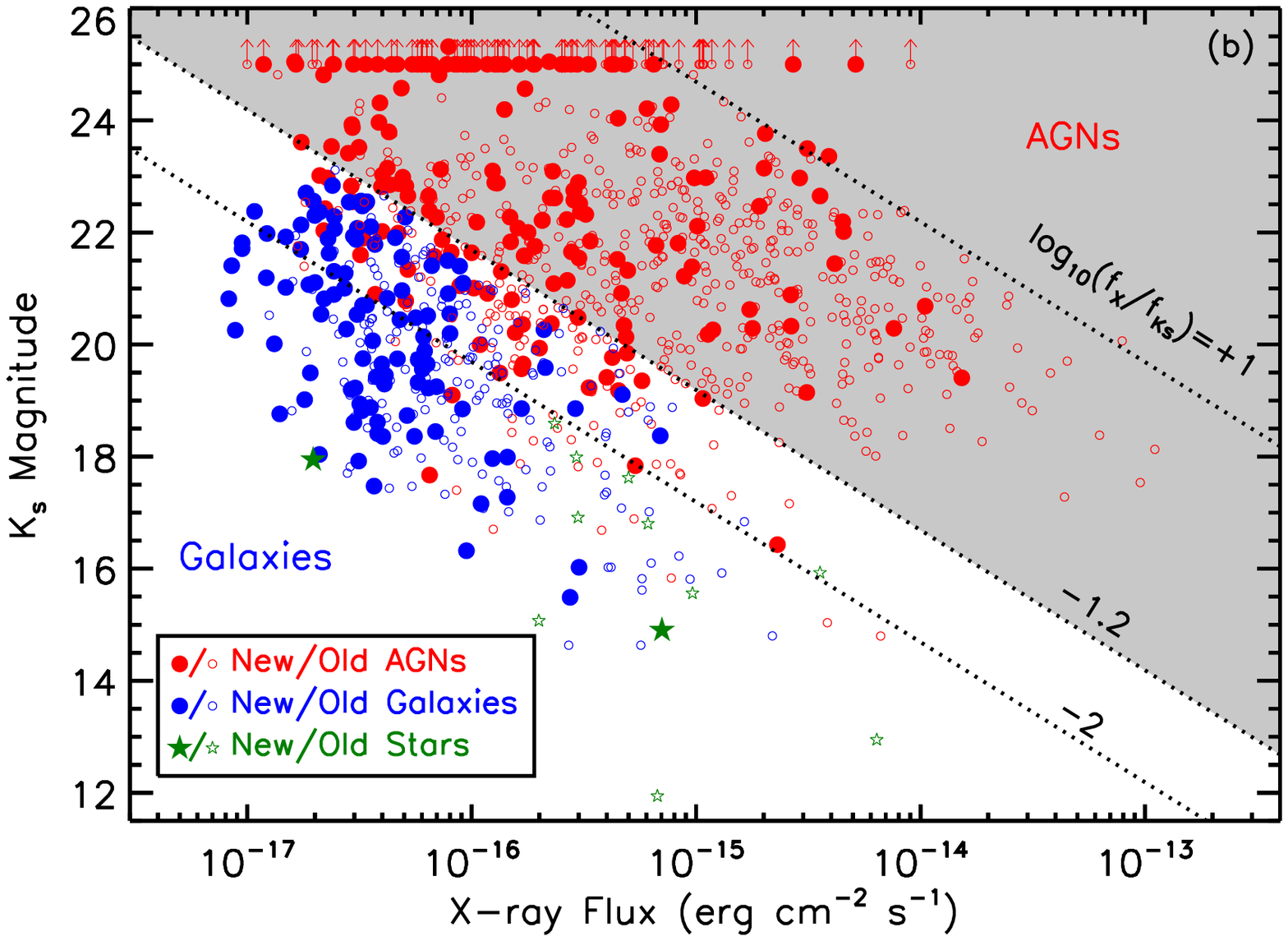}
}
\caption{
X-ray flux vs.\ (a) WFI $R$-band magnitude or (b) TENIS $K_s$-band
magnitude for the main-catalog sources. We used, in order of priority,
the full-band (91\%), soft-band (8\%), or hard-band (1\%)
detected fluxes.
Sources having limits on the magnitudes are indicated
by the arrows.
Red, blue, and green symbols represent AGNs, galaxies, and stars,
respectively. Filled symbols indicate newly detected sources, while
open symbols are sources that were present in the \citet{Xue2011} 4~Ms CDF-S catalogs.
The diagonal lines show constant X-ray to $R$- or $K_s$-band flux ratios.
We adopted $\log(f_{\rm X}/f_{R})>-1$ or $\log(f_{\rm X}/f_{Ks})>-1.2$ as two of the six criteria
to classify AGNs; the AGN regions are shaded in both panels.
}
\label{fig-fox}
\end{figure*}

We classified an X-ray source as an AGN if it satisfies one of the 
following six criteria: (1) $L_{\rm X,int}\ge3\times10^{42}$~\lum\
(luminous X-ray sources),
(2) $\Gamma_{\rm eff}\le1.0$ (hard X-ray sources),\footnote{For the 479 sources detected in either 
the soft or hard band but not both, the mode values of the band ratios from 
{\sc behr} were adopted which were then converted to $\Gamma_{\rm eff}$. 
To ensure reliable AGN identification,
we did not use these 
$\Gamma_{\rm eff}$ values for the AGN selection. Instead, for a source to be 
classified as an AGN based on its $\Gamma_{\rm eff}$,
we required the source to be detected
in the hard band and the 90\% confidence-level upper limit on $\Gamma_{\rm eff}$  
to be less than 1 which was derived from the lower limit on the band ratio.} (3) X-ray-to-optical flux ratio of 
$\log(f_{\rm X}/f_{R})>-1$, where $f_{\rm X}$ is, in order of priority,
the full-band, soft-band, or hard-band
detected flux, and $f_{R}$ is the $R$-band flux, 
(4) spectroscopically classified as AGNs, (5) X-ray-to-radio luminosity
ratio of $L_{\rm X,int}/L_{\rm 1.4GHz}\ge2.4\times10^{18}$, (6) X-ray-to-NIR flux ratio of 
$\log(f_{\rm X}/f_{Ks})>-1.2$. The first five criteria were described in detail in Section~4.4 of 
\citet{Xue2011}. For the last criterion, we chose an empirical threshold to classify X-ray excess 
sources as AGNs with the available AGN sample classified from the previous five criteria and the 
X-ray and TENIS \hbox{$K_s$-band} flux information (see Figure~\ref{fig-fox}b);
12 new AGNs were classified based on this additional criterion.

In total, we identified 711 AGNs from the main catalog, and 
most (86\%) of them were classified based on two or more criteria.
There are only five AGNs (XIDs 416, 494, 517, 523, 718) classified solely based on the 
first criterion ($L_{\rm X,int}\ge3\times10^{42}$~\lum). We caution that it
is probable that the X-ray emission from some of these five sources could instead come from
intense star formation (star-formation rate $\gtrsim300$~$M_\sun$~yr$^{-1}$; e.g., \citealt{Lehmer2016});
however, their estimated 
star-formation rates are only $\approx0.1$--$70$~$M_\sun$~yr$^{-1}$ \citep{Skelton2014}.
Excluding the 12 stars, the remaining 285 sources are considered as normal galaxies 
(including the few off-nuclear sources). The distributions of 
the X-ray fluxes versus WFI $R$-band magnitudes and TENIS $K_s$-band
magnitudes (the third and sixth classification criteria) for the \hbox{main-catalog}
sources are displayed in Figure~\ref{fig-fox}, and the regions expected to be occupied by AGNs are highlighted. 
The X-ray source classification is presented in Column~70 of the \hbox{main-catalog} table.

\begin{figure}
\centerline{
\includegraphics[scale=0.48]{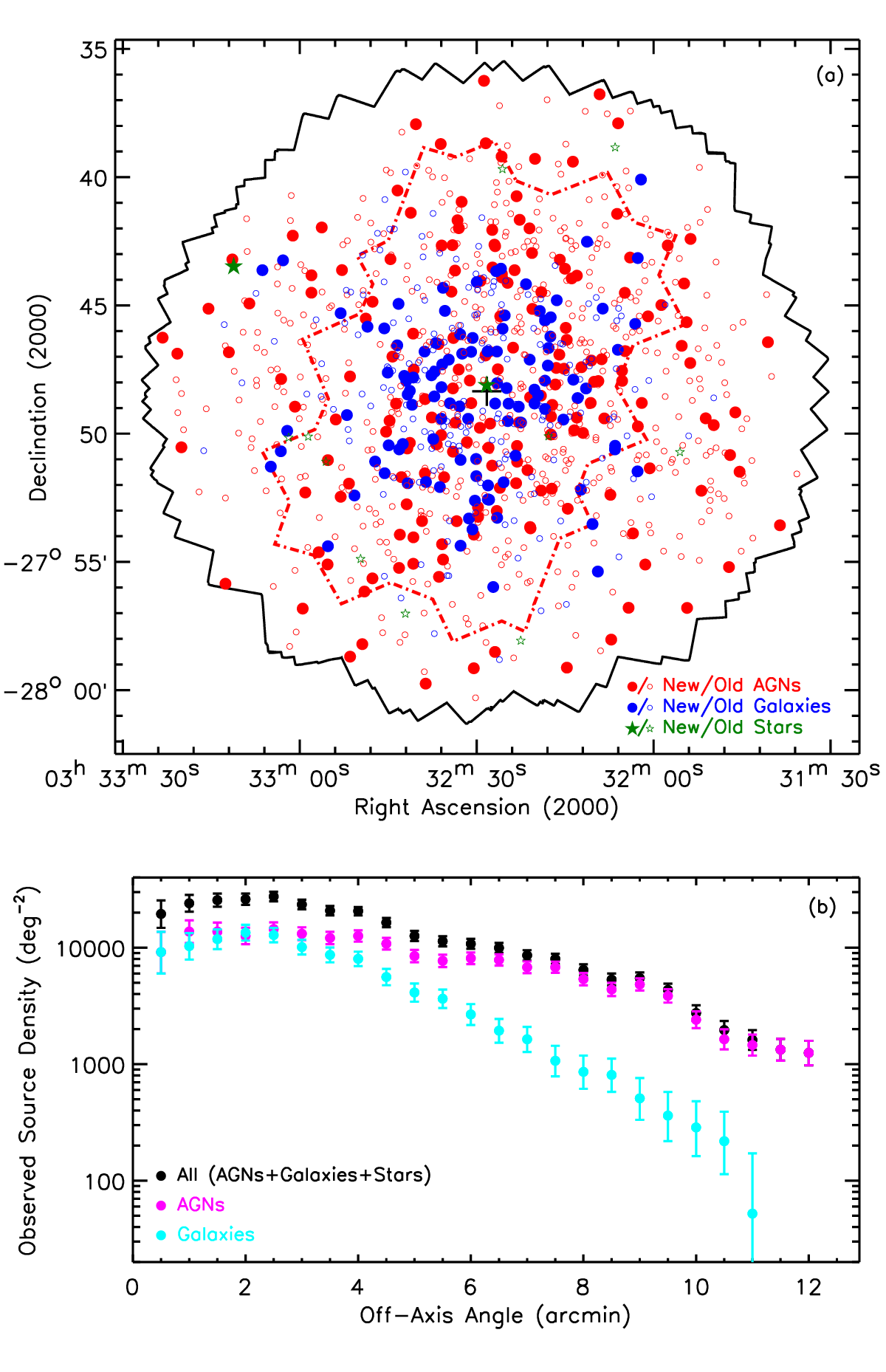}
}
\caption{
(a) Spatial distribution of the main-catalog sources.
Red, blue, and green symbols represent AGNs, galaxies, and stars,
respectively. Filled symbols indicate newly detected sources, while
open symbols are sources that were present in the \citet{Xue2011} 4~Ms CDF-S
catalogs.
The average aim point, CDF-S boundary, and
GOODS-S region are shown, as was done in Figure 1.
(b) Observed X-ray source sky density as a function of \hbox{off-axis} angle in 
off-axis angle bins of 1\arcmin. The error bars are the $1\sigma$ Poisson uncertainties \citep{Gehrels1986} on
the source density in each bin. The AGN and galaxy density distributions
are also shown. 
}
\label{fig-srcdist}
\end{figure}

\begin{figure*}
\centerline{
\includegraphics[scale=0.85]{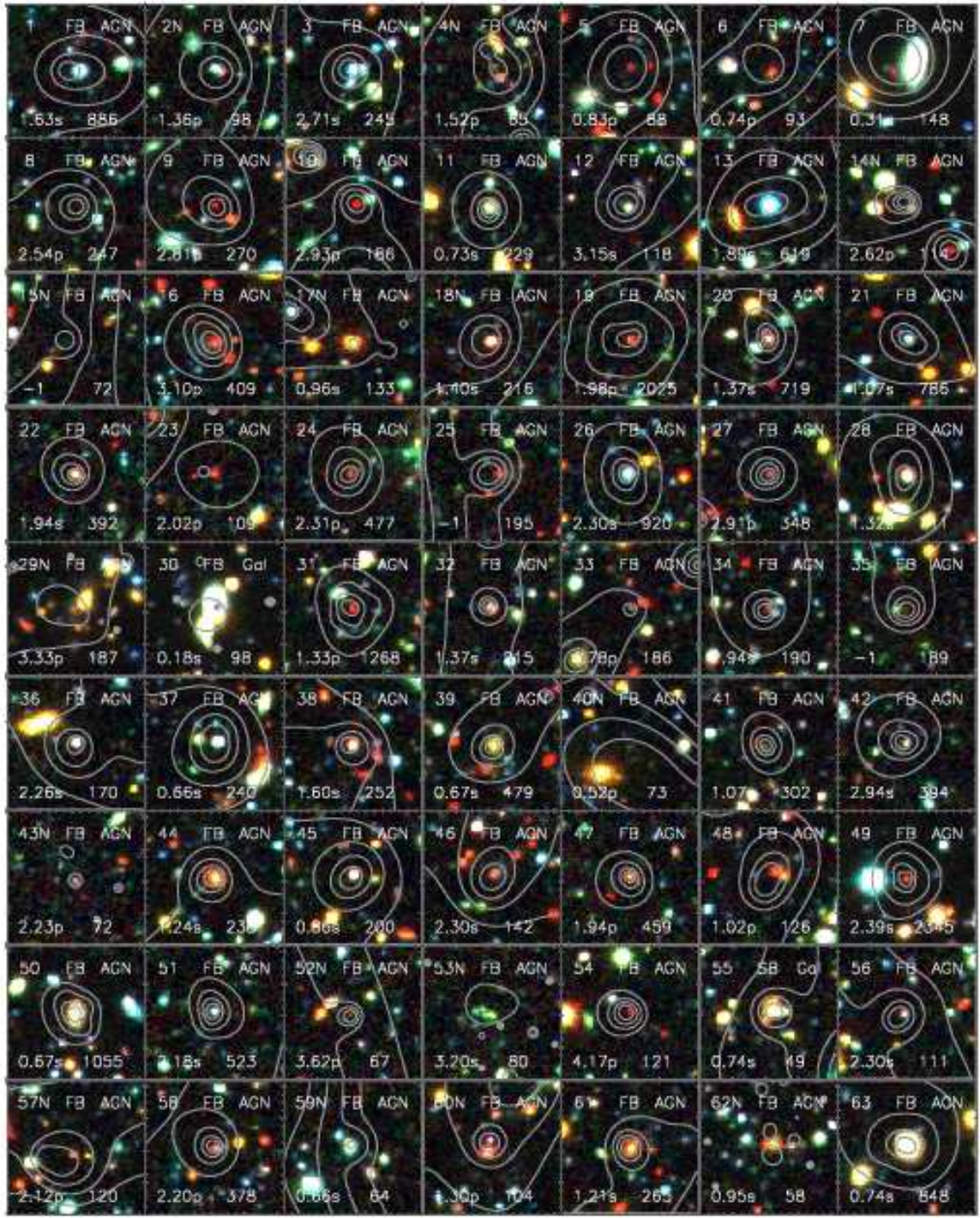}
}
\caption{
Postage-stamp images for the main-catalog sources overlaid
with adaptively smoothed X-ray contours.
The images are color composites of the MUSYC $B$-band (blue),
WFI $R$-band (green), and TENIS $J+K_s$-band (red) images.
Each image is 25\arcsec
on a side, with the X-ray source located at the center.
The X-ray contours were created using,
in order of priority, the full-band, soft-band,
or hard-band smoothed X-ray image ($25\arcsec\times25\arcsec$) 
in which the source is detected,
and they are logarithmically scaled. 
If a source is faint (thus not apparent in the smoothed image) 
or if there is a brighter source nearby (the contour levels centered on the
brighter source),
there may be no X-ray contours around the image center.
The main-catalog ID number (a letter ``N'' is attached if the source 
is newly detected compared to the 4~Ms \cdfs\ catalogs), X-ray band,
and source classification (AGN, galaxy, or star) are given
at the top of each image; the numbers at the bottom are
the adopted redshift (``$-1$'' if not available; marked with
``s'' if it is a spectroscopic redshift or ``p'' if a photometric
redshift) and the net source counts
in the corresponding X-ray band.
Only the first page is shown here for illustrative
purposes;
the entire set of 16 pages of images is available in the online version
of the journal.
}
\label{fig-postimg}
\end{figure*}

\begin{figure*}
\centerline{
\includegraphics[scale=0.85]{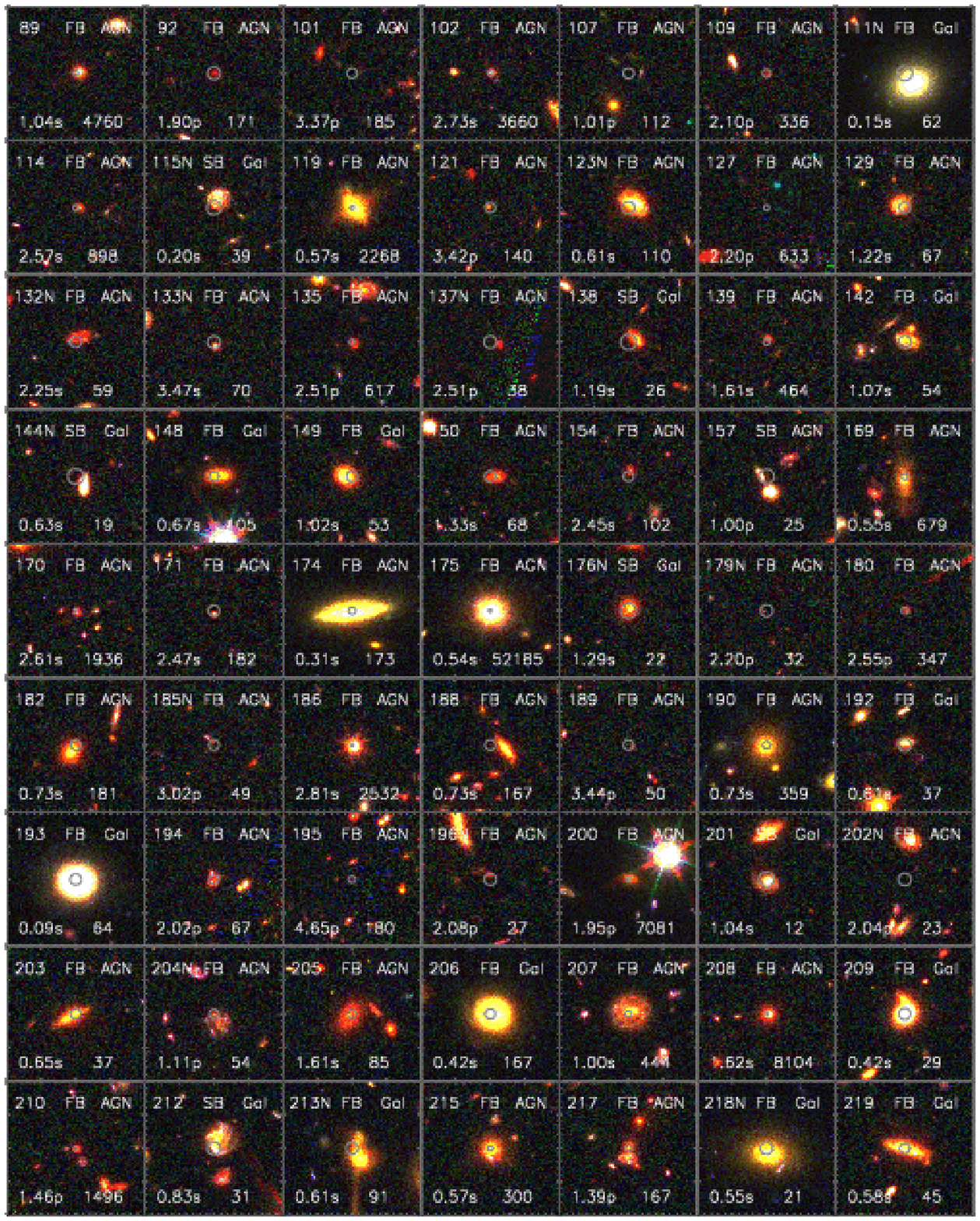}
}
\caption{
Postage-stamp images for the main-catalog sources that have GOODS-S and CANDELS coverage.
The images are color composites of the GOODS-S $b$-band (blue),
GOODS-S $z$-band (green), and CANDELS F160W-band (red) images.
Each image is 12\arcsec
on a side, with the X-ray source indicated by a small central circle of radius
equal to the 1$\sigma$ X-ray positional uncertainty.
The main-catalog ID number (a letter ``N'' is attached if the source
is newly detected compared to the 4~Ms \cdfs\ catalogs), X-ray band,
and source classification (AGN, galaxy, or star) are given
at the top of each image; the numbers at the bottom are
the adopted redshift (``$-1$'' if not available; marked with
``s'' if it is a spectroscopic redshift or ``p'' if a photometric
redshift) and the net source counts
in the corresponding X-ray band.
Only the first page is shown here for illustrative
purposes;
the entire set of 11 pages of images is available in the online version
of the journal.
}
\label{fig-postimghst}
\end{figure*}

Figure~\ref{fig-srcdist}a shows the spatial distribution of the \hbox{main-catalog} 
sources, which are
color-coded as AGNs, galaxies, and stars. Figure~\ref{fig-srcdist}b displays the observed source
sky density as a function of the off-axis angle.
These apparent source densities have not been corrected for detection incompleteness or 
Eddington bias; the number-count results, taking into account these effects,
are presented in Section~\ref{sec-ncounts} below.
The source densities decrease at larger off-axis angles
due to the sensitivity degradation in the outer regions (e.g., see Section 7). 
In Figure~\ref{fig-postimg}, we show ``postage-stamp'' images for the \hbox{main-catalog} 
sources overlaid
with adaptively smoothed X-ray contours. 
The images are color composites of the MUSYC $B$-band,
WFI $R$-band, and TENIS $J+K_s$-band images. 
The X-ray contours were created using,
in order of priority, the full-band, soft-band,
or hard-band smoothed \xray\ image in which the source is detected, and the wide range of
source sizes represents the PSF broadening with off-axis angle.
The source classification, adopted redshift, and net source counts are also indicated in each image.
In Figure~\ref{fig-postimghst}, we show postage-stamp images for the \hbox{main-catalog}
sources that have GOODS-S and CANDELS coverage. The images are color composites of the GOODS-S $b$-band,
GOODS-S $z$-band, and CANDELS F160W-band images.

\subsection{Comparison with the 4 Ms CDF-S and Other Source Catalogs} \label{sec-compare4ms}
We matched the main-catalog sources to sources in the 4~Ms \cdfs\ main and supplementary catalogs \citep{Xue2011}.
A matching radius of 4\arcsec\ was used, and we visually inspected the X-ray images for the 
unrecovered 4~Ms sources and included two additional matches at large off-axis angles
where the positional offsets are $\approx4$--6\arcsec.
The 7 Ms \cdfs\ main source catalog recovered 704 of the 740 4~Ms \hbox{main-catalog} sources and
13 of the 36 4~Ms supplementary catalog sources. 
Of the 36 4~Ms \cdfs\ main-catalog sources that are not present in the 7~Ms main catalog, 10 are included in the
7~Ms \cdfs\ supplementary catalog. We inspected the remaining 26 missing 4~Ms sources in detail,
and they belong 
to one of the following four categories:
\begin{enumerate}
\item
Nine sources have no multiwavelength counterparts, and they are weak X-ray sources
with $\approx8$--60 detected
counts in the 4~Ms \cdfs\ catalog.\footnote{The source with $\approx60$ counts
from 0.5--8.0~keV is 
at an off-axis angle of \hbox{$\approx9\arcmin$}, and it is a weak detection considering 
the large PSF size in the outer region of the \cdfs\ ($P_{\rm B,4Ms}=0.002$).} 
These are likely false detections in the 4~Ms \cdfs.

\item
Seven sources are located at large off-axis angles \hbox{(7--11\arcmin)}, and have a 
companion X-ray source nearby (within 3.5--10\arcsec) which is detected in the 7~Ms \cdfs\
(i.e., two X-ray sources in the 4~Ms \cdfs\ and only one in the 7~Ms \cdfs).
They have \hbox{$\approx16$--130} detected counts in the 4 Ms \cdfs\
catalog, and two of them have no multiwavelength counterparts. 
None of these seven off-axis
sources were detected in the 250~ks E-CDF-S catalogs \citep{Xue2016}.
Most of these sources are likely false detections in the 4 Ms \cdfs\ introduced when
a single off-axis source was detected in two X-ray bands at different positions 
(separated by 3.5--10\arcsec) and was treated as two sources.
A few of them might be real sources that are blended
with the companion source in the 7~Ms \cdfs.

\item
Eight sources have faint counterparts (GOODS-S \hbox{$z_{850}\approx24$--27}) 
in the 4 Ms catalog, and
they have \hbox{$\approx8$--40} detected counts. They are not detected in the 7 Ms \cdfs\
probably due to source variability and/or background fluctuation.
A few of these sources could also be false detections considering their 
counterparts are faint and the probability of a chance association is relatively high.

\item
Two sources were detected by {\sc wavdetect} in the 7~Ms \cdfs, but they did not pass the 
$P_{\rm B}$ threshold cut, and their counterparts are not sufficiently bright to be
included in the 7 Ms supplementary catalog. They have 26 and 34
detected counts in the 4 Ms catalog, respectively. Their nature is similar
to those sources in category (3).

\end{enumerate}
In total, $\approx16$ of the 26 missing 4~Ms \cdfs\ main-catalog sources are probably spurious detections,
which constitute $\approx2\%$ of the 4~Ms \cdfs\ main catalog. The expected total number of 
spurious detections was $\approx12$ in Section~6.2 of \citet{Xue2011}, consistent with our 
assessment here.
Of the 23 ($36-13=23$) missing 4~Ms \cdfs\ supplementary catalog sources, 
one is included in the 7~Ms \cdfs\ supplementary catalog, and the other 
22
are likely real sources that fall below the 7~Ms \cdfs\ detection threshold due 
to source variability and/or background fluctuations (e.g., affected by Eddington bias);
a minor fraction of them could be spurious detections.

\begin{figure}
\centerline{
\includegraphics[scale=0.5]{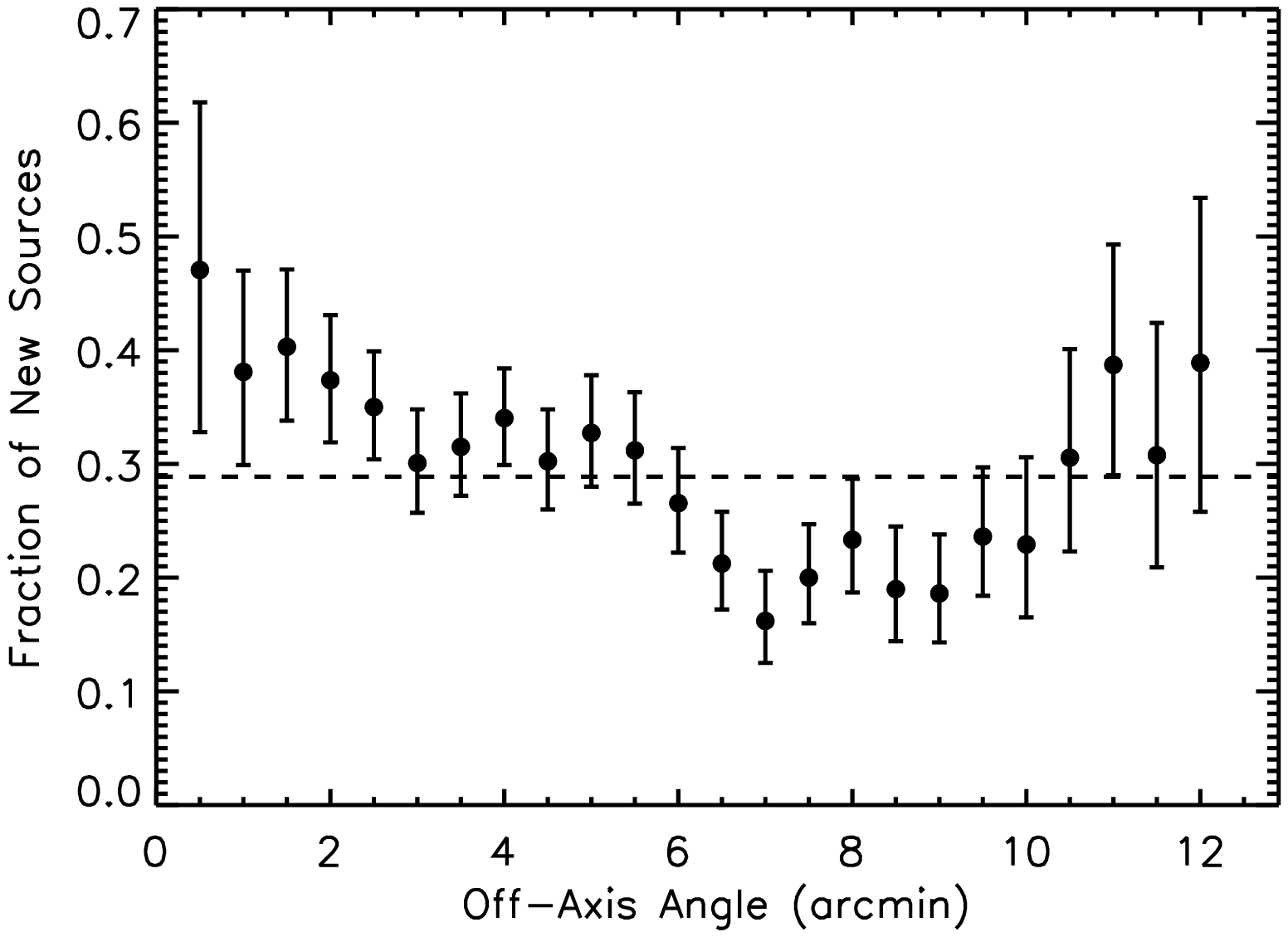}
}
\caption{
Fraction of new sources as a function of off-axis angle in
off-axis angle bins of 1\arcmin, excluding
the three new sources outside the 4~Ms \cdfs\ footprint. 
The error bars are the $1\sigma$ Poisson uncertainties \citep{Gehrels1986}. The horizontal dashed line represents the total fraction of new sources ($291/1008=29\%$).
The rise of the new-source fraction beyond an off-axis angle of $\approx10\arcmin$
is likely caused by the dominating population of new sources detected in
the more sensitive 2--7 keV~band adopted in the 7~Ms \cdfs\ rather than the
2--8 keV~band in the 4~Ms \cdfs.
}
\label{fig-posfrac}
\end{figure}

There are 291 sources in the main catalog that are new detections compared to the 4~Ms \cdfs\ catalogs.
The new sources are distributed over the entire \cdfs\ field (Figure~\ref{fig-srcdist}), and they
are noted in the postage-stamp images (Figure~\ref{fig-postimg}).
Three of the new sources (XIDs 528, 996, 1008) lie outside the footprint of the 4~Ms \cdfs;
they are all luminous AGNs and were detected 
in the 250~ks E-CDF-S catalogs \citep{Lehmer2005,Xue2016}.
We present
the fractions of new sources at different off-axis angles in Figure~\ref{fig-posfrac}, excluding
the three new sources outside the 4~Ms \cdfs\ footprint.
The fraction of new sources is $\approx40\%$ in the field center, and it decreases 
to $\approx20\%$ at large
off-axis angles; this behavior is likely due to the greater sensitivity improvement in the 
central region from the 4~Ms \cdfs\ to the 7~Ms \cdfs\ (see Section~7.3 of \citealt{Xue2011}).
Beyond an off-axis angle of $\approx10\arcmin$, there is a weak rise of the new-source fraction 
toward larger off-axis angles. Most of the new sources in this outer region are detected
in the hard band ($18/21=86\%$), while the hard-band detection fraction among all the 
291 new sources is only 42\%. One possible explanation is 
that the new sources detected in the outer region are largely
due to the more sensitive 2--7~keV~band adopted in the 7~Ms \cdfs\ rather than the 
\hbox{2--8~keV} 
band in the 4~Ms \cdfs\ (see Footnote 16 of \citealt{Xue2016} and our sensitivity discussion
in Section~\ref{sec-bkg} below). Although there are also such new sources detected at smaller
off-axis angles, the source number is smaller than the number of new sources detected due to 
improved sensitivity (increased exposure). They only become a dominant population 
at large off-axis angles where the number of new sources detected due to increased exposure is
small, and this extra population of new sources causes the rise of the 
new-source fraction.

\begin{figure}
\centerline{
\includegraphics[scale=0.5]{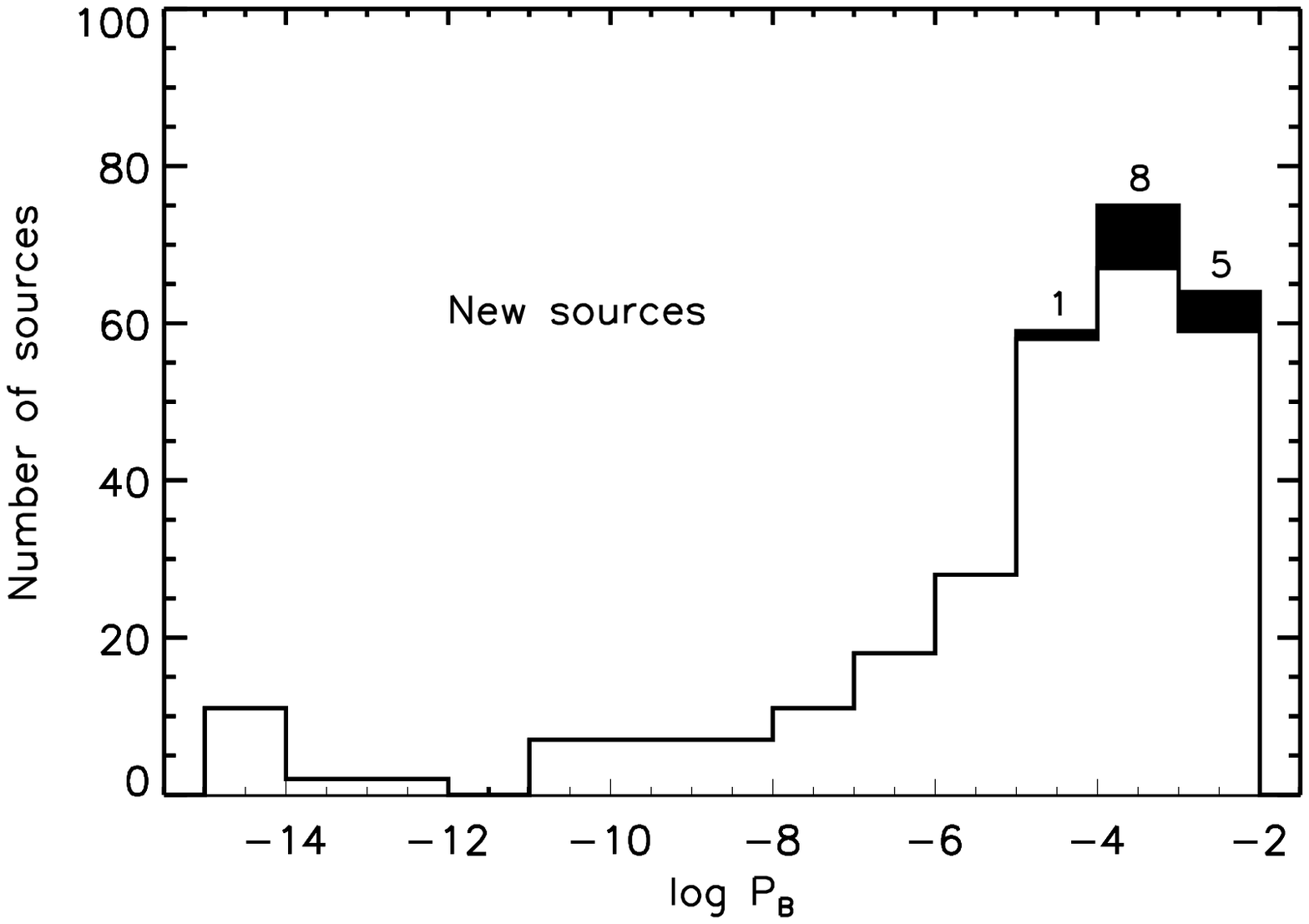}
}
\caption{
Similar to Figure~\ref{fig-pbhist} but for the 291 new sources, showing
the distribution of the AE binomial no-source probability ($P_{\rm B}$).
The shaded regions indicate sources that have no multiwavelength
counterparts, with the numbers of such sources shown on the top of
each bin. 
}
\label{fig-pbhistnew}
\end{figure}

In terms of source classification, a smaller fraction of the new sources are AGNs
($184/291=63\%\pm5\%$) compared to that for the entire catalog ($711/1008=71\%\pm3\%$), indicating the rise
of galaxy population toward lower X-ray fluxes
(e.g., \citealt{Bauer2004,Ranalli2005,Lehmer2012}; Section~\ref{sec-ncounts} below).
There are also two new stars detected.
The $P_{\rm B}$ distribution for the new sources is shown in Figure~\ref{fig-pbhistnew}. Compared to
the distribution for the entire catalog (Figure~\ref{fig-pbhist}), the new sources are less significantly
detected with larger $P_{\rm B}$ values overall. Most of the sources without multiwavelength
counterparts (14 out of the 16) are new sources and are likely to be spurious detections.

\begin{figure}
\centerline{
\includegraphics[scale=0.5]{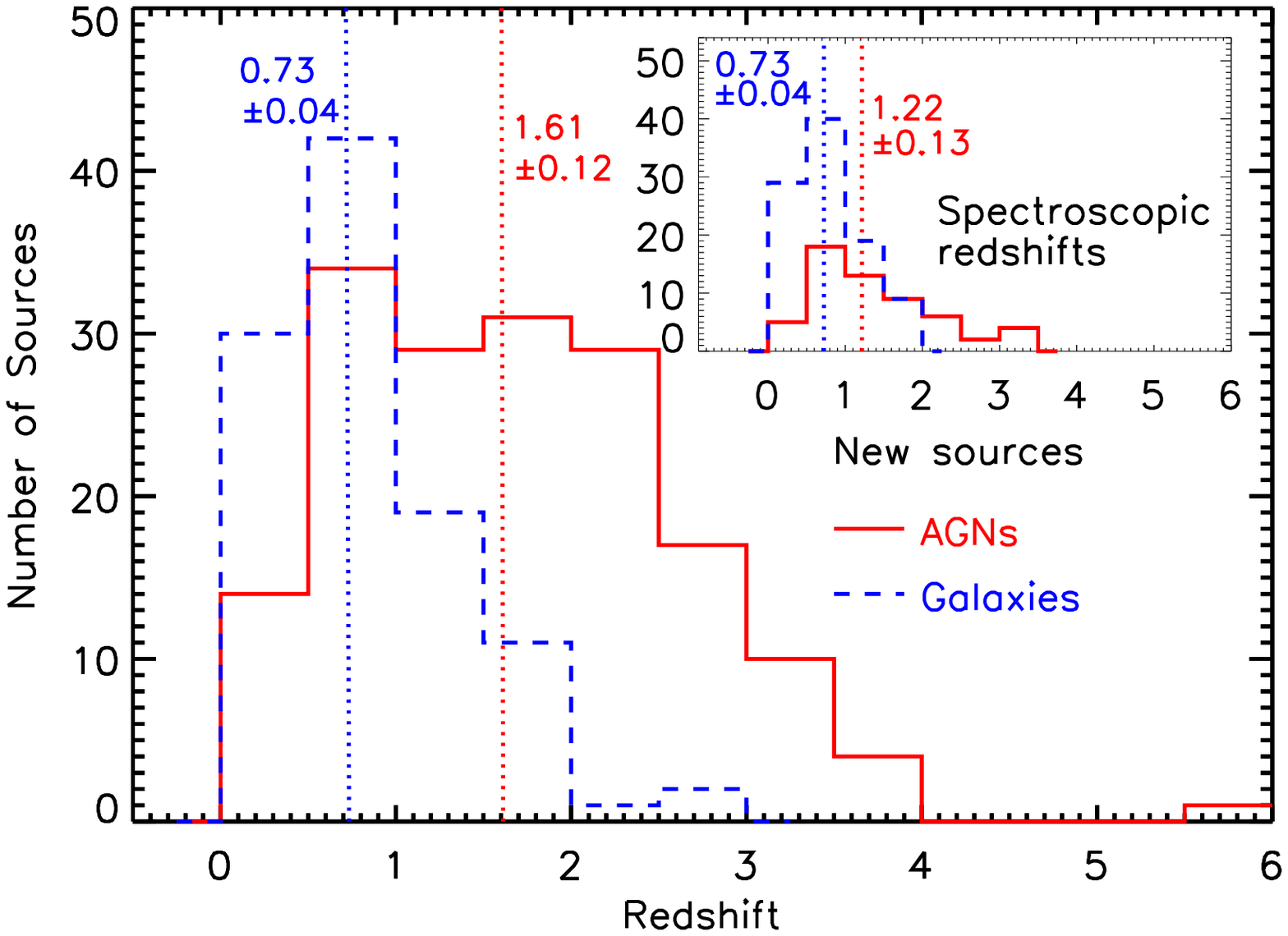}
}
\caption{
Similar to Figure~\ref{fig-zhist}a but for the new sources, showing
the redshift distributions.
The vertical dotted lines indicate the
median redshift for every distribution, and the corresponding
median values and their
1$\sigma$ uncertainties (derived via bootstrapping) are listed.
The median redshifts for the new AGNs and galaxies are consistent with those for
the entire AGN and galaxy samples (Figure~\ref{fig-zhist}a) within the uncertainties.
}
\label{fig-zhistnew}
\end{figure}

\begin{figure*}
\centerline{
\includegraphics[scale=0.5]{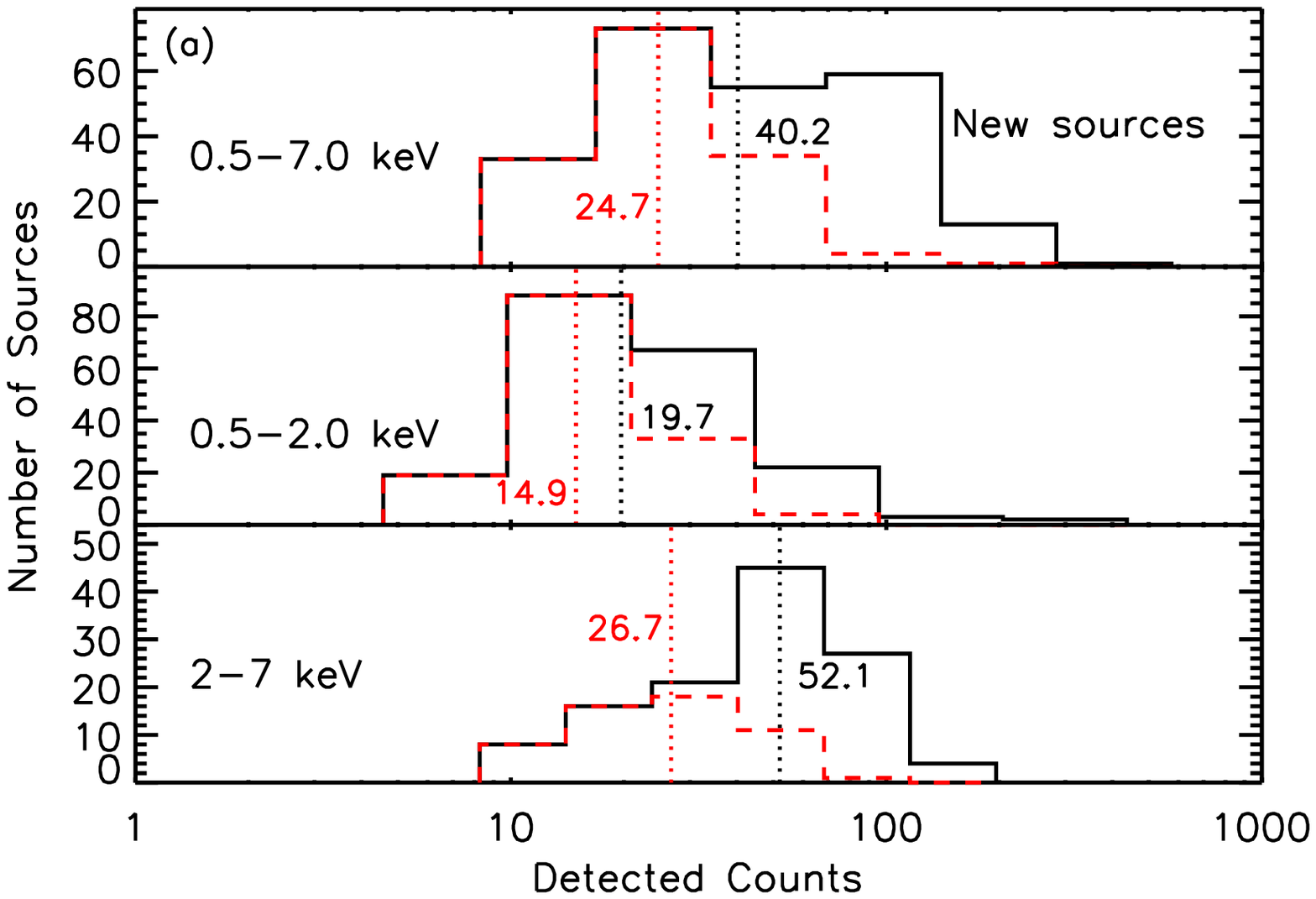}
\includegraphics[scale=0.5]{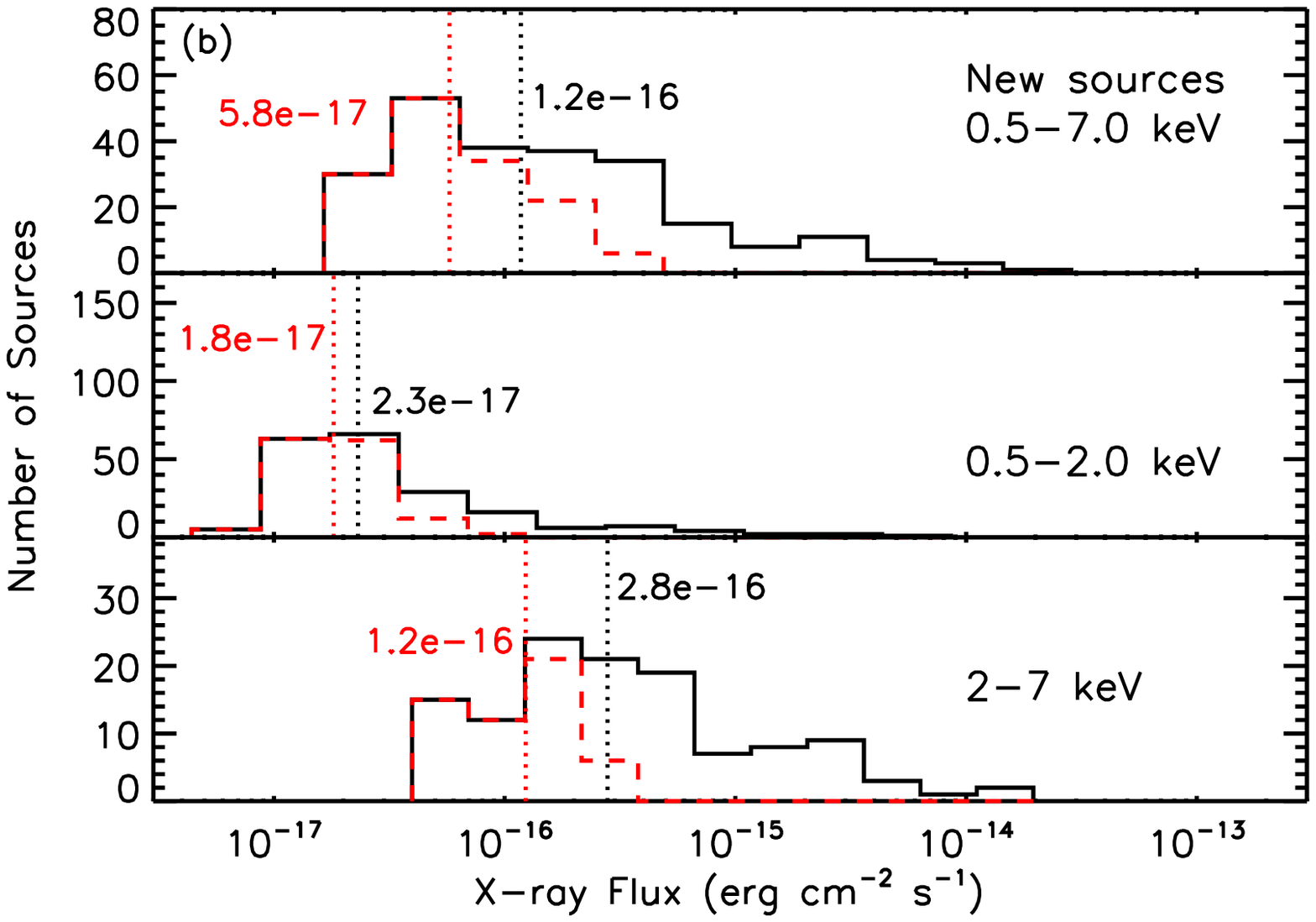}
}
\caption{
(a) Similar to Figure~\ref{fig-cnthist} but for the new sources, showing the
distributions of detected source counts.
The black-solid histograms show the count
distributions for all the sources in the \cdfs\ field, while
the red-dashed histograms show the
distributions for the subgroups of
sources within $6\arcmin$ of the average aim point.  
The vertical dotted lines indicate the
median numbers of counts for every distribution, and the corresponding
median values are listed.
(b) Similar to Figure~\ref{fig-fluxhist} but for the new sources, showing 
the distributions of X-ray fluxes.
The black-solid histograms show the flux
distributions for all the sources in the \cdfs\ field, while
the red-dashed histograms show the
distributions for the subgroups of
sources within $6\arcmin$ of the average aim point. 
The vertical dotted lines indicate the median fluxes
for every distribution, and the corresponding
median values are listed.
}
\label{fig-cnthistnew}
\end{figure*}

\begin{figure*}
\centerline{
\includegraphics[scale=0.5]{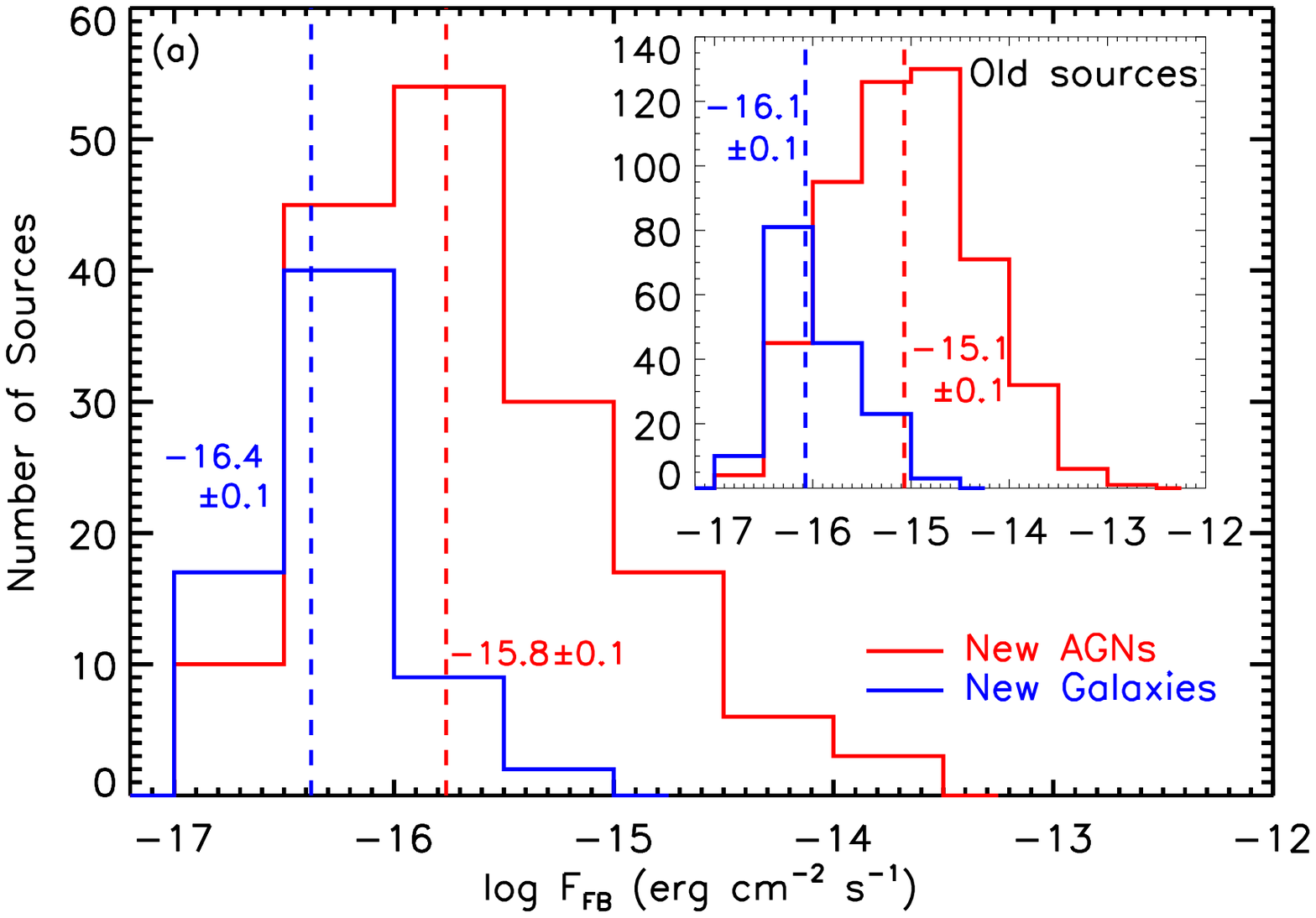}
\includegraphics[scale=0.5]{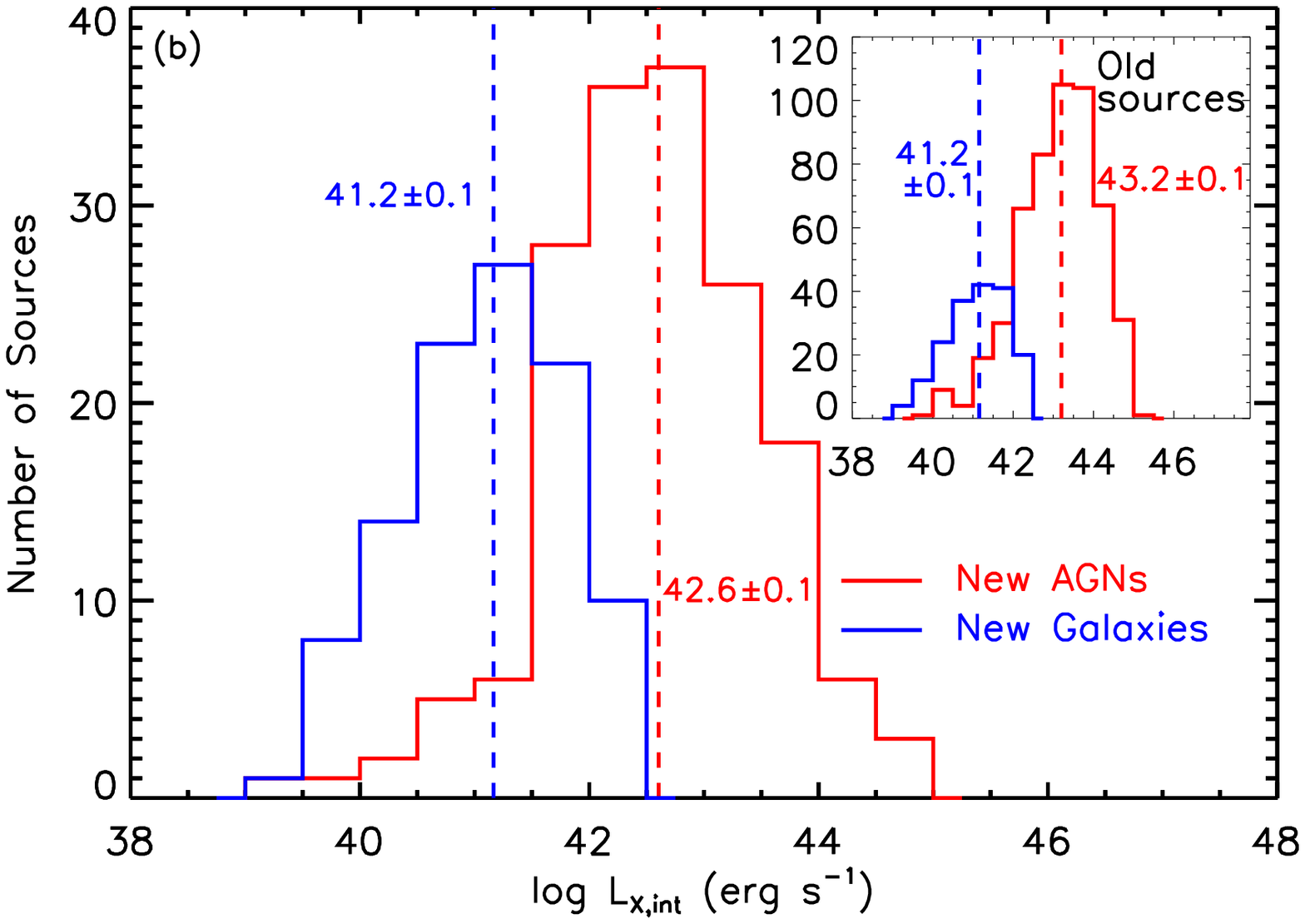}
}
\caption{
Distributions of (a) full-band fluxes and (b)
intrinsic 0.5--7.0~keV luminosities for the newly detected main-catalog sources.
In each panel,
the inset displays the corresponding distributions for the
main-catalog sources that were present in the \citet{Xue2011} 4~Ms CDF-S catalogs.
The red and blue histograms indicate the distributions for
AGNs and galaxies, respectively. The vertical dashed lines
show the median values of the corresponding distributions, and
the numbers display the corresponding median values and their
1$\sigma$ uncertainties (derived via bootstrapping).
Sources with upper limits on the full-band fluxes (92 sources; 9\%)
are not included in
panel (a). The 22 sources which lack redshifts and
the 12 stars are
not included in panel (b).
}
\label{fig-comflux}
\end{figure*}

The redshift distributions for the new AGNs and galaxies are displayed in Figure~\ref{fig-zhistnew}.
The median redshifts for the new sources do not differ from those for 
the entire AGN and galaxy samples
(Figure~\ref{fig-zhist}a) after accounting for the uncertainties. 
The source-count and flux distributions for the 291 new sources are shown 
in Figure~\ref{fig-cnthistnew}.
Compared to the distributions for the entire catalog (Figures~\ref{fig-cnthist} and \ref{fig-fluxhist}), the newly
detected sources have fewer source counts
and lower \xray\ fluxes, regardless of their location in the field.
The new sources and the sources already present in the 4~Ms \cdfs\ catalogs (``old sources'') are represented by
different symbols in the luminosity versus redshift plot (Figure~\ref{fig-lz}). The new sources appear 
to be less luminous overall. To better compare the X-ray fluxes and luminosities between the new
and old sources, we present flux and luminosity histograms for the new and old AGNs and galaxies in
Figure~\ref{fig-comflux}.
The new AGNs in the main catalog
have a lower median flux and median luminosity than the old AGNs.
The new galaxies have a slightly lower median flux than the old galaxies, while
their median luminosities are comparable.

The new AGNs appear to be more X-ray absorbed than the old AGNs.
The median absorption column density estimates for the new and old AGNs are 
$(7.0\pm1.8)\times 10^{22}$~cm$^{-2}$ and $(4.2\pm0.6)\times 10^{22}$~cm$^{-2}$, respectively.
A Kolmogorov-Smirnov (K-S) test suggests that the two column-density distributions
differ significantly, with a K-S probability of 0.009.
We caution that the absorption column densities were estimated with simplistic assumptions 
(Section~\ref{sec-xphotmetric})\footnote{The spectral shapes ($\Gamma_{\rm eff}$) and thus intrinsic
absorption column densities
are especially uncertain for the 
479 sources detected in either the soft or hard band but not both, where the best-guess estimates 
of the band ratios were adopted.} and may be not reliable for such 
statistical comparisons.

We matched the main-catalog sources to sources in 
the 250~ks E-CDF-S main and supplementary catalogs \citep{Xue2016}
using a matching radius of 4\arcsec. One additional \hbox{off-axis} 
match with an offset of 5\farcs9 was found via visual inspection. 
In total, 307 \hbox{main-catalog} sources were detected in the \hbox{E-CDF-S} main and supplementary catalogs.
We also matched the \hbox{main-catalog} sources to sources in the 
3~Ms \xmm\ \cdfs\ catalogs \citep{Ranalli2013}. For each main-catalog
source, a \xmm\
counterpart was searched for within a radius that is three times the quadratic sum of 
the 1$\sigma$ \chandra\ and \xmm\ positional errors. In total,
292 main-catalog sources have \xmm\ counterparts. There are cases where multiple \chandra\ sources
were matched to a single \xmm\ source (i.e., the \xmm\ counterpart IDs for different 
7~Ms \cdfs\ sources are the same), probably due to source blending in the \xmm\ catalogs.
The counterpart IDs in the 4~Ms \cdfs, 250~ks E-CDF-S, and 3~Ms \xmm\ \cdfs\
are presented in Columns~71--73 of the main-catalog table.

There have also been specialized searches for faint X-ray sources in
the CDF-S using optical/NIR source positions as priors \citep[e.g.,][]{Fiore2012,Giallongo2015,Cappelluti2016}.
We do
not compare our source catalogs to those detections due to the
significantly different source-detection approaches adopted.

There are 19 main-catalog sources having \xmm\ counterparts but no 4~Ms \cdfs\ or 250~ks E-CDF-S
counterparts. Most (15) of these \xmm\ counterparts were also matched to other \chandra\ sources
(i.e., multiple \chandra\ sources matched to a single \xmm\ source), and they
were detected in the 4~Ms \cdfs\ or 250~ks E-CDF-S (i.e.,
not new sources detected by \xmm). However,
there are four main-catalog sources (XIDs 4, 571, 942, 954) having \xmm\ counterparts that
appear to be missed by the 4~Ms \cdfs\ and 250~ks E-CDF-S catalogs. 
Three of them (except XID 954) are hard X-ray sources with $\Gamma_{\rm eff}\approx-1.5$--0.3,
and they are at large off-axis angles ($\approx7$--11\arcmin) detected with 66--156 full-band
counts in the 7~Ms \cdfs;
these qualities help to explain why
they were missed by previous \chandra\ surveys but were detected by \xmm.
The 7~Ms \cdfs\ hard-band (2--7 keV) fluxes for these three sources are consistent with 
the 2--10~keV \xmm\ fluxes within a factor of $\approx1$--2.\footnote{Considering statistical
uncertainties, cross-calibration uncertainties, and the difference in the energy bands, there is no
significant hard X-ray flux variability for these objects.}
The other source, XID 954, is at an off-axis angle of 9\farcm2 and 
detected with $\approx70$ full-band
and $\approx60$ \hbox{soft-band} counts in the 7~Ms \cdfs. The effective exposure time in the 4~Ms \cdfs\
at the source location is $\approx45\%$ of that in the 7~Ms \cdfs, which likely explains the 
non-detection of this weak off-axis source in the 4~Ms \cdfs. However, in the 7~Ms \cdfs,
XID 954 was not detected in the hard band with an upper limit on the 2--7 keV flux 
of $2.5\times10^{-16}$~\flux, while in the \xmm\ catalog, it has no signal in the 0.5--2.0 keV
band and the 2--10 keV measured flux is $3.0\times10^{-15}$~\flux. It was not classified as an AGN 
in the 7~Ms \cdfs, and the X-ray emission likely originates from a $z=0.129$ galaxy with 
$L_{\rm X}=1.6\times10^{40}$~\lum. The substantial hard X-ray signal in the 3~Ms \xmm\ data 
probably came from some outburst phenomenon, for example, an obscured stellar tidal disruption event
or a heavily obscured AGN revealing temporarily some of its hard \xray\ emission due to 
reduced absorption. The nature of this source is worth
further investigation.
Two other candidate new sources proposed in the 3~Ms \xmm\ \cdfs\ catalogs (Section 8.4 of 
\citealt{Ranalli2013}), with \xmm\ IDs 85 and 1149, remain undetected in the 7~Ms \cdfs.

\subsection{Notes on Individual Objects}  \label{sec-notes}
\subsubsection{Variable Sources}
One of the main-catalog sources, XID 725, was discovered as a fast X-ray transient 
shortly after the relevant observation was taken \citep{Luo2014a}. Most ($\approx90\%$) of the source counts 
($\approx120$)
arrived within a timespan of $\approx5$~ks during observation 16454 on 2014 October 1.
The source did not produce any detectable X-ray emission before or after this transient event in the 7~Ms
\cdfs. The counterpart is a faint galaxy (${\rm F125W}=27.1$) at a photometric redshift of $\approx2.1$. 
The nature of the source is unclear and detailed investigation is presented in
F.~E.~Bauer et al. (in preparation). 
Our source catalog was constructed using the merged 7~Ms data, and thus the X-ray properties (e.g., flux
and luminosity) of this source were averaged over the entire dataset. 

Another main-catalog source, XID 403, was discovered as a highly variable X-ray source during the 
first half of the new \cdfs\ observations \citep{Luo2014b}. It was not detected in the 4~Ms \cdfs,
yet it was significantly detected in the recent 3~Ms \cdfs\ observations with $\approx250$ full-band counts,
brightening by a factor of $>13$. The counterpart is a $R=24.4$ galaxy at a photometric redshift of 
$\approx1.5$. The X-ray source is very soft and it was not detected in the hard band 
($\Gamma_{\rm eff}\approx3.0$). The X-ray light
curve shows a gradual decline over the timespan of years. This X-ray source is probably associated with
a stellar tidal disruption event. If confirmed, it is the highest redshift tidal disruption event
discovered so far,
owing to the exceptionally deep exposure of the 7~Ms \cdfs. 
Another possibility is that it is an AGN that varies strongly on long timescales
\citep[e.g., PHL~1092;][]{Miniutti2012}. 
Detailed investigation of this source 
is presented in W.~Wang et al. (in preparation). The X-ray properties provided in the main catalog
(e.g., flux and luminosity)
are the average values over the entire 7~Ms dataset.
 
\subsubsection{Sources Without Multiwavelength Counterparts} \label{sec-notes-blank}
Among all the sources without multiwavelength counterparts, only one (XID 912) is highly 
significantly detected with
a minimum {\sc wavdetect} false-positive probability of $10^{-8}$ and an 
AE no-source probability of $P_{\rm B}\approx10^{-13}$
(Figure~\ref{fig-pbhist}); it is also present in the 4~Ms \cdfs\ catalog.
This source 
has 73.2 soft-band counts, and it was not detected in the hard band. The effective power-law photon index
is $\approx2.9$, and a simple \hbox{power-law} 
fit to the X-ray spectrum resulted in a photon index of 
$3.1\pm0.5$. The X-ray photons were relatively evenly distributed among the 7~Ms observations, although some
low-amplitude variability is apparent. If we compute the $P_{\rm B}$ value for each observation individually, 
the source would
be considered detected in observations 582 and 8596 but not in the other observations. 
We inspected the multiwavelength images (Section~\ref{sec-ID}) and did not find any radio/IR/NIR/optical counterpart at the 
\xray\ source position. 
The X-ray source is within the CANDELS coverage but it is close to the edge of that field, and 
it lies outside the GOODS-S footprint.
We examined the SEDS IRAC 5.8~$\mu$m and the Far-Infrared Deep 
Extragalactic Legacy Survey (FIDEL) MIPS 24~$\mu$m and 70~$\mu$m images, and there is no apparent source
at the X-ray source position.
There is a bright ($R\approx18$), low-redshift (0.105) \xray\ detected 
galaxy (XID 916) located $7\farcs5$ away from 
this unidentified \xray\ source. The \xray\ photometry of the two sources is not affected significantly
as the separation 
allows $\approx85\%$ ECF source-extraction apertures to be used for both sources. 
The nature of this unidentified \xray\ source remains unclear.
One possible explanation is that it is an off-nuclear source associated with the nearby galaxy, approximately
15~kpc from the center, although the {\it HST} CANDELS image does not reveal such a large extent of the
galaxy. Another possibility is that this is a high-redshift dusty AGN where the observed-frame NIR/IR emission
(\hbox{rest-frame} optical/NIR) is heavily obscured, and it requires longer-wavelength sensitive observations (e.g.,
ALMA) for a detection. The soft observed X-ray spectral shape is inconsistent with the latter scenario, but the 
limited X-ray photon statistics cause significant uncertainties in the estimated spectral shape. 

\subsection{Main-Catalog Details}  \label{sec-maincat}

We present the main \chandra\ source catalog in Table~\ref{tbl-mcat}. The details of 
the table columns are given below.

\begin{enumerate}

\item
Column~1: the source sequence number (XID) assigned in order of increasing right ascension.

\item 
Columns~2 and 3: the right ascension and declination of the source, respectively.
See Section~\ref{sec-detection}.

\item
Column~4: 
the logarithm of the final $P_{\rm B}$ value. We set $\log P_{\rm B}=-99$ when
$P_{\rm B}=0$. The threshold for being included in the main catalog is $P_{\rm B}<0.007$.
See Section~\ref{sec-detection}.

\item
Column~5:
the logarithm of the minimum {\sc wavdetect} false-positive probability, with $-8$, $-7$, $-6$, and $-5$ 
ranging from most significantly detected to least significantly detected. See Section~\ref{sec-detection}.

\item
Column~6:
the 1$\sigma$ ($\approx68\%$ confidence-level)
positional uncertainty in units of arcseconds derived using Equation~\ref{eq-dpos}.
The 90\% and 95\% confidence-level 
positional uncertainties are approximately 1.6 and 2.0 times the 1$\sigma$ positional uncertainty.
See Section~\ref{sec-dpos}.

\item
Column~7:
the off-axis angle in units of arcminutes, which is the separation between the source position
and the average aim point of the 7~Ms \cdfs\ (Section~2.1).

\item
Columns~8--16:
the aperture-corrected net source counts and the corresponding $1\sigma$ lower and upper 
uncertainties in the full, soft, and hard bands, respectively. For sources undetected 
in a given band, the source-count column lists the 90\% confidence-level upper limit on the source
counts while the two associated uncertainty columns are set to ``$-1.0$''. See Section~\ref{sec-xphotmetric}.

\item
Column~17:
photometric notes on individual sources.
Sources covered by less than 20 of the 102 \cdfs\ observations are marked with ``E'',
and sources in crowded regions are marked with ``C''.
The other sources have this column set to ``...''.
See Section~\ref{sec-xphotmetric}.

\item
Column~18:
the catalog from which the primary counterpart was selected, being, in order of priority, one of the following 
six catalogs: 
CANDELS, GOODS-S, GEMS, TENIS, WFI, and SEDS. There are 710, 26, 187, 49, 4, and 16
primary counterparts from these six catalogs, respectively. 
The right ascension, declination, and magnitude of the primary counterpart are included
in Columns~21--41 below.
Sources with no counterparts have this column set to ``...''.
See Section~\ref{sec-ID}.

\item
Column~19:
the positional offset between the X-ray source and the primary counterpart in 
units of arcseconds. Sources with no counterparts have this column set to ``$-1.00$''.

\item
Column~20:
counterpart notes on individual sources. 
Sources with their counterparts selected manually are marked with ``Manual'',
sources matched to the same counterparts are marked with ``Pair'',
sources that are candidates for being off-nuclear sources are marked with ``Off-nuclear'',
and sources that are candidates for being extended jet/lobe emission are marked with ``Jet''.
There are six sources marked as ``Manual'', three as ``Manual$+$Off-nuclear'',
one as ``Off-nuclear'', one as ``Pair$+$Off-nuclear'', two as ``Pair'',
one as ``Pair$+$Manual$+$Off-nuclear'', one as ``Pair$+$Jet'', and one as 
``Pair$+$Off-nuclear/Jet''.
The other sources have this column set to ``...''.
See Section~\ref{sec-ID}.

\item
Columns~21--41: 
the right ascension, declination, and magnitude of the counterparts in the
WFI, GOODS-S, GEMS, CANDELS, TENIS, SEDS, and VLA catalogs, respectively.
The AB magnitudes for the VLA 1.4 GHz sources were converted from the radio flux densities
($m=-2.5\log f_\nu-48.6$).
Sources with no counterparts in a given catalog have the corresponding columns set to 
``$-1.00$''.
See Section~\ref{sec-ID}.

\item 
Columns~42--44:
the spectroscopic redshift, quality flag (``Secure'' or ``Insecure''), 
and the catalog from which the redshift was collected (numbered 1--26; see Section~\ref{sec-redshift}
for the references).
The spectroscopic redshifts for stars were set to zero.
Sources without spectroscopic redshifts have these three columns set to 
``$-1.000$'', ``...'', and ``$-1$'', respectively.
See Section~\ref{sec-redshift}.

\item
Columns~45--50:
the photometric redshifts from \citet{Luo2010}, \citet{Rafferty2011}, \citet{Hsu2014}, 
\citet{Skelton2014}, \citet{Santini2015}, and \citet{Straatman2016}, respectively.
Sources which lack photometric redshifts in a given catalog have the corresponding column set to ``$-1.00$''.
See Section~\ref{sec-redshift}.

\item
Column~51:
the adopted redshift.
Sources which lack redshifts have this column set to ``$-1.00$''.
See Section~\ref{sec-redshift}.

\item
Column~52:
the origin of the adopted redshift, being ``zSpec'' for spectroscopic redshifts and 
``L10'', ``R11'', ``H14'',  ``S14'', and ``S16'' for photometric redshifts from
\citet{Luo2010}, \citet{Rafferty2011}, \citet{Hsu2014},
\citet{Skelton2014}, and \citet{Straatman2016}, respectively.
Sources which lack redshifts have this column set to ``...''.
See Section~\ref{sec-redshift}.

\item
Columns~53--54:
the $1\sigma$ lower and upper uncertainties on the adopted photometric redshifts.
Sources which lack redshifts or have adopted spectroscopic redshifts 
have these columns set to ``$-1.00$''.
See Section~\ref{sec-redshift}.

\item
Columns~55--57:
the effective exposure times derived from the exposure maps 
in the full, soft, and hard bands, respectively.
See Section~\ref{sec-imagesandemaps}.

\item
Columns~58--60:
the X-ray band ratio and its $1\sigma$ lower and upper uncertainties. 
Band ratios for sources detected in either the soft band or the hard band but not both are
the mode values from {\sc behr} (not upper or lower limits but best-guess estimates) 
and the corresponding uncertainty columns were set to ``$-1.000$''.
Sources detected only in the full band have these columns set to ``$-1.000$''.
See Section~\ref{sec-xphotmetric}.

\item
Columns~61--63:
the effective power-law photon index ($\Gamma_{\rm eff}$) and its $1\sigma$ lower and upper uncertainties.
Sources detected in either the soft band or the hard band but not both have their uncertainty columns
set to ``$-1.00$''. 
We adopted $\Gamma_{\rm eff}=1.4$ for sources detected only in the full band, and the 
uncertainty columns are set to ``$-1.00$''.
See Section~\ref{sec-xphotmetric}.

\item
Columns~64--66:
the X-ray fluxes in the full, soft, and hard bands, respectively.
Negative values indicate 90\% confidence-level upper limits on the fluxes which were
derived from the upper limits on the source counts.
See Section~\ref{sec-xphotmetric}.

\item
Column~67:
the apparent rest-frame 0.5--7.0 keV luminosity, which has not been corrected for any intrinsic
absorption. Sources which lack redshifts or are identified as stars have this column set to ``$-1.00$''.
See Section~\ref{sec-xphotmetric}.

\item
Column~68:
the intrinsic absorption column density estimated based on the deviation between
the effective photon index and the assumed intrinsic photon index of 1.8.
Sources which lack redshifts or are identified as stars have this column set to ``$-1.00$''.
See Section~\ref{sec-xphotmetric}.

\item
Column~69:
the absorption-corrected intrinsic \hbox{0.5--7.0~keV} luminosity.
Sources which lack redshifts or are identified as stars have this column set to ``$-1.00$''.
See Section~\ref{sec-xphotmetric}.

\item
Column~70:
the X-ray source type: ``AGN'', ``Galaxy'', or ``Star''.
See Section~\ref{sec-classify}.

\item
Column~71:
the matched 4~Ms \cdfs\ source ID number \citep{Xue2011}. A letter ``S'' is added to the 
ID number if the matched source is from the supplementary catalog. 
Sources which lack 4~Ms \cdfs\ counterparts have this column set to ``...''.
See Section~\ref{sec-compare4ms}.

\item
Column~72:
the matched 250~ks E-CDF-S source ID number \citep{Xue2016}. A letter ``S'' is added to the
ID number if the matched source is from the supplementary catalog.
Sources which lack 250~ks \hbox{E-CDF-S} counterparts have this column set to ``...''.
See Section~\ref{sec-compare4ms}.

\item
Column~73:
the matched 3~Ms \xmm\ \cdfs\ source ID number \citep{Ranalli2013}. 
Sources which lack 3~Ms \xmm\ \cdfs\ counterparts have this column set to ``$-1$''.
See Section~\ref{sec-compare4ms}.

\end{enumerate}

\section{Supplementary Near-infrared Bright {\emph{Chandra}}
Source Catalog}
The supplementary source catalog contains information for the 47 X-ray sources 
that have $0.007\le P_{\rm B}<0.1$ and also 
bright ($K_s\le23$) TENIS counterparts (Section~\ref{sec-detection}).
We created the supplementary catalog following the same procedures as 
for the main catalog, except that in the supplementary catalog, a source is considered to be detected
when its $P_{\rm B}$ value is less than 0.1 (instead of 0.007
as in the main catalog). The \xray\ positional uncertainties were calculated
following Equation~\ref{eq-dpos}. The primary counterparts of the X-ray sources were set to 
be their TENIS counterparts, and we then searched for their 
optical through radio counterparts in the other multiwavelength catalogs (see Section~\ref{sec-ID}).
Thirty supplementary sources have spectroscopic redshifts, and another 16 
have photometric redshifts from \citet{Hsu2014} or \citet{Straatman2016}.
The remaining supplementary source, supplementary XID 3, does not have a redshift estimate,
as it lies outside the GOODS-S and CANDELS regions and it 
does not appear to have any optical counterparts despite having 
a TENIS counterpart
with $K_s=22.9$.
The median redshift of the entire supplementary sample 
is $1.17\pm0.10$, similar to that for the main-catalog sources.
There are 25 AGNs classified in the supplementary catalog, and the other 22 sources are likely
normal galaxies. The fraction of AGNs in the supplementary catalog ($25/47=53\%$) 
is smaller than that in the main catalog ($711/1008=71\%$), because fainter X-ray sources
generally have a higher galaxy fraction 
(e.g., \citealt{Bauer2004,Ranalli2005,Lehmer2012}; Section~\ref{sec-ncounts} below)
and our
selection of supplementary sources is biased toward galaxies by requiring 
bright TENIS counterparts for the X-ray sources 
(e.g., see the sixth criterion of AGN classification in 
Section~\ref{sec-classify}).

The spatial distribution of the supplementary sources is displayed in 
Figure~\ref{fig-srcdistsupp}. Compared to the \citet{Xue2011} 4~Ms CDF-S
catalogs, 36 sources were newly detected, including 21 AGNs and 15 galaxies. 
We present the supplementary \chandra\ source catalog in Table~\ref{tbl-scat}; the details of
the table columns are the same as those for the main catalog (Section~\ref{sec-maincat}).

\begin{figure}
\centerline{
\includegraphics[scale=0.48]{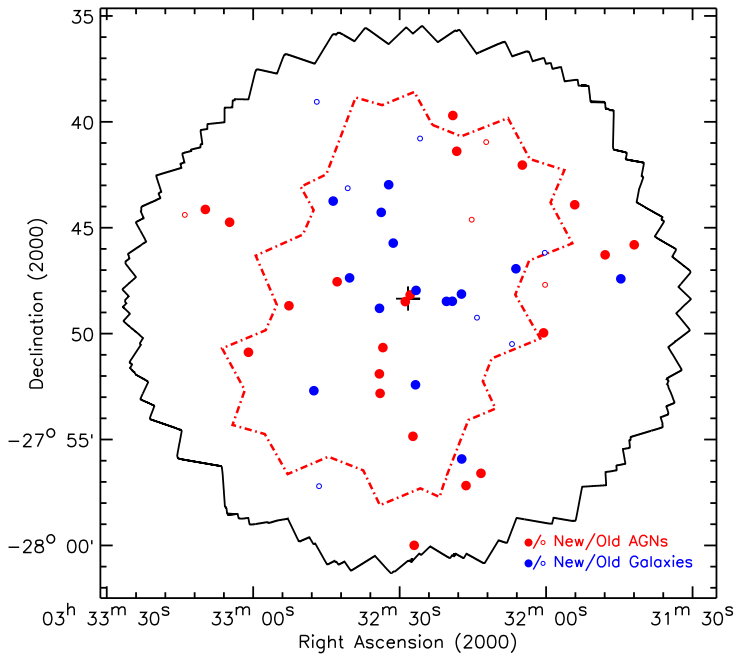}
}
\caption{
Spatial distribution of the supplementary catalog sources.
Red and blue symbols represent AGNs and galaxies, 
respectively. Filled symbols indicate newly detected sources, while
open symbols are sources that were present in the \citet{Xue2011} 4~Ms CDF-S
catalogs.
The average aim point, CDF-S boundary, and
GOODS-S region are indicated, as was done in Figure 1.
}
\label{fig-srcdistsupp}
\end{figure}

\section{Completeness and Reliability Analysis} \label{sec-comprel}
Given the nature of our two-stage source-detection approach,
it is not straightforward to assess the completeness and reliability of our main source 
catalog.
Therefore, we resorted to simulations for such an assessment, as done routinely
among previous \xray\ surveys 
\citep[e.g.,][]{Cappelluti2007,Cappelluti2009,Puccetti2009,Xue2011,Xue2016}.
We followed Section~6.1 of \citet{Xue2011} to generate simulated 7~Ms \cdfs\ observations 
with an input catalog of simulated sources;
each observation has the same exposure time, aim point, roll angle, and aspect solution file
as the corresponding real \cdfs\ observation.
We then created images in the three X-ray bands from the merged simulated event file, and
we ran {\sc wavdetect} on the images at a false-positive probability threshold of 10$^{-5}$
to obtain a candidate source list. We subsequently utilized AE to compute
photometric properties and $P_{\rm B}$ values for these candidate-list sources.

\begin{figure*}
\centerline{
\includegraphics[scale=0.83]{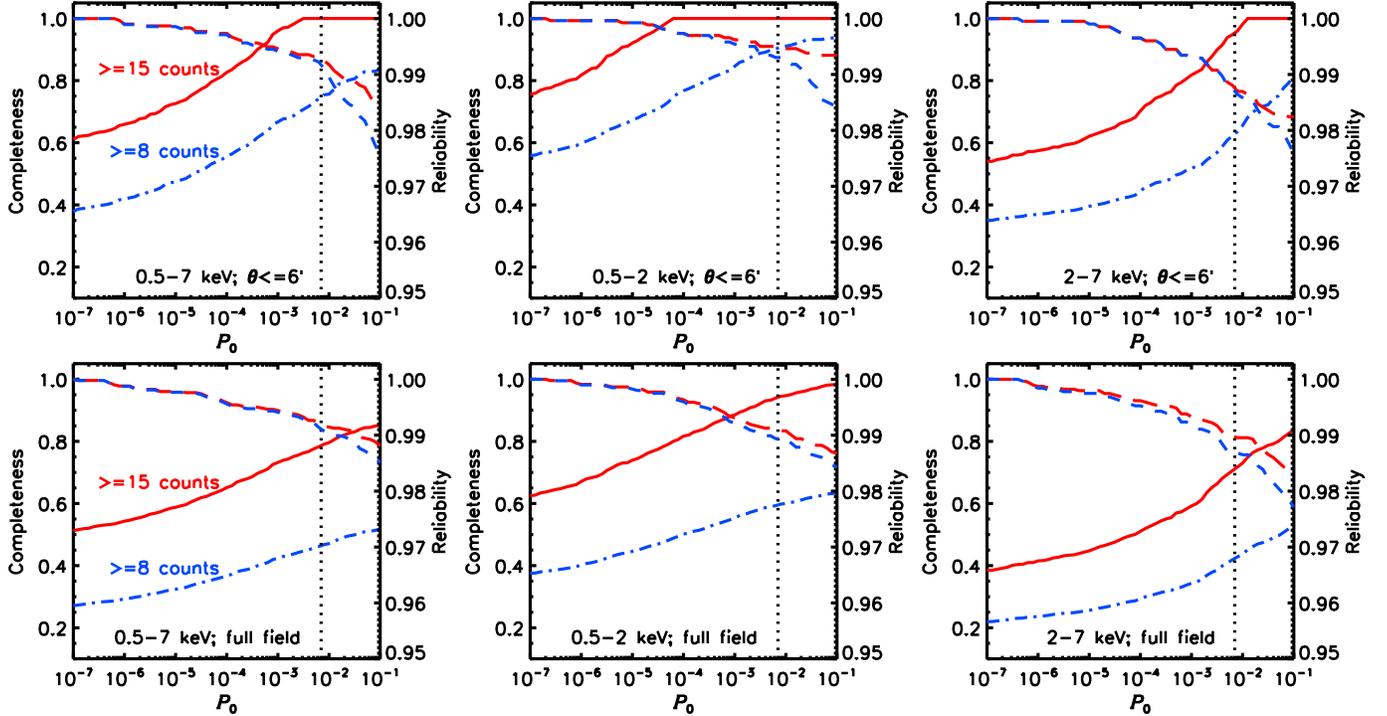}
}
\figcaption{
Completeness (left $y$-axis; solid and dashed-dot curves) and
reliability
(right $y$-axis; long-dashed and short-dashed curves) as a function of the 
$P_{\rm B}$ threshold value, $P_0$, for sources with $\ge 15$~counts 
(red solid and long-dashed curves)
and $\ge 8$~counts (blue dashed-dot and short-dashed curves)
in the full, soft, and hard bands, respectively. The top panels show 
the completeness and reliability curves for the central $6\arcmin$-radius region
and the bottom panels show those for the entire \cdfs\ field.
The vertical dotted lines indicate our adopted main-catalog $P_{\rm B}$
threshold of 0.007.
\label{fig-comp-rel}
}
\end{figure*}

By comparing the input sources and detected sources from the simulation, we can
assess the completeness and reliability of our main catalog.
For a given source-count limit, the completeness is defined as the fraction of the 
input sources detected above the count limit (the source recovery fraction), 
while the reliability is defined as 1 minus the ratio between 
the number of spurious sources (not in the input source list) above the count limit and 
the number of input sources above the count limit.
As we further filtered the candidate sources with
a $P_{\rm B}$ threshold cut during the 
second stage of our \hbox{source-detection} approach, the completeness and reliability
also have a dependence on the adopted $P_{\rm B}$ threshold value, denoted as $P_0$.
Figure~\ref{fig-comp-rel} shows
the completeness and reliability as a function of $P_0$
within the central $6\arcmin$-radius region
and over the entire \cdfs\ field, for sources with at least 15 and 8 counts in
the full, soft, and hard bands, respectively;
the minimum number of detected counts in the soft and hard bands for 
our main-catalog sources is $\approx8$ (Table~\ref{tbl-cnt}). 

According to Figure~\ref{fig-comp-rel}, 
the detection completeness increases
and the reliability decreases when the $P_{\rm B}$ threshold value is raised, as expected.
For a larger source-count limit (15 vs.\ 8), the completeness and reliability are both
better, although the difference in reliability is negligible when the $P_{\rm B}$ threshold 
value is small (e.g., $P_0=10^{-5}$).
Within the central $6\arcmin$-radius region, our source detection achieves a better 
completeness overall compared to that in the entire \cdfs\ field.
At our adopted main-catalog $P_{\rm B}$ threshold value of 0.007,
the completeness levels within the central $6\arcmin$-radius region are
100.0\% and 74.7\% (full band), 100.0\% and 90.5\% (soft band),
and 95.3\% and 62.8\% (hard band) for sources with $\ge 15$ and $\ge 8$~counts, respectively.
Across the entire \cdfs\ field, the completeness levels are 78.6\% and 46.4\% (full band), 94.2\% and 59.5\% (soft band),
and 71.1\% and 42.3\% (hard band) for sources with $\ge 15$ and $\ge 8$~counts, respectively.
At our adopted
$P_{\rm B}$ threshold, the reliability levels range from 98.7\% to 99.5\% for all the  
cases (the central $6\arcmin$-radius region or the entire field, $\ge 15$ or $\ge 8$~counts,
in one of the three X-ray bands) 
in Figure~\ref{fig-comp-rel}, 
suggesting that there are $\approx7$, 6, and 5 spurious 
detections with $\ge 15$~counts in the full,
soft, and hard bands, and
$\approx7$, 7, and 5 spurious detections with $\ge 8$~counts in the full,
soft, and hard bands, respectively.
The total number of spurious detections estimated from simulations is thus $\approx19$.
Most of these spurious sources should not 
have a multiwavelength counterpart; there are 16 such sources in our main catalog 
(Section~\ref{sec-ID}).

\begin{figure}
\centerline{
\includegraphics[scale=0.55]{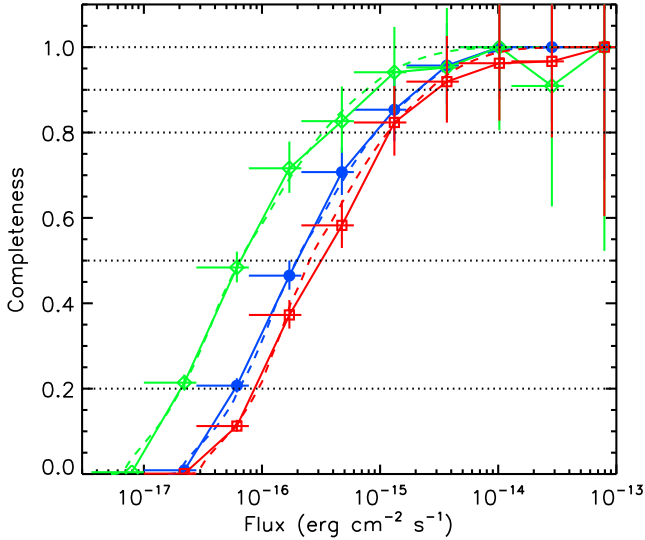}
}
\figcaption{Completeness as a function of source flux
in the full (blue filled circles), soft (green open diamonds), and hard (red open squares) 
bands, given our adopted main-catalog $P_{\rm B}$ threshold value of 0.007.
The solid lines link the corresponding completeness data points.
Overlaid are the survey solid angle vs.\ sensitivity curves (dashed curves)
normalized to the maximum solid angle (see Figure~\ref{fig-senhist} below).
The horizontal dotted lines denote five specific completeness levels (100\%,
90\%, 80\%, 50\%, and 20\%).
\label{fig-comp-flux}}
\end{figure}

At our adopted main-catalog $P_{\rm B}$ threshold
value of 0.007, the detection
completeness as a function of source flux
is presented in
Figure~\ref{fig-comp-flux}.
The completeness versus flux curves are consistent with the 
survey solid angle versus flux-limit curves (Figure~\ref{fig-senhist} below) derived in 
Section~\ref{sec-bkg} below, indicating that the  
completeness at a given flux
is dominated by the \cdfs\ area fraction that is sensitive for detecting sources
at this flux limit.
In Table~\ref{tbl-comp} we list
the source fluxes at four specific completeness
levels (90\%, 80\%, 50\%, and 20\%) in the full, soft, and hard bands, respectively.

\section{Background and Sensitivity Analysis} \label{sec-bkg}
We created background and sensitivity maps following the procedure described in
Section~7 of \citet{Xue2011}. Briefly, we masked out the main and supplementary
catalog sources from the raw images in the three X-ray bands, and we filled in
the masked regions with random counts that are consistent with the local background level.
The resulting background maps in the three bands were used to determine the 
detection sensitivity at each pixel location, which is the flux limit required for a source 
to be selected by our AE $P_{\rm B}$ criterion.
Given the background level at each pixel location, we derived the minimum number of 
source counts required for a detection using Equation~\ref{eq-pb}
and our adopted $P_{\rm B}$ threshold value of 0.007. Utilizing the exposure maps (Section~3.1) 
and
assuming a simple power-law model with $\Gamma=1.4$, we 
converted the limiting count rates to limiting fluxes and produced sensitivity maps
for the main catalog in the three X-ray bands.

The background properties for the 7~Ms \cdfs\ are summarized in
Table~\ref{tbl-bkg}.
The observed \cdfs\ background consists of several components, primarily
the unresolved cosmic background, particle background, and instrumental
background \citep[e.g.,][]{Markevitch2001,Markevitch2003}; 
we do not attempt to separate these components for our analysis here.
The mean numbers of background counts per pixel are small (0.17--0.60),
indicating that many of the \cdfs\ pixels did not receive any X-ray photons over the 
7~Ms exposure. Indeed, in the full-band, soft-band, and hard-band raw images, 
$\approx59\%$, 84\%, and 68\% of the pixels have zero counts.
The mean background count rate in the soft band (0.055~counts~Ms$^{-1}$~pixel$^{-1}$)
is $\approx13\%$--$17\%$ smaller compared to those for the 2~Ms \cdfs\
\citep{Luo2010} and 4~Ms \cdfs\ \citep{Xue2011}, probably due to the decline of 
the ACIS effective area below 2~keV (from build-up of contaminant
on the ACIS optical blocking filters),\footnote{For example, the ACIS-I soft-band effective
area has dropped by $\approx10\%$ from Chandra Cycle 11 (proposal cycle for 
the 4~Ms \cdfs) to Cycle 15 (proposal cycle for
the 7~Ms \cdfs). The 7~Ms
\cdfs\ observations span a broad range in time
and not all the background components 
are affected by the decline of
the ACIS effective area. \label{footnote-area}} and 
the increased sensitivity of the 7~Ms \cdfs\ that resolves
a larger fraction of the cosmic background,
and/or variations of the particle and instrumental
background components over the past several years. The ratio between the total numbers of
background and source counts in the soft band is approximately the same ($\approx4.2$) for the
2~Ms, 4~Ms, and 7~Ms \cdfs. In the full and hard bands, the mean background count rates
are smaller ($\approx25\%$) than those for the 2~Ms and 4~Ms \cdfs, 
mainly because we adopted a smaller upper energy bound of 7~keV (instead of 8~keV) here.
The ratio between the total numbers of background and source counts in the 
hard band ($\approx14$; 2--7 keV) is significantly lower than those ($\approx20$; 2--8 keV) 
for the 2~Ms and 4~Ms \cdfs,
confirming that it is advantageous
to search for sources in the 2--7~keV band instead of the 2--8~keV band where
the background has a larger contribution.

We are able to achieve unprecedented X-ray sensitivity in the 7~Ms \cdfs.
The lowest estimated flux limits achievable 
are $\approx1.5\times10^{-17}$, $4.8\times10^{-18}$, and $2.1\times10^{-17}$~\flux\
in the full, soft, and hard bands, respectively, and the 
average achievable flux limits over the central $\approx1$~arcmin$^2$ region
are $\approx1.9\times10^{-17}$, $6.4\times10^{-18}$, and $2.7\times10^{-17}$~\flux. 
The lowest detected fluxes in the main catalog are actually around
these limits, being 
$\approx1.7\times10^{-17}$, $7.7\times10^{-18}$, and $3.5\times10^{-17}$~\flux\ in the three 
bands. 
Compared to the average soft-band flux limit in the central $\approx1$~arcmin$^2$ region
of the 4~Ms \cdfs\ ($9.1\times10^{-18}$~\flux; \citealt{Xue2011}), the 7~Ms \cdfs\
sensitivity has been improved by a factor of $1.42$.
The full-band and hard-band sensitivities are not directly comparable, 
as the energy ranges are different.
If we simply scale the 4~Ms \cdfs\ 0.5--8.0~keV and 2--8~keV flux limits in the central region to the
full (0.5--7.0~keV) and hard (2--7~keV) bands assuming a $\Gamma=1.4$ power-law spectrum, the full-band
sensitivity has been improved by a factor of 1.52 ($2.9\times10^{-17}$ vs.\ $1.9\times10^{-17}$~\flux)
and the hard-band sensitivity has been improved by a factor of 1.76 ($4.8\times10^{-17}$
vs.\ $2.7\times10^{-17}$~\flux). The sensitivity improvement due to increased exposure 
in the full or hard band should be smaller 
than that in the soft band, because of the lower background level in the soft band (e.g., Table~\ref{tbl-bkg}).
The larger full- and \hbox{hard-band}
improvement factors obtained for the 7~Ms \cdfs\ are due to the more sensitive 
0.5--7.0~keV and \hbox{2--7~keV} bands adopted for source detection where the 
background levels are lower compared to the 0.5--8.0~keV and 2--8~keV bands
(see Footnote 16 of \citealt{Xue2016}
and our comparison of the background-to-source count ratios
in the previous paragraph).

\begin{figure}
\centerline{
\includegraphics[scale=0.5]{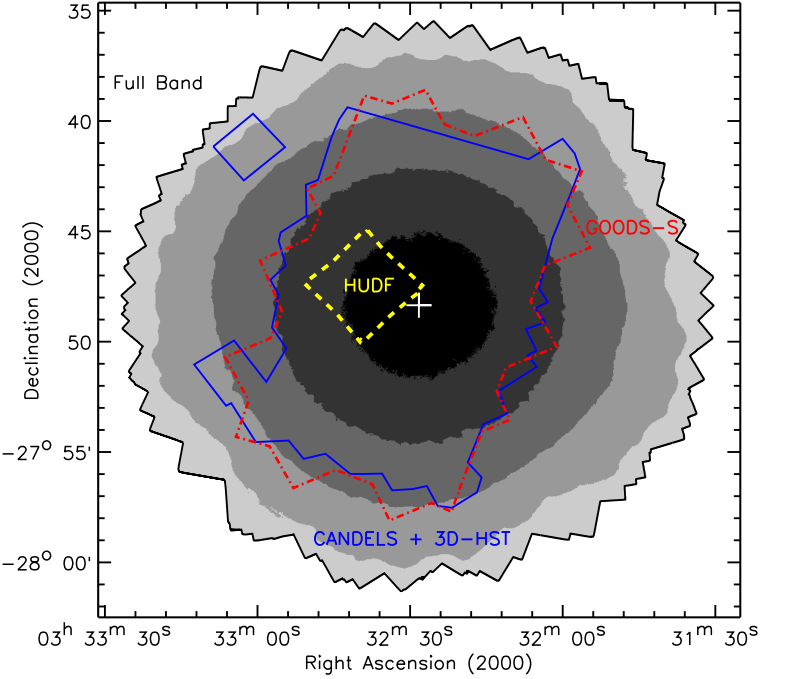}
}
\figcaption{
Full-band sensitivity map for the main source catalog.
The gray-scale levels, from black to light gray, represent regions with flux limits
of \hbox{$<3\times10^{-17}$}, \hbox{$3\times10^{-17}$--$8\times10^{-17}$},
\hbox{$8\times10^{-17}$--$2\times10^{-16}$}, \hbox{$2\times10^{-16}$--$10^{-15}$},
and \hbox{$>10^{-15}$~\flux}, respectively.
The regions and the plus sign are the same as those in
Figure~\ref{fig-rawimg}.
\label{fig-fbsen}}
\end{figure}

\begin{figure}
\centerline{
\includegraphics[scale=0.5]{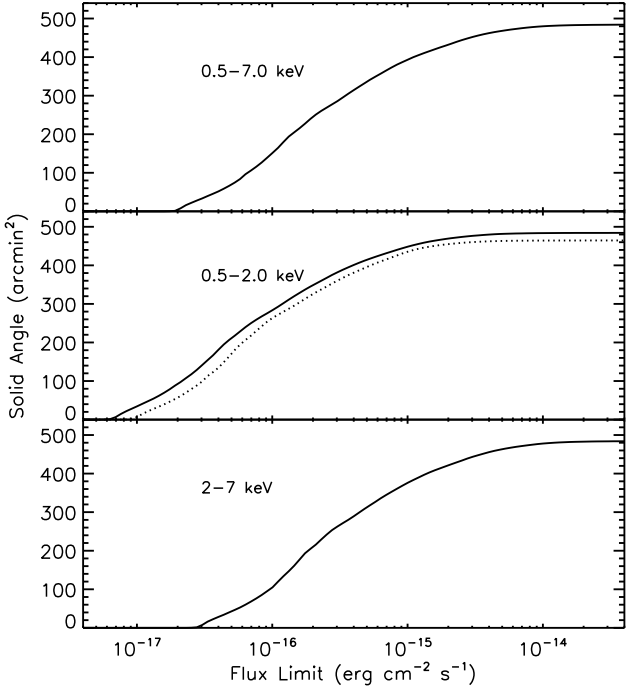}
}
\figcaption{
Survey solid angle as a function of flux limit in the
full (top), soft
(middle), and hard (bottom) bands. For comparison, the soft-band
relation for the 4~Ms \cdfs\ \citep{Xue2011} is displayed as the 
dotted curve in the middle panel (the full-band and hard-band results
are not directly comparable due to the different choices of the energy bands).
The 4~Ms \cdfs\ curve can be approximately scaled to the 7~Ms one by dividing 
the flux limits by a scaling factor of $\approx1.38$.
\label{fig-senhist}}
\end{figure}

Beyond the small central region, the sensitivity drops with increasing off-axis angle.
The full-band sensitivity map is displayed in Figure~\ref{fig-fbsen}; flux limits in
different ranges are shaded with different gray-scale levels.
It is possible to detect sources with fluxes somewhat smaller than 
the sensitivity limits at their
locations, due to the difference between the $\Gamma_{\rm eff}$ values of the sources and the 
assumed value of 1.4
when computing the flux limits. There are 11 soft-band, 28 \hbox{full-band}, and
2 \hbox{hard-band} sources with fluxes $\approx1\%$--13\% below their corresponding flux limits.
Given the spatial dependence of the sensitivity, we display in Figure~\ref{fig-senhist}
the survey solid angle versus flux limit in the full, soft,
and hard bands. The flux limits for $\approx50\%$ of the \cdfs\ area are 
$\approx2.0\times10^{-16}$, $6.5\times10^{-17}$, and $2.5\times10^{-16}$~\flux\ in the full, soft, and 
hard bands, respectively,
approximately an order of magnitude larger than those for the central 
$\approx1$~arcmin$^2$ region, and these values are consistent 
with the 50\% completeness flux limits presented in Table~\ref{tbl-comp}. 

We also compared the 7~Ms \cdfs\ \hbox{soft-band} 
sensitivity--area
curve to the one for the 4~Ms \cdfs\ \citep{Xue2011} in Figure~\ref{fig-senhist}.
The 4~Ms \cdfs\ curve can be approximately scaled to the 7~Ms one by dividing
the flux limits by a scaling factor of $\approx1.38$, indicating that 
the 7~Ms \cdfs\ 
sensitivity has been improved by a factor of $\approx1.38$ on average.
Such an improvement in sensitivity is expected given the factor of 
$\approx1.75$ increase in exposure (see Section~7.3 of \citealt{Xue2011}) and the 
decrease of the ACIS-I soft-band effective
area over the past years (see Footnote~\ref{footnote-area}).

\section{Number Counts for the 7~Ms CDF-S}  \label{sec-ncounts}
\begin{figure*}
\centerline{
\includegraphics[scale=0.6]{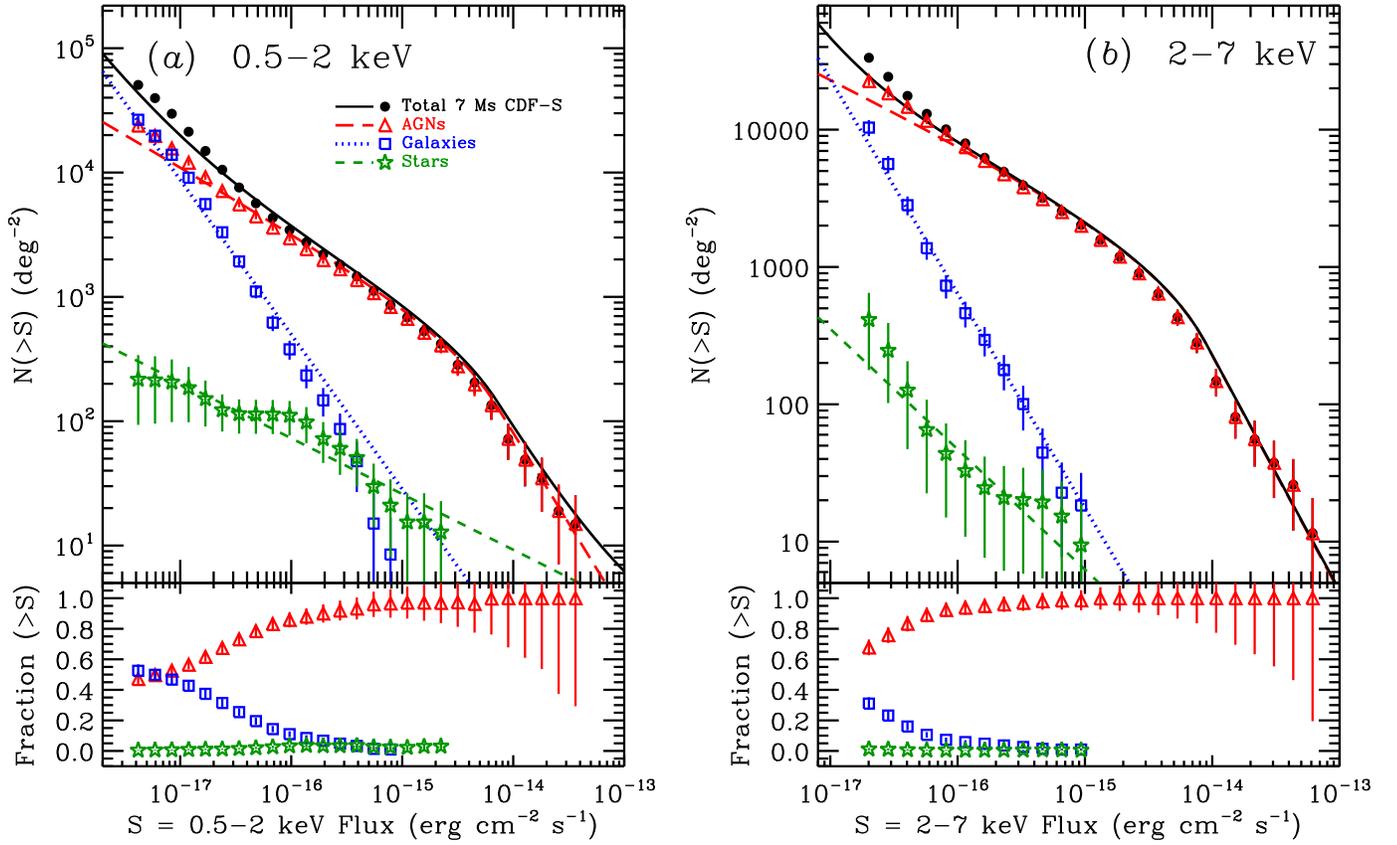}
}
\figcaption{
Cumulative number counts (number of sources brighter than a given flux) 
for the main source catalog (filled circles) in the 
(a) soft band and (b) hard band. The number counts are further 
broken down 
into AGNs (open red triangles), 
normal galaxies (open blue squares), 
and Galactic stars (open green stars) with their corresponding $1\sigma$ errors displayed. 
The red long-dashed, blue dotted, and green short-dashed curves represent
the number-count models based on the best-fit differential number-count models 
(Equation~\ref{eq-ncounts} and Table~\ref{tbl-ncounts}) for AGNs, galaxies, and stars, respectively,
and the black curves show the number-count models for all X-ray sources (sum of the three
components). The bottom panels show the fractional contributions from AGNs, galaxies, 
and stars to the total number counts. Galaxies provide significant contributions to the 
total number counts near the soft-band flux limit and they overtake AGNs at soft-band fluxes smaller 
than $\approx6.0\times10^{-18}$~\flux. 
\label{fig-ncounts}}
\end{figure*}

We computed the cumulative number of
sources, \hbox{$N(>S)$}, brighter than a given intrinsic flux, $S$, for the soft and hard
bands following the procedure described in Section~2 of \citet{Lehmer2012}.
Simulations based on our \hbox{source-detection} method were performed to obtain the 
source recovery functions,
which account for detection incompleteness (see the Appendix of \citealt{Lehmer2012} for details). 
A Bayesian approach 
was employed to obtain the 
\hbox{flux-probability} distributions, which account for Eddington
bias (see Section~2.1 of \citealt{Lehmer2012} for details).
Unlike the creation of the sensitivity maps (Section~\ref{sec-bkg}) where simple
count-rate-to-flux conversion factors were used, we consider here the conversion to be
probabilistic. A Bayesian prior based on the differential number-count models 
($dN/dS|_{\rm model}$) is utilized in 
determining the flux-probability distributions, and following \citet{Lehmer2012},
we adopted double or single power-law forms for the number-count models of AGNs, 
normal galaxies, and Galactic stars: 
\[\frac{dN}{dS}^{\rm AGN} = \left\{
\begin{array}{c c}
K^{\rm AGN} (S/S_{\rm ref})^{-\beta_1^{\rm AGN}}  & (S \le f^{\rm AGN}_{\rm b})
\\
K^{\rm AGN} (f_{\rm b}/S_{\rm ref})^{\beta_2^{\rm AGN} - \beta_1^{\rm AGN}}(S/S_{\rm
ref})^{-\beta_2^{\rm AGN}} &  (S > f^{\rm AGN}_{\rm b}) \\ \end{array} \right.
\]

\begin{eqnarray}
\frac{dN}{dS}^{\rm gal} & = & K^{\rm gal} (S/S_{\rm
ref})^{-\beta^{\rm gal}} \nonumber \\
\frac{dN}{dS}^{\rm star} & = & K^{\rm star} (S/S_{\rm
ref})^{-\beta^{\rm star}},
\label{eq-ncounts}
\end{eqnarray}
where $S_{\rm ref}=10^{-14}$~\flux.
As shown later in Figure~\ref{fig-ncounts}, these power-law models 
provide acceptable descriptions of the overall shapes of the cumulative number counts. 

To derive the best-fit model parameters in Equation~\ref{eq-ncounts}, a maximum-likelihood technique
was utilized to maximize the total likelihood of 
obtaining the main source catalog and its source-count distribution
(see Section~2.2 of \citealt{Lehmer2012} for details).
The best-fit parameters and the corresponding 1$\sigma$ uncertainties
are listed in Table~\ref{tbl-ncounts}.
After determining the Bayesian prior for the flux-probability distributions, 
the cumulative number counts for the soft and 
hard bands were computed 
based on the main-catalog sources, broken down into AGNs, galaxies, and stars.
In Figure~\ref{fig-ncounts}, we present the cumulative number counts for the three
source populations as well as all the X-ray sources in the soft and hard bands, respectively.  
The number-count models derived from the best-fit differential models
(Equation~\ref{eq-ncounts} and Table~\ref{tbl-ncounts}) are also displayed, showing
general agreement with the data.

Our number-count calculations implement a probabilistic approach, in which we
calculated, at each location in our images, the probability of detection as a
function of flux and photon index (see Section~2.1 of \citealt{Lehmer2012}).  As
such, there are sufficiently large areas of the survey in which sources with
fluxes below the sensitivity limits quoted in Section~\ref{sec-bkg} could be detected. The
flux limits for
$\approx 10$~arcmin$^2$ effective areas derived from this approach are
$4.2\times10^{-18}$~\flux\ in the soft band and $2.0\times10^{-17}$~\flux\ in the hard band.
Unlike the AGN number counts, where a break flux around $10^{-15}$~\flux\ is evident, 
the galaxy number counts follow a single power-law form and continue to rise 
down to the lowest fluxes observed.
At the faint end, the galaxy number counts rise more sharply 
with power-law slopes of 
$\approx1.2$--1.6 ($\beta^{\rm gal}-1$) than the AGN number counts with faint-end
\hbox{power-law} slopes of $\approx0.5$.

The bottom panels of Figure~\ref{fig-ncounts} show the fractional contributions from AGNs, galaxies,
and stars to the total number counts. The contribution from galaxies increases toward lower
fluxes, especially in the soft band which is more sensitive than the hard band.
Similar trends have been observed in previous studies of number counts in deep X-ray
surveys \citep[e.g,][]{Bauer2004,Ranalli2005,Lehmer2012}, and it has been suggested that 
the galaxy number counts will overtake the AGN number counts just below the 
4~Ms \cdfs\ soft-band flux limit \citep[e.g.,][]{Lehmer2012}. Indeed, thanks to the
unprecedented sensitivity of the 7~Ms \cdfs,
we observe for the first time that the galaxy number counts exceed the AGN
number counts at soft-band fluxes smaller than
$\approx6.0\times10^{-18}$~\flux. 
At the soft-band flux limit ($4.2\times10^{-18}$~\flux),
the AGN density is 
$\approx23\,900$~deg$^{-2}$, $47\%\pm4\%$ of the total X-ray source density ($\approx50\,500$~deg$^{-2}$), 
while the galaxy density
reaches $\approx26\,600$~deg$^{-2}$, $52\%\pm5\%$ of the total, indicating that 
normal galaxies start to 
dominate the X-ray source population at the faintest flux levels. At this flux limit,
the entire sky contains $\approx1.0$ billion AGNs and $\approx1.1$ billion galaxies.
We caution that due to cosmic variance, the \cdfs\ field might not be a representative patch of the entire 
sky, 
and small field-to-field variations
of the \hbox{X-ray} source distribution have been observed between the CDF-S and other
survey fields \citep[e.g.,][]{Bauer2004,Luo2008}.
 
Among the main-catalog sources, there are 264 X-ray galaxies (excluding AGNs) 
with $K_s\le22$, which
constitute $4.0\%\pm0.2\%$ of the 6651 $K_s\le22$ TENIS sources within the 7~Ms \cdfs\
field of view. The fraction of X-ray detected galaxies increases toward better X-ray 
sensitivity; 
within the innermost 2\arcmin-radius region, there are 31 $K_s\le22$ X-ray galaxies,
$24\%\pm4\%$ of the 129 $K_s\le22$ TENIS sources in this region.  
With the number-count results, we further
investigated the fraction of galaxies that are
detectable in deep \hbox{X-ray} surveys as a function of the X-ray 
limiting flux, accounting for detection incompleteness and Eddington bias.
At the \hbox{soft-band} flux limit ($4.2\times10^{-18}$~\flux), the density of
$K_s\le22$ X-ray galaxies is $\approx22\,300$~deg$^{-2}$ (derived similarly to
those in Figure~\ref{fig-ncounts}).
The TENIS catalog is complete at $K_s\le22$,
and we verified its completeness by comparing its $K_s$-band galaxy number counts to 
previous number-count results \citep[e.g.,][]{Saracco2001}. Excluding 295 stars 
from the 6651 TENIS sources identified from spectroscopic catalogs, the $K_s\le22$
galaxy density is $\approx47\,300$~deg$^{-2}$. Therefore, 
the X-ray detected $K_s\le22$ galaxies account for $\approx 47\%$ of all $K_s\le22$ galaxies
at the \hbox{soft-band} flux limit.
The fraction of $K_s\le22$ galaxies detected in the \hbox{X-rays}
decreases rapidly with increasing soft-band
limiting flux, being $\approx20\%$ at a limiting flux of $10^{-17}$~\flux\ and
$\approx2\%$ at a limiting flux of $5\times10^{-17}$~\flux.

Using the number-count estimates derived above, we 
integrated the X-ray source fluxes to obtain the 
fraction of the cosmic X-ray background (CXRB) that is resolved into point sources at 
the 7~Ms \cdfs\ flux limit. 
Our analysis followed the same procedure described in Section~3.4 of \citet{Lehmer2012}.
We adopted total CXRB intensities of $(8.15\pm0.58)\times10^{-12}$~\flux\ and
$(1.49\pm0.20)\times10^{-11}$~\flux\ in the soft and hard bands, respectively 
\citep{Hickox2006,Kim2007b}. The resolved CXRB fractions, after including 
the contributions from the relatively rare X-ray bright-source population \citep{Kim2007b},
are $80.9\%\pm4.4\%$ and $92.7\%\pm13.3\%$ in the soft and hard bands, respectively.
For comparison, the resolved CXRB fractions at the 4~Ms \cdfs\ flux limit \citep{Lehmer2012}
are $75.7\%\pm4.3\%$ and $82.4\%\pm13.0\%$ in the soft band and the 2--8 keV band, respectively.

\section{Summary}

We have presented X-ray source catalogs for the deepest \chandra\ X-ray
survey, the 7~Ms \cdfs. The main points from this work are summarized below:

\begin{enumerate}

\item
The entire 7~Ms \cdfs\ consists of 102 individual observations 
covering a total area of 484.2~arcmin$^{2}$. The cleaned net
exposure time is 6.727~Ms.
See Section~2.

\item
The main \chandra\ source catalog contains 1008 \xray\ sources that were
detected by {\sc wavdetect} with a \hbox{false-positive} probability threshold
of $1\times10^{-5}$ and filtered by AE with a binomial 
no-source probability ($P_{\rm B}$) threshold
of 0.007. These sources were detected in up to three X-ray
bands: 0.5--7.0~keV (full band), \hbox{0.5--2.0~keV} (soft band),
and 2--7~keV (hard band).
See Section~3.2.

\item
The supplementary \chandra\ source catalog contains 47 X-ray sources that were
detected by {\sc wavdetect} with a false-positive probability threshold
of $1\times10^{-5}$, have $0.007\le P_{\rm B}<0.1$, and are matched to
bright ($K_s\le23$) NIR 
counterparts. 
See Section~3.2.

\item
The absolute astrometry of the 7~Ms \cdfs\ was registered to the TENIS
NIR astrometric frame. The \xray\ source positions were determined based on
their centroid or matched-filter positions.  
For the main-catalog sources, the 1$\sigma$ positional 
uncertainties range from $0\farcs11$ to $1\farcs28$,
with a median value of $0\farcs47$.
See Sections~2.2, 3.2, and 4.1.

\item
We identified optical/NIR/IR counterparts for 992 (98.4\%) of the main-catalog sources,
with an average false-match rate of $\approx1.6\%$. 
Basic counterpart
information in the optical through radio catalogs for the 992 sources are provided.
Most of the 16 sources without
multiwavelength counterparts are likely spurious detections. 
See Section~4.2.

\item
We collected redshifts from public catalogs for 986 main-catalog sources, 
including 
653 spectroscopic redshifts and 333 photometric redshifts.
The photometric redshifts are of high quality in general. 
The median redshift for these X-ray sources is $1.12\pm0.05$
with an interquartile range of 0.67--1.95.
See Section~4.3.

\item
For the main catalog, the median numbers of source counts in the 
full, soft, and hard bands are 
98.9, 47.7, and 94.6, respectively.
There are 456 sources with $>100$ full-band counts,
and 90 with $>1000$ \hbox{full-band} counts.
The median X-ray fluxes in the full, soft, and hard bands are
$3.1\times10^{-16}$, $6.5\times10^{-17}$, and $5.7\times10^{-16}$~\flux, respectively.
There are 613 sources with \hbox{absorption-corrected} intrinsic
0.5--7.0~keV luminosities above $10^{42}$~\lum\ and 108 sources above $10^{44}$~\lum.
See Section~4.4.

\item
We identified 711 AGNs (71\%) from the main catalog based on their X-ray and multiwavelength
properties. Besides 12 Galactic stars, the remaining 285 sources are likely normal galaxies.
See Section~4.5.

\item
We detected 291 new X-ray sources in the 7~Ms \cdfs\ main catalog
compared to the 4~Ms \cdfs\ catalogs, three of which
were outside the footprint of the 4~Ms \cdfs.
A smaller fraction of the new sources are classified as AGNs
(63\%) compared to that for the entire catalog (71\%).
The median redshifts for the new AGNs and galaxies are comparable to those 
for the entire AGN and galaxy samples.
The new AGNs 
have a lower median flux and median luminosity than the old AGNs.
The new galaxies have a slightly lower median flux than the old galaxies, while
their median luminosities are comparable.
See Section~4.6.

\item
Simulations suggest that our main catalog is highly reliable 
with $\approx7$, 7, and 5 spurious detections in the full,
soft, and hard bands, respectively.
The completeness levels within the central $6\arcmin$-radius region are
74.7\%, 90.5\%, and 62.8\% for sources with $\ge 8$~counts in the full,
soft, and hard bands, respectively, and they are higher for sources with larger
numbers of counts.
See Section~6.

\item
The mean numbers of background counts per pixel are still small (0.17--0.60 in
the three bands) in the 7~Ms \chandra\ exposure; most of the pixels have zero counts.
The low background level and deep exposure result in unprecedented X-ray sensitivity for
the 7~Ms \cdfs. 
The
average flux limits over the central $\approx1$~arcmin$^2$ region
reach $\approx1.9\times10^{-17}$, $6.4\times10^{-18}$, and $2.7\times10^{-17}$~\flux\
in the full, soft, and hard bands, respectively, and
the flux limits for $\approx50\%$ of the \cdfs\ area are
$\approx2.0\times10^{-16}$, $6.5\times10^{-17}$, and $2.5\times10^{-16}$~\flux.
Compared to the average \hbox{soft-band} flux limit in the central $\approx1$~arcmin$^2$ region
of the 4~Ms \cdfs, the 7~Ms \cdfs\ 
sensitivity has been improved by a factor of $1.42$.
See Section~7.

\item
We computed cumulative number counts
down to the soft-band flux limit of $4.2\times10^{-18}$~\flux\ and 
\hbox{hard-band} flux limit of $2.0\times10^{-17}$~\flux.
The number counts are 
broken down into AGNs, normal galaxies, 
and Galactic stars. After correcting for detection incompleteness and Eddington bias,
the AGN density is $\approx23\,900$~deg$^{-2}$ ($47\%\pm4\%$ of the total source density) 
and the galaxy density 
is $\approx26\,600$~deg$^{-2}$ ($52\%\pm5\%$ of the total)
at the soft-band flux limit. We observe, for the first time, that
normal galaxies start to
dominate the X-ray source population 
at the faintest flux levels ($\lesssim6.0\times10^{-18}$~\flux).
The resolved CXRB fractions computed using the number-count estimates are
$80.9\%\pm4.4\%$ and $92.7\%\pm13.3\%$ in the soft and hard bands, respectively.
See Section~8.

\end{enumerate}

The 7~Ms \cdfs\ will 
serve as a multi-decade \chandra\ legacy for advancing 
deep-survey science projects, owing to 
its unique combination of great depth and high angular resolution.
Detailed science results for the 7 Ms \cdfs\ are presented in
additional papers \citep[e.g.,][]{Lehmer2016,Vito2016,Yang2016}.

~\\

We acknowledge financial support from
the National Thousand Young Talents program of China (B.L., Y.Q.X.),
National Natural Science Foundation of China
grant 11673010 (B.L.),
Ministry of Science and Technology of China
grant 2016YFA0400702 (B.L.),
CXC grant GO4-15130A (B.L., W.N.B., F.V., G.Y.),
973 Program 2015CB857004 (Y.Q.X., M.S.) and 2015CB857005 (J.X.W.),
NSFC-11473026 (Y.Q.X., M.S.), NSFC-11421303 (Y.Q.X., M.S., J.X.W.), 
Strategic Priority Research Program of CAS grant XDB09000000 (Y.Q.X., M.S., J.X.W.), 
Fundamental Research
Funds for the Central Universities (Y.Q.X., M.S.),
CAS Frontier Science Key Research Program QYZDJ-SSW-SLH006 (Y.Q.X., M.S., J.X.W.),
STFC grant ST/L00075X/1 (D.M.A., I.S.),
CONICYT-Chile grants Basal-CATA PFB-06/2007 (F.E.B.), 
FONDECYT Regular 1141218 (F.E.B.), "EMBIGGEN" Anillo ACT1101 (F.E.B.),
the Ministry of Economy, Development, and Tourism's Millennium Science
Initiative through grant IC120009, awarded to the Millennium Institute
of Astrophysics, MAS (F.E.B.),
ERC Advanced grant DUSTYGAL 321334 (I.S.),
and the Royal Society/Wolfson Merit Award (I.S.).

\begin{deluxetable}{lrcccccc}

\tablewidth{0pt}
\tablecaption{Journal of 7 Ms Chandra Deep Field-South Observations}
\tablehead{
\colhead{Obs. ID}                                 &
\colhead{Obs. Start}                                 &
\colhead{Exposure}                             &
\multicolumn{2}{c}{Aim Point}                 &
\colhead{Roll Angle}                                 &
\colhead{Obs.}                                 &
\colhead{Pipeline}                             \\
\cline{4-5}
\noalign{\smallskip}
\colhead{}                                 &
\colhead{(UT)}                         &
\colhead{Time (ks)}               &
\colhead{$\alpha$ (J2000.0)}                &
\colhead{$\delta$ (J2000.0)}                &
\colhead{(deg)}         &
\colhead{Mode}         &
\colhead{Version}
}
\startdata
   1431-0&       1999 Oct 15, 17:38&   25.1&    03 32 29.31&  $-$27 48 22.2&   47.3&  VF&     8.4.5\\
   1431-1&       1999 Nov 23, 02:30&   93.2&    03 32 29.31&  $-$27 48 22.2&   47.3&   F&     8.4.5\\
    441&       2000 May 27, 01:18&   53.5&    03 32 26.91&  $-$27 48 19.4&  166.7&   F&     8.4.5\\
    582&       2000 Jun 03, 02:38&  127.6&    03 32 26.97&  $-$27 48 18.5&  162.9&   F&     8.4.5\\
   2406&       2000 Dec 10, 23:35&   28.7&    03 32 28.33&  $-$27 48 36.5&  332.2&   F&     8.4.5\\
   2405&       2000 Dec 11, 08:14&   56.3&    03 32 28.82&  $-$27 48 43.5&  331.8&   F&     8.4.5\\
   2312&       2000 Dec 13, 03:28&  123.7&    03 32 28.28&  $-$27 48 36.9&  329.9&   F&     8.4.5\\
   1672&       2000 Dec 16, 05:07&   95.0&    03 32 28.73&  $-$27 48 44.5&  326.9&   F&     8.4.5
\enddata
\tablecomments{
The 7 Ms CDF-S consists of 102 observations, with a total cleaned exposure time of 6.727~Ms.
The average aim point for the merged observations,
weighted by the individual exposure times, is $\alpha_{\rm J2000.0}=
03^{\rm h}32^{\rm m}28\fs27$, $\delta_{\rm J2000.0}=-27\degr48\arcmin21\farcs8$.
The observations were performed using Faint (F) or Very Faint (VF)
modes.
(This table is available in its entirety in the online journal. A portion is shown here
for guidance regarding its form and content.)
}
\label{tbl-obs}
\end{deluxetable}

\begin{deluxetable}{lcccccc}
\tabletypesize{\small}
\tablewidth{0pt}
\tablecaption{Summary of {\it Chandra} Source Detections \label{tbldet}}

\tablehead{
\colhead{} &
\colhead{Number of} &
\multicolumn{5}{c}{Detected Counts Per Source} \\
\cline{3-7}
\colhead{Band (keV)} &
\colhead{Sources} &
\colhead{Maximum} &
\colhead{Minimum} &
\colhead{Median} &
\colhead{$NMAD$\tablenotemark{a}} &
\colhead{Mean}
}

\startdata
Full (0.5--7.0)   & 916 & 56916.2 & 11.2 & $98.9\pm6.1$ &104.4 & $571.6\pm93.2$ \\
Soft (0.5--2.0)  & 871 & 38817.0 & 6.1 & $47.7\pm2.0$ & 47.6 & $343.5\pm65.9$  \\
Hard (2--7)  & 622 & 18137.8 & 9.2 & $94.6\pm6.0$ & 90.6 &$356.0\pm46.1 $
\enddata
\tablenotetext{a}{Normalized median absolute deviation, defined as
$NMAD=1.48\times {\rm median}(|\rm counts-{\rm median}(\rm counts)|)$.}
\label{tbl-cnt}
\end{deluxetable}

\begin{deluxetable}{lcccc}
\tabletypesize{\small}
\tablewidth{0pt}
\tablecaption{Sources Detected in One Band but not Another \label{tblundet}}
\tablehead{
\colhead{Detection Band} &
\multicolumn{4}{c}{Nondetection Energy Band} \\
\cline{2-5}
\colhead{(keV)} &
\colhead{Full} &
\colhead{Soft} &
\colhead{Hard} &
\colhead{Either}
}
\startdata
Full (0.5--7.0)  & \ldots & 129 & 302 & 22\\
Soft (0.5--2.0)  & ~~~~84~~~~ & \ldots & 364 & 84\\
Hard (2--7)   & ~~~~~8~~~~ & 115  & \ldots & ~8
\enddata
\tablecomments{For example, there were 129 sources detected in the full band
but not in the soft band, and there were 22 sources detected in the
full band but not in either the soft band or the hard band.}
\label{tbl-det}
\end{deluxetable}

\begin{deluxetable}{lllcccccccccc}
\tabletypesize{\scriptsize}
\tablewidth{0pt}
\tablecaption{Main {\it Chandra} Source Catalog}

\tablehead{
\colhead{XID}                    &
\colhead{RA}       &
\colhead{Dec}       &
\colhead{$\log P_{\rm B}$}       &
\colhead{{\sc wavdetect}}       &
\colhead{Pos Err}       &
\colhead{Off-axis}       &
\colhead{FB}          &
\colhead{FB Low Err}          &
\colhead{FB Upp Err}          &
\colhead{SB}          &
\colhead{SB Low Err}          &
\colhead{SB Upp Err}          \\
\colhead{(1)}         &
\colhead{(2)}         &
\colhead{(3)}         &
\colhead{(4)}         &
\colhead{(5)}         &
\colhead{(6)}         &
\colhead{(7)}         &
\colhead{(8)}         &
\colhead{(9)}        &
\colhead{(10)}        &
\colhead{(11)}        &
\colhead{(12)}        &
\colhead{(13)}
}

\startdata
       1    & 52.899178 & $-27.859588$ & $-99.0$ &$-8$ &  0.53 &    12.04 &    886.7 &     38.2& 39.3 & 604.3  & 28.2 & 29.4  \\
       2    & 52.911023 & $-27.892965$ & $-10.2$ &$-8$ &  1.08 &    12.15 &    \phantom{0}98.9 &     20.1& 21.2 & \phantom{0}67.3  & 12.3 & 13.5  \\
       3    & 52.917119 & $-27.796253$ & $-99.0$ &$-8$ &  0.66 &    10.67 &    245.2 &     22.0& 23.1 & 152.6  & 14.9 & 16.1  \\
       4    & 52.919726 & $-27.773984$ & \phantom{0}$-4.5$ &$-5$ &  1.01 &    10.69 &    \phantom{0}65.6 &     18.2& 19.4 & \phantom{0}29.8  & $-1.0$ & $-1.0$  \\
       5    & 52.920710 & $-27.743110$ & $-14.4$ &$-8$ &  0.97 &    11.12 &    \phantom{0}88.7 &     13.8& 15.0 & \phantom{0}32.3 & \phantom{0}7.5 & \phantom{0}8.7  
\enddata

\tablecomments{
The full table contains 73 columns of
information for the 1008 \hbox{X-ray} sources.
(This table is available in its entirety in the online journal. A portion is shown here
for guidance regarding its form and content.)}
\label{tbl-mcat}
\end{deluxetable}

\begin{deluxetable}{lllcccccccccc}
\tabletypesize{\scriptsize}
\tablewidth{0pt}
\tablecaption{Supplementary NIR Bright {\it Chandra} Source Catalog}

\tablehead{
\colhead{XID}                    &
\colhead{RA}       &
\colhead{Dec}       &
\colhead{$\log P_{\rm B}$}       &
\colhead{{\sc wavdetect}}       &
\colhead{Pos Err}       &
\colhead{Off-axis}       &
\colhead{FB}          &
\colhead{FB Low Err}          &
\colhead{FB Upp Err}          &
\colhead{SB}          &
\colhead{SB Low Err}          &
\colhead{SB Upp Err}          \\
\colhead{(1)}         &
\colhead{(2)}         &
\colhead{(3)}         &
\colhead{(4)}         &
\colhead{(5)}         &
\colhead{(6)}         &
\colhead{(7)}         &
\colhead{(8)}         &
\colhead{(9)}        &
\colhead{(10)}        &
\colhead{(11)}        &
\colhead{(12)}        &
\colhead{(13)}
}
\startdata
       1    & 52.925294 & $-27.763536$ & $-1.8$ &$-5$ &  1.09 &    10.53 &    47.9 &     20.4& 21.6 & 20.5  & $-1.0$ & $-1.0$  \\
       2    & 52.936641 & $-27.790331$ & $-1.5$ &$-6$ &  1.03 &    9.66 &    39.4&     23.9& 25.0 & 17.8  & 12.8 & 14.0  \\
       3    & 52.950038 & $-27.771467$ & $-1.8$ &$-6$ &  0.88 &    9.14 &    51.3 &     26.1& 27.3 & 24.0  & $-1.0$ & $-1.0$  \\
       4    & 52.975983 & $-27.732096$ & $-1.7$ &$-5$ &  0.77 &    8.74 &   64.7 &     27.0& 28.1 & 17.8  & 13.8 & 15.0  \\
       5    & 53.001139 & $-27.795137$ & $-1.9$ &$-5$ &  0.76 &    6.23 &   45.9 &  $-1.0$& $-1.0$ & 22.8 & 10.3 & 11.4
\enddata

\tablecomments{
The full table contains 73 columns of
information for the 47 supplementary \hbox{X-ray} sources.
(This table is available in its entirety in the online journal. A portion is shown here
for guidance regarding its form and content.)}
\label{tbl-scat}
\end{deluxetable}

\begin{deluxetable}{lccc}
\tabletypesize{\small}
\tablewidth{0pt}
\tablecaption{Flux Limits and Completeness Levels}

\tablehead{
\colhead{Completeness}                    &
\colhead{$f_{\rm 0.5-7\ keV}$}       &
\colhead{$f_{\rm 0.5-2\ keV}$}       &
\colhead{$f_{\rm 2-7\ keV}$}     \\
\colhead{(\%)} &
\colhead{(\flux)} &
\colhead{(\flux)} &
\colhead{(\flux)} 
}
\startdata
90 & $2.1\times 10^{-15}$ & $9.1\times 10^{-16}$ & $3.0\times 10^{-15}$ \\
80 & $9.1\times 10^{-16}$ & $3.7\times 10^{-16}$ & $1.2\times 10^{-15}$ \\
50 & $2.0\times 10^{-16}$ & $6.6\times 10^{-17}$ & $3.2\times 10^{-16}$ \\
20 & $5.9\times 10^{-17}$ & $2.1\times 10^{-17}$ & $8.7\times 10^{-17}$ 
\enddata
\label{tbl-comp}
\end{deluxetable}

\begin{deluxetable}{lcccc}

\tabletypesize{\small}
\tablecaption{Background Parameters}
\tablehead{
\colhead{Band (keV)}                                 &
\multicolumn{2}{c}{Mean Background}                 &
\colhead{Total Background}                                 &
\colhead{Count Ratio}                                \\
\cline{2-3}
\colhead{}                                 &
\colhead{(counts pixel$^{-1}$)}                         &
\colhead{(counts Ms$^{-1}$ pixel$^{-1}$)}               &
\colhead{(10$^5$ counts)}                &
\colhead{(background/source)} \\
\colhead{(1)}         &
\colhead{(2)}         &
\colhead{(3)}         &
\colhead{(4)}         &
\colhead{(5)}         
}
\tablewidth{0pt}
\startdata
Full (0.5--7.0)  & 0.600 & 0.184  & 43.2 & \phantom{0}8.2  \\
Soft (0.5--2.0)   & 0.173 & 0.055  & 12.5 & \phantom{0}4.2  \\
Hard (2--7)   & 0.427 & 0.127  & 30.7 & 13.8  
\enddata
\tablecomments{Column 1: X-ray band. Column 2: mean number of background counts per pixel
averaged across the background map (Section~\ref{sec-bkg}).
Column 3: mean number of background counts per pixel (Column 2)
divided by the mean effective exposure time averaged across the exposure map (Section~3.1);
the mean effective exposure times are 3.26~Ms, 3.18~Ms, and 3.37~Ms 
for the full, soft, and hard bands, respectively.
Column 4: total number of background counts measured from the background map.
Column 5: ratio of the total number of background 
counts (Column 4) to the total number of source counts in the main and supplementary catalogs.
}
\label{tbl-bkg}
\end{deluxetable}

\begin{deluxetable}{lcccccccc}

\tablecaption{Differential Number-Count Prior Best-Fit Model Parameters}
\tablehead{
\colhead{Band (keV)}                                 &
\multicolumn{4}{c}{AGNs}                 &
\multicolumn{2}{c}{Galaxies}                 &
\multicolumn{2}{c}{Stars}                 \\
\cmidrule(lr){2-5}
\cmidrule(lr){6-7}
\cmidrule(lr){8-9}
\colhead{}                                 &
\colhead{$K^{\rm AGN}_{14}$} &
\colhead{$\beta_1^{\rm AGN}$} &
\colhead{$\beta_2^{\rm AGN}$} &
\colhead{$f_{\rm break}^{\rm AGN}$} &
\colhead{$K^{\rm gal}_{14}$} &
\colhead{$\beta^{\rm gal}$} &
\colhead{$K^{\rm star}_{14}$} &
\colhead{$\beta^{\rm star}$} \\
\colhead{(1)}         &
\colhead{(2)}         &
\colhead{(3)}         &
\colhead{(4)}         &
\colhead{(5)}         &
\colhead{(6)}         &
\colhead{(7)}         &
\colhead{(8)}         &
\colhead{(9)}         
}
\tablewidth{0pt}
\startdata
Soft (0.5--2.0) & 161.96~$\pm$~7.10 & $1.52 \pm 0.03$ & $2.45 \pm 0.30$ & 7.1$^{+2.5}_{-2.4}$ & 2.01~$\pm$~0.10 & $2.24 \pm 0.06$&4.14~$\pm$~0.30 & 1.45$^{+0.13}_{-0.12}$\\
Hard (2--7) & 453.70~$\pm$~20.61 & $1.46 \pm 0.03$ & $2.72 \pm 0.24$ & $8.9 \pm 1.7$ & 0.75~$\pm$~0.09 & $2.56 \pm 0.15$&0.73~$\pm$~0.13 & 1.88$^{+0.36}_{-0.35}$
\enddata
\tablecomments{
Column 1: X-ray band. Columns 2--5: normalization in units of 
$10^{14}$~deg$^{-2}$~[\flux]$^{-1}$, faint-end slope, bright-end slope, and break flux
in units of $10^{-15}$ \flux\ for the AGN double power-law differential number-count model.
Columns 6--7: normalization in units of
$10^{14}$~deg$^{-2}$~[\flux]$^{-1}$ and power-law slope for the galaxy power-law
differential number-count model. Columns 8--9: normalization in units of
$10^{14}$~deg$^{-2}$~[\flux]$^{-1}$ and power-law slope for the star power-law
differential number-count model.}
\label{tbl-ncounts}
\end{deluxetable}

\end{document}